\documentclass[3p]{elsarticle} 

\myfooter[R]{First version: September 25, 2022. This version: February 8, 2024.}

\usepackage{amscd}      
\usepackage{amsbsy}     
\usepackage{amsmath}    
\usepackage{amsfonts}   
\usepackage{amsthm}     
\usepackage{enumerate}  
\usepackage{fancybox}   
\usepackage{geometry}   
\usepackage{setspace}   

\usepackage[linesnumbered,ruled,longend]{algorithm2e} 

\usepackage{tikz}       
\usetikzlibrary{arrows,calc,decorations,shapes}

\usepackage{cancel}     

\usepackage{caption}    
\usepackage{fancyhdr}   
\usepackage{float}      
\usepackage{graphics}   
\usepackage{graphicx}   
\usepackage{subcaption} 

\usepackage{bbm}        
\usepackage[OT1]{fontenc}

\usepackage{bookmark}   
\usepackage{hyperref}   
\usepackage{cleveref}   

\usepackage{hhline}     
\usepackage{longtable}  
\usepackage{multirow}   
\usepackage{pdflscape}  
\usepackage{tabularx}   

\renewcommand\footnoterule{\kern2pt\hrule width \textwidth height.25pt\kern4pt}         
\numberwithin{equation}{section}	                                                    

\DeclareMathOperator{\FX}{FX}
\DeclareMathOperator{\FXForward}{F}

\DeclareMathOperator{\LGD}{LGD}

\newcommand{\xva}{\text{xVA}}
\newcommand{\xVA}{\xva}

\DeclareMathOperator{\CVA}{CVA}

\DeclareMathOperator{\DVA}{DVA}

\DeclareMathOperator{\FVA}{FVA}
\DeclareMathOperator{\FCA}{FCA}
\DeclareMathOperator{\FBA}{FBA}

\DeclareMathOperator{\EPE}{EPE}

\DeclareMathOperator{\WWR}{WWR}

\newcommand{\EPEFVA}[2]{\EPE(#1;#2)}
\newcommand{\EPEFVAWWR}[2]{\EPE^{\WWR}(#1;#2)}
\newcommand{\EPEFVAIndep}[2]{\EPE^{\perp}(#1;#2)}
\newcommand{\FVAWWR}{\FVA^{\WWR}}
\newcommand{\FVAIndep}{\FVA^{\perp}}

\DeclareMathOperator{\PnL}{P\&L}

\newcommand{\half}{\frac{1}{2}}
\newcommand{\bigOh}{\mathcal{ O}}       
\renewcommand{\d}{{\rm d}}
\newcommand{\e}{{\rm e}}                
\newcommand{\E}{\mathbb{ E}}            
\newcommand{\F}{\mathcal{F}}            
\newcommand{\G}{\mathcal{G}}            
\renewcommand{\H}{\mathcal{H}}          
\newcommand{\I}{\mathbbm{1}}            

\newcommand{\N}{\mathcal{N}}            
\renewcommand{\P}{\mathbb{ P}}          
\newcommand{\Q}{\mathbb{ Q}}            

\newcommand{\R}{\mathbb{ R}}            

\def\dt{{\d}t}
\def\du{{\d}u}
\def\dv{{\d}v}

\def\dx{{\d}x}

\newcommand{\pderiv}[2]{\frac{\partial#1}{\partial #2}}

\newcommand{\maxOperator}[1]{\left( #1 \right)^+}

\newcommand{\indicator}[1]{\I_{\left\{#1\right\}}}
\newcommand{\sign}[1]{\mathrm{sgn}\left(#1\right)}
\newcommand{\expPower}[1]{\e^{#1}}
\newcommand{\expBrace}[1]{\exp{\left\{#1\right\}}}
\newcommand{\equalDistr}{\stackrel{\text{d}}{=}}

\newcommand{\rdef}{=:}
\newcommand{\ldef}{:=}

\newcommand{\var}{\mathbb{V}\text{ar}}      
\newcommand{\cov}{\mathbb{C}\text{ov}}      

\newcommand{\condExp}[2]{\E\left[ \left. #1 \right| \F(#2)\right]}

\newcommand{\condProbSmall}[2]{\P_{#2}\left[ #1 \right]}
\newcommand{\condExpSmall}[2]{\condExpSmallGeneric{#1}{#2}{}}
\newcommand{\condVarSmall}[2]{\condVarSmallGeneric{#1}{#2}{}}
\newcommand{\condCovSmall}[3]{\condCovSmallGeneric{#1}{#2}{#3}{}}

\newcommand{\condExpSmallGeneric}[3]{\E_{#2}^{#3}\left[ #1 \right]}
\newcommand{\condVarSmallGeneric}[3]{\var_{#2}^{#3}\left(#1\right)}
\newcommand{\condCovSmallGeneric}[4]{\cov_{#3}^{#4}\left(#1,#2\right)}

\newcommand{\etal}{\textit{et al. }}
\newcommand{\zeroRomanUpperCase}[1]{\ifcase #1 \relax 0 \else {\MakeUppercase{\romannumeral #1}}\fi}
\newcommand{\romanNumeralUpperCase}[1]{\text{\tiny{\zeroRomanUpperCase{#1}}}}

\newtheorem*{rem}{Remark}                   
\newtheorem{result}{Result}                 

\newcommand{\resultFigureSize}{0.49\linewidth}

\newcommand{\strike}{K}

\newcommand{\liquidity}{\ell}

\newcommand{\moment}{m}
\newcommand{\momentTruncNorm}{\overline{\moment}}
\newcommand{\notional}{N}
\newcommand{\shortRate}{r}

\newcommand{\zcb}{P}

\newcommand{\spread}{s}

\newcommand{\borrowingSpread}{\spread_b}

\newcommand{\taylor}{T}
\newcommand{\taylorTrunc}[2]{\taylor_{#1}^{#2}}
\newcommand{\tradeVal}{V}

\newcommand{\brownian}{W}

\newcommand{\dct}{\alpha}

\newcommand{\swapType}{\delta}
\newcommand{\errorTerm}{\varepsilon}
\newcommand{\error}[1]{\errorTerm_{\romanNumeralUpperCase{#1}}}
\newcommand{\errorWWRPartOne}{\errorTerm^{\WWR,1}}
\newcommand{\errorWWRPartTwo}{\errorTerm^{\WWR,2}}
\newcommand{\errorWWRPartThree}{\errorTerm^{\WWR,3}}
\newcommand{\errorIndep}{\errorTerm^{\perp}}

\newcommand{\intensity}{\lambda}

\newcommand{\corr}{\rho}
\newcommand{\vol}{\sigma}
\newcommand{\default}{\tau}
\newcommand{\normPDF}{\phi}

\newcommand{\normCDF}{\Phi}

\title{Efficient Wrong-Way Risk Modelling for Funding Valuation Adjustments}

\begin{document}

\author[1,2]{Thomas van der Zwaard\corref{cor1}}
\ead{T.vanderZwaard@uu.nl}
\author[1,2]{Lech A.~Grzelak}
\ead{L.A.Grzelak@uu.nl}
\author[1]{Cornelis W.~Oosterlee}
\ead{C.W.Oosterlee@uu.nl}
\cortext[cor1]{Corresponding author at Mathematical Institute, Utrecht University, Utrecht, the Netherlands.}
\address[1]{Mathematical Institute, Utrecht University, Utrecht, the Netherlands}
\address[2]{Rabobank, Utrecht, the Netherlands}

\begin{abstract}
    \noindent Wrong-Way Risk (WWR) is an important component in Funding Valuation Adjustment ($\FVA$) modelling.
    Yet, the standard assumption is independence between market risks and the counterparty defaults and funding costs.
    This typical industrial setting is our point of departure, where we aim to assess the impact of WWR without running a full Monte Carlo simulation with all credit and funding processes.
    We propose to split the exposure profile into two parts: an independent and a WWR-driven part.
    For the former, exposures can be re-used from the standard $\xva$ calculation.
    We express the second part of the exposure profile in terms of the stochastic drivers and approximate these by a common Gaussian stochastic factor.
    Within the affine setting, the proposed approximation is generic, is an add-on to the existing $\xVA$ calculations and provides an efficient and robust way to include WWR in $\FVA$ modelling.
    Case studies for an interest rate swap and a representative multi-currency portfolio of swaps illustrate that the approximation method is applicable in a practical setting.
    We analyze the approximation error and use the approximation to compute WWR sensitivities, which are needed for risk management.
    The approach is equally applicable to other metrics such as Credit Valuation Adjustment.
\end{abstract}

\begin{keyword}
    Gaussian approximation \sep Wrong-Way Risk (WWR) \sep Funding Valuation Adjustment ($\FVA$) \sep computational finance \sep risk management
\end{keyword}
\maketitle

{\let\thefootnote\relax\footnotetext{The views expressed in this paper are the personal views of the authors and do not necessarily reflect the views or policies of their current or past employers. The authors have no competing interests.}}

\section{Introduction}  \label{sec:introduction}

Funding Valuation Adjustments ($\FVA$) are used in financial derivatives pricing to include the funding costs of uncollateralized deals.
When transactions are not fully collateralized, $\FVA$ captures the funding cost of hedging the market risk of these transactions.
Wrong-Way Risk (WWR) should be included in $\FVA$ modelling, and can be understood as an increase in the funding risk due to increased market risk (exposure).
For a qualitative discussion about $\FVA$, WWR, and its importance, see, for example,~\cite{ZwaardGrzelakOosterlee202210}.
Going forward, $\FVA$ WWR is referred to as WWR.
As a starting point, assume that a Valuation Adjustment ($\xva$) calculation is in place where WWR is not included.
We develop an efficient methodology to compute $\FVA$ WWR, but avoid a full Monte Carlo simulation including additional risk factors driving the counterparty defaults and funding spread, similar to~\cite{Moni201411}.
In this fashion, it is possible to assess WWR effects when the existing $\xva$ calculation cannot simulate stochastic credit and funding correlated to the existing set of risk factors.

The existing literature focuses on WWR modelling and its efficient computation.
Kenyon \etal tackle WWR with a model-independent approach by rewriting the $\xva$ expressions into separate components to assess the various contributions to correlation effects~\cite{KenyonBerrahouiPoncet202210}.
Different WWR modelling approaches can be compared using this framework.
Brigo and Pallavicini touch upon the topic in a symmetric funding case in a Monte Carlo BSDE setting~\cite{BrigoPallavicini201405}.
Green indicates the natural choice is a stochastic funding spread, where the dependence with the underlying asset is introduced through Gaussian correlation~\cite{Green201511}, e.g., see also~\cite{Valsecchi202104}, with an idiosyncratic/systemic decomposition of the stochastic funding spread.
Moni adds correlation among credit spreads, funding spreads and market risk factors using a polynomial delta-gamma approximation~\cite{Moni201411}.
This approach avoids a full Monte Carlo simulation where the credit and funding spreads are also simulated.
Alternatively, one can focus on extreme events like Turlakov~\cite{Turlakov201303} or consider worst-case $\FVA$ like Singh and Zhang~\cite{SinghZhang202005}.

To put $\FVA$ WWR in perspective, we compare it with Credit Valuation Adjustment ($\CVA$) WWR.
$\CVA$ WWR is introduced through a dependence between exposure and default probabilities, which will increase the $\CVA$.
Literature on WWR modelling for $\CVA$ for non-credit derivatives can be divided into three approaches.~\footnote{For credit derivatives, a separate stream of literature exists, but in this paper we do not focus on this class of derivatives.}
Firstly, there is simulation, which models interest rates, default intensities and their dependence through either a deterministic relationship or a set of correlated SDEs~\cite{FengOosterlee201610,HullWhite201108}.
Secondly, a copula approach where the multivariate distribution of exposure and defaults is modelled through a copula~\cite{CernyWitzany201802,Cherubini201306}. 
Finally, a worst-case method provides an upper bound for the $\CVA$ without requiring exact knowledge of the dependence structure of exposures and default dynamics~\cite{GlassermanYang201611,KenyonGreen201610}. 
These approaches could also be extended to $\FVA$ WWR.

In this paper, we essentially follow Green~\cite{Green201511} and Moni~\cite{Moni201411}, but we propose a Gaussian WWR approximation, which allows for efficient and robust WWR computation in a generic fashion.
We extend our previous work~\cite{ZwaardGrzelakOosterlee202210} by focusing on the quantitative computational challenges and avoiding simulating extra (correlated) dynamics for credit and funding spreads.
We assume that all short-rate dynamics are of the affine form.
Credit and funding spread dynamics are chosen and correlated to the existing risk factor dynamics that drive the exposure.
We split the exposure profile into two parts: an independent and a WWR-driven part.
For the first part, already computed exposures are used.
We express the second part of the exposure profile in terms of the stochastic drivers, and approximate these by a common Gaussian stochastic factor.
The approximation also provides intuition on WWR.
We analyze the approximation quality and indicate in which situations the approximation works well.
We use a brute-force Monte Carlo approach as a benchmark, where all risk factors are simulated.
Case studies are presented for an uncollateralized interest rate (IR) swap, as well as a representative multi-currency portfolio of swaps.
We show that the approximation method is applicable in a practical setting due to its generic nature.
Hence, the approximation can be applied to portfolios, including other products and asset classes.
The approach is introduced for $\FVA$, but is equally applicable to other metrics such as $\CVA$.

When computing $\FVA$, calculating the relevant sensitivities is particularly important from a risk-management perspective.
The cross-gamma risk between funding spreads and exposure is a sensitivity which is introduced by including WWR in $\FVA$ modelling.
The effects are two-fold: the WWR premium is compensation for re-hedging at expensive moments and recognizing when to re-hedge to have low hedging costs~\cite{ZwaardGrzelakOosterlee202210}.
Furthermore, the WWR sensitivities are needed in the $\xva$ $\PnL$ explain process~\cite{ZwaardGrzelakOosterlee202102} to keep the unexplained $\PnL$ as low as possible.
The proposed Gaussian approximation enables a fast approximation of the sensitivities so that hedges in the market can be set up early.

In Section~\ref{sec:wwr}, we introduce the modelling framework with the $\FVA$ equation, correlation assumptions, funding spreads and WWR as in~\cite{ZwaardGrzelakOosterlee202210}.
Then, in Section~\ref{sec:approximation}, we propose our Gaussian approximation.
We analyze the approximation error in Section~\ref{sec:approxError}.
Numerical experiments are presented in Section~\ref{sec:numericalResults}, and finally, in Section~\ref{sec:conclusion}, we conclude.

\section{FVA and Wrong-Way Risk}  \label{sec:wwr}
The framework is summarized in Section~\ref{sec:setupSummary}, covering the $\FVA$ equation, correlation assumptions, funding spread definitions and exposure split between the independent and WWR part.
Funding costs and benefits are assumed to be asymmetric: when borrowing funds, a spread over the risk-free rate is paid, but when lending out funds, only the risk-free rate is earned.
As a consequence $\FVA(t) = \FCA(t)$, since $\FBA(t) = 0$.~\footnote{Here, $\FCA$ and $\FBA$ represent the Funding Cost Adjustment and Funding Benefit Adjustment, respectively.}
The framework is summarized in Section~\ref{sec:setupSummary}.

In Section~\ref{sec:fvaEquationFundingSpread}, we extend the framework to prepare for the Gaussian approximation of the WWR exposure from Section~\ref{sec:approximation}.
We use Taylor series expansions of the interest rate (IR) discount factors and credit adjustment factors that appear in the exposure formulae.

\subsection{The setup} \label{sec:setupSummary}
The framework presented in this section is identical to the setup and results from~\cite{ZwaardGrzelakOosterlee202210}.
All relevant results are re-iterated for convenience.

We assess the impact of WWR on the $\FVA$ for a portfolio of uncollateralized derivatives, $\tradeVal$, between counterparty $C$ and institution $I$, with maturity $T$.
Values are denoted from $I$'s perspective.
The borrowing spread, $\borrowingSpread >0$, is an instantaneous spread over the risk-free rate $\shortRate$, and will be the funding costs used to compute $\FVA$.
Default times $\default_z$, $z\in\{I, \ C\}$, are modelled as the first jump of a Cox process~\cite{BrigoMercurio2006}, driven by hazard rate $\intensity_z$.

We look at the processes $\intensity_I$ and $\intensity_C$, and their correlation structure with the IR risk factor $\shortRate$, introduced via correlated Brownian increments. 
Next, the $\FVA$ equation is given and split into an independent and a WWR part. 
The same split is done for the corresponding exposures. 
Furthermore, we assume a funding spread of a particular form, such that exposures can be rewritten to a convenient form in Section~\ref{sec:fvaEquationFundingSpread}. 

\subsubsection{Default processes, model dynamics and correlations} \label{sec:SDE}
We work with affine short-rate dynamics~\cite{BrigoMercurio2006} for interest rate $\shortRate$ and hazard rates $\intensity_I$ and $\intensity_C$, where the integrated dynamics fit in the following generic setup:
\begin{align} \displaystyle
  \overline{z}(u)
    &= x_{z}(u) + b_{z}(u), \ \
  x_{z}(u)
    = \mu_{z}(t, u) + y_{z}(t,u), \ \
  \int_{t}^{u} x_{z}(v) \dv
    = M_{z}(t, u) + Y_{z}(t,u). \label{eq:shortRateFramework}
\end{align}
Here, $\overline{z} \in \{ \shortRate,\ \intensity_I, \ \intensity_C\}$ and $z \in \{\shortRate,\ I,\ C\}$.
Furthermore, $b_{z}(u)$, $\mu_{z}(t, u)$, and $M_{z}(t, u)$ are deterministic quantities.
Respectively, $y_{z}(t,u)$ and $Y_{z}(t,u)$ are stochastic and integrated stochastic processes, with $\condExp{y_{z}(t,u)}{t} = \condExp{Y_{z}(t,u)}{t} = 0$.

We introduce a dependence between the credit and IR processes through correlated Brownian increments in $y_{z}(t,u)$, as in~\cite{GarciaMunoz201312}, rather than using a copula, as in~\cite{BrigoPallaviciniPapatheodorou201107}. 
For IR derivatives, the main WWR driver will be the dependence between the funding spread and the IR exposure~\cite{ZwaardGrzelakOosterlee202210}.
Hence, we choose to work with independent defaults of the counterparties $I$ and $C$.
This will slightly restrict correlations between the other risk factors for the correlation matrix to remain symmetric positive definite.
Survival probabilities can be factorized due to this independent default assumption, which we will use in Section~\ref{sec:fvaEquation}.

The correlation assumptions can also be formulated in terms of the Brownian motions $\brownian_z(t)$ of the stochastic processes:
\begin{align} \displaystyle
  \brownian_{\shortRate}(t) \brownian_I(t) = \corr_{\shortRate,I} t, \
  \brownian_{\shortRate}(t) \brownian_C(t) = \corr_{\shortRate,C} t, \
  \brownian_I(t) \brownian_C(t) = \corr_{I,C} t = 0. \nonumber
\end{align}
Correlations $\corr_{\shortRate,z}$, $z \in \{I,\ C\}$, can be calibrated to historical data or can be mapped to alternative data in the case of illiquid counterparties~\cite{Green201511}.
The latter correlation, $\corr_{I,C}$, is zero due to the independent default assumption.

\begin{rem}[Credit derivatives]
    Default times $\default_I$ and $\default_C$ are assumed to be conditionally independent on $\F(t)$, i.e., $\corr_{I,C}(t) = 0$.
    This assumption cannot be justified when considering credit derivatives, where the dependence between counterparties is a crucial component that must be captured in any modelling approach.
    Therefore, credit derivatives cannot be handled straight away, even though dynamics for the credit processes are already specified.
\end{rem}

\subsubsection{FVA equation} \label{sec:fvaEquation}

For the $\FVA$ of portfolio $\tradeVal$, the following expression can be derived, see~\cite{ZwaardGrzelakOosterlee202210}, under the assumption of conditional independence of defaults, and using the instantaneous spread $\borrowingSpread(u)$:
\\[1.0ex]
\noindent\fbox{\parbox[][90pt][c]{0.98\textwidth}{
    \begin{align} \displaystyle
      \FVA(t)
        &= \E \left[ \left. \int_t^{T\wedge\default_I\wedge\default_C} \expPower{-\int_{t}^{u} \shortRate(v)\dv} \borrowingSpread(u) \maxOperator{\tradeVal(u)} \du \right| \G(t) \right] \label{eq:fca0}\\
        &=  \int_t^{T} \condExp{ \expPower{-\int_{t}^{u} \intensity_I(v) + \intensity_C(v)\dv}\expPower{-\int_{t}^{u} \shortRate(v)\dv} \borrowingSpread(u) \maxOperator{\tradeVal(u)} }{t} \du\nonumber \\
        &\rdef  \int_t^{T} \EPEFVA{t}{u} \du,  \label{eq:fca1}
    \end{align}
}}\\[1.0ex]
where $\EPEFVA{t}{u}$ denotes the time $t$ value of the time $u$ Expected Positive Exposure, which is a credit- and funding-adjusted discounted positive exposure.~\footnote{In typical notation, the credit and funding components are not part of the $\EPE$, but they are included here for ease of notation.}
Furthermore, filtration $\F(t)$ is the `standard' default-free filtration; $\G(t) := \F(t) \otimes \H_I(t) \otimes \H_C(t)$ is the enriched filtration with all available market information; $\H_z(t) = \sigma\left(\left\{\default_z \leq s \right\}:s \leq t\right)$ is the filtration generated by the default time $\default_z$ for $z \in \{I,\ C\}$, see~\cite{BrigoMercurio2006}.
Furthermore, $\maxOperator{x} = \max \{x, 0\}$.

From Equation~\eqref{eq:fca0} it is clear that if one of the parties defaults before $T$, funding only needs to take place up to the default date.
This translates into credit adjustment factors $\expPower{-\int_{t}^{u} \intensity_I(v)\dv} < 1$  and $\expPower{-\int_{t}^{u}\intensity_C(v)\dv} < 1$ for the potential default of $I$ and $C$, which can be interpreted as survival probabilities.
These credit adjustment factors can change the $\FVA$ magnitude significantly, with increased effects for lower credit quality.
Hence, including default times $\default_I$ and $\default_C$ in the $\FVA$ definition and modelling is an important consideration.
$\FVA$ including these credit adjustment factors is also known as `contingent' $\FVA$ and is consistent with a closeout based on the risk-free value~\cite{Gregory202007}.
In addition, these credit adjustment factors influence the dependence structure, introducing additional WWR effects.
Therefore, including or excluding the default times is also relevant from a hedging perspective.

\begin{rem}[Asymmetric funding assumption]
    The assumption of asymmetric funding can be made to avoid double counting of funding benefits from $\FBA$ and $\DVA$~\cite{Gregory202007}.
    In general, this assumption leads to a recursive and non-linear pricing equation, where the derivative price depends on the future funding requirement, such that the $\xva$ is the solution of a BSDE~\cite{PallaviciniPeriniBrigo201112}.
    The consequence is that $\FVA$ can no longer be considered as an additive $\xva$ term.
    However, the BSDE formulation is not strictly needed unless the closeout is on the risky pre-default value~\cite{BichuchCapponiSturm202003}.
    Therefore, we can safely assume asymmetric funding.
    When funding and borrowing rates are equal (symmetric funding), the pricing equation is no longer recursive and $\FVA$ becomes additive.
\end{rem}

\subsubsection{Funding spread} \label{sec:fundingSpread}
When choosing a funding spread, a key requirement is that it is in line with the ability of a financial institution to fund itself in the market.
A stochastic funding spread takes into account the credit of $I$, through hazard rate $\intensity_I(t)$.
We assume the information on $I$'s credit is extracted from the CDS market.
Instantaneous borrowing spread $\borrowingSpread(t)$ is now defined as~\cite{Green201511}:
\begin{align} \displaystyle
  \borrowingSpread(t)
    &= \LGD_I \intensity_I(t), \label{eq:fundingSpread}
\end{align}
with a deterministic loss given default $\LGD_I$, which is a standard assumption~\cite{OKane2008}.
For a stochastic loss given default, additional WWR will be introduced, but this is outside the scope of this paper as it is not common practice.
Similar as in~\cite{ZwaardGrzelakOosterlee202210}, we assume $\intensity_I(t)$ in Equation~\eqref{eq:fundingSpread} is stochastic, with model dynamics of the form~\eqref{eq:shortRateFramework}.
Furthermore, $\intensity_I(t)$ is correlated with the other underlyings, which results in WWR.
We split the borrowing spread into a deterministic part, $\mu_{S}(t, u)$, and a stochastic part, $y_{I}(t,u)$, as in Equation~\eqref{eq:shortRateFramework}:
\begin{align} \displaystyle
  \borrowingSpread(u)
    &= \LGD_I\left[x_I(u) + b_I(u)\right] \nonumber \\
    &= \LGD_I \left[\mu_{I}(t, u) + b_I(u)\right]  + \LGD_I y_{I}(t,u) \nonumber \\
    &\rdef \mu_{S}(t, u) + \LGD_I y_{I}(t,u). \label{eq:fundingSpreadCredit}
\end{align}
A (deterministic) liquidity adjustment $\liquidity(t)$ could be added to Equation~\eqref{eq:fundingSpread} to reflect the bond-CDS basis in the borrowing spread~\citep{ZwaardGrzelakOosterlee202210}.
This adjustment will naturally be absorbed by $\mu_{S}(t, u)$ in Equation~\eqref{eq:fundingSpreadCredit}, and none of the results to follow will change.

\subsubsection{Splitting the FVA equation} \label{sec:fvaSplit}
Going forward, we write $\condExp{\cdot}{t} = \condExpSmall{\cdot}{t}$ for ease of notation.
For exposure $\EPEFVA{t}{u}$ in Equation~\eqref{eq:fca1}, we consider a generic form:
\begin{align} \displaystyle
  \EPEFVA{t}{u}
    & = \condExpSmall{f(t,u;\intensity_I,\intensity_C) g(t,u;\shortRate) h(t,u;\shortRate,\tradeVal) }{t}, \label{eq:epe1}
\end{align}
where $f(\cdot)$, $g(\cdot)$ and $h(\cdot)$ depend on the $\FVA$ modelling assumptions.
For the funding spread from Section~\ref{sec:fundingSpread}, $f(\cdot)$ in Eq.~\eqref{eq:epe2} is defined as
\begin{align} \displaystyle
    f(t,u;\intensity_I,\intensity_C)
      &= \expPower{-\int_{t}^{u} \intensity_I(v) + \intensity_C(v)\dv} \borrowingSpread(u), \nonumber
\end{align}
so that
\begin{align} \displaystyle
    \condExpSmall{f(t,u;\intensity_I,\intensity_C)}{t}
      &= \zcb_{I}(t,u) \zcb_{C}(t,u)\mu_{S}(t, u) + \LGD_I \condExpSmall{\expPower{-\int_{t}^{u} \intensity_I(v) + \intensity_C(v)\dv} y_I(t,u)}{t},  \nonumber
\end{align}
with Zero Coupon Bond (ZCB) $\zcb_{z}(t,u) = \condExpSmall{\expPower{-\int_{t}^{u} \overline{z}(v)\dv}}{t}$.
Due to the independence assumption from Section~\ref{sec:SDE}, $\zcb_{I}(t,u)$ and $\zcb_{C}(t,u)$ are independent.
Furthermore, $g(t,u;\shortRate) = 1$, and $h(t,u;\shortRate,\tradeVal) = \expPower{-\int_{t}^{u} \shortRate(v)\dv}\maxOperator{\tradeVal(u)}$, such that $\condExpSmall{h(t,u;\shortRate,\tradeVal)}{t}$ is simply the discounted positive exposure, which is readily available from an existing $\xva$ calculation.

Eq.~\eqref{eq:epe1} can be rewritten by decomposing it into covariances:
\\[1.0ex]
\noindent\fbox{\parbox[][50pt][c]{0.98\textwidth}{
\begin{align} \displaystyle
  \EPEFVA{t}{u}
    &= \EPEFVAIndep{t}{u} + \EPEFVAWWR{t}{u}, \label{eq:epe2} \\
  \EPEFVAIndep{t}{u}
    &\ldef \condExpSmall{f}{t} \condExpSmall{g}{t} \condExpSmall{h}{t}, \nonumber \\
  \EPEFVAWWR{t}{u}
    &\ldef \condExpSmall{ h}{t} \condCovSmall{f}{g}{t} + \condExpSmall{\left(h - \condExpSmall{h}{t} \right)\left(f g - \condExpSmall{f}{t} \condExpSmall{g}{t} - \condCovSmall{f}{g}{t}\right)}{t}. \nonumber
\end{align}
}}\\[1.0ex]
The overall exposure $\EPEFVA{t}{u}$ is then split into an independent exposure, $\EPEFVAIndep{t}{u}$, and a WWR exposure, $\EPEFVAWWR{t}{u}$, which is driven by the correlation assumptions and captures cross-dependence.
Rather than decomposing into covariances, the expectation could be decomposed into correlations and the corresponding standard deviations, see~\cite{KenyonBerrahouiPoncet202210}.
However, for us it appears more convenient to work with the covariance decomposition.

Equivalently, substituting Equation~\eqref{eq:epe2} into Equation~\eqref{eq:fca1} yields
\\[1.0ex]
\noindent\fbox{\parbox[][27pt][c]{0.98\textwidth}{
\begin{align} \displaystyle
  \FVA(t)
    &= \int_t^{T} \EPEFVAIndep{t}{u}\du + \int_t^{T}\EPEFVAWWR{t}{u} \du
    \rdef \FVAIndep(t) + \FVAWWR(t),  \label{eq:fca2}
\end{align}
}}\\[1.0ex]
such that $\FVA$ is split in a similar fashion as the exposure.

\subsection{FVA exposure for a given funding spread} \label{sec:fvaEquationFundingSpread}
Given the funding spread definition, the functions $f(\cdot)$, $g(\cdot)$ and $h(\cdot)$ have an explicit form.
The independent and WWR exposure from Equation~\eqref{eq:epe2} can be rewritten using these functional forms.
We use Taylor series approximations to rewrite these expressions in a convenient form with the stochastic drivers $y_{z}(t,u)$ and $Y_{z}(t,u)$ as introduced in Equation~\eqref{eq:shortRateFramework}.
This step is done to prepare for the Gaussian approximation in Section~\ref{sec:approximation}.

\subsubsection{Taylor series approximations} \label{sec:taylor}
The affine dynamics from Section~\ref{sec:SDE} imply that the ZCB under the model of choice can be written as
\begin{align} \displaystyle
  \zcb_{z}(t,T)
    &= \expPower{A_{z}(t,T) - x_{z}(t)B_{z}(t,T) - \int_{t}^{T} b_{z}(v) \dv}
    = \expPower{\overline{A}_{z}(t,T) - x_{z}(t)B_{z}(t,T)}, \label{eq:zcb}
\end{align}
where $z \in \{\shortRate,\ I,\ C\}$ and $\overline{A}_{z}(t,T) \ldef A_{z}(t,T) - \int_{t}^{T} b_{z}(v) \dv$.
We introduce the notation $H_z(t,u)$ for a deterministic factor, to split $\expPower{-\int_{t}^{u} \overline{z}(v) \dv}$ into a deterministic and stochastic part:
\begin{align} \displaystyle
  \expPower{-\int_{t}^{u} \overline{z}(v) \dv}
    &= H_z(t,u) \expPower{-Y_z(t,u) }, \label{eq:HzSingle}
\end{align}
with $z \in \{\shortRate,\ I,\ C\}$ and $Y_z(t,u)$ as defined in Equation~\eqref{eq:shortRateFramework}.
$H_z(t,u)$ can be interpreted as a discount factor for $z$.
This modelling framework is flexible regarding various choices for the affine short-rate dynamics.
We keep the derivations generic and express the results in terms of the stochastic processes $y_{z}(t,u)$ and their integrated version $Y_z(t,u)$.
For notational convenience, the product of $n$ deterministic factors $H_z(t,u)$ is denoted as
\begin{align} \displaystyle
  H_{z_1, \ldots, z_n}(t,u)
    &:= H_{z_1}(t,u)\cdots H_{z_n}(t,u). \label{eq:HzMultiple}
\end{align}
Next, we denote the Taylor series expansion $\taylor(x)$ of $\expPower{-x}$ as
\begin{align} \displaystyle
  \taylor(x)
    &\ldef \sum_{j=0}^{\infty} \frac{(-x)^j}{j!}
    = \sum_{j=0}^{n} \frac{(-x)^j}{j!} + \sum_{j=n+1}^{\infty} \frac{(-x)^j}{j!}
    \rdef \taylorTrunc{0}{n}(x) + \taylorTrunc{n+1}{\infty}(x). \label{eq:taylor0}
\end{align}
Combining the other notation from this section with the Taylor expansion yields:
\begin{align} \displaystyle
  \expPower{-\int_{t}^{u} \shortRate(v)\dv}
    &= H_{\shortRate}(t,u) \taylor(Y_{\shortRate}(t,u))
    = H_{\shortRate}(t,u) \left[\taylorTrunc{0}{n_{\shortRate}}(Y_{\shortRate}(t,u)) + \taylorTrunc{n_{\shortRate} + 1}{\infty}(Y_{\shortRate}(t,u))\right], \label{eq:taylor3}\\
  \expPower{-\int_{t}^{u} \intensity_I(v) + \intensity_C(v)\dv}
    &= H_{I,C}(t,u) \left[\taylorTrunc{0}{n_{\intensity}}(Y_I(t,u) + Y_C(t,u)) + \taylorTrunc{n_{\intensity} + 1}{\infty}(Y_I(t,u) + Y_C(t,u))\right]. \label{eq:taylor4}
\end{align}
These two expansions will be used in the approximations that follow in Sections~\ref{sec:fvaEquationFundingSpreadWithoutWWR} and~\ref{sec:fvaEquationFundingSpreadWithWWR}.

\subsubsection{Independent exposure} \label{sec:fvaEquationFundingSpreadWithoutWWR}
Combining the funding spread assumption from Equation~\eqref{eq:fundingSpreadCredit} with the notation in Section~\ref{sec:taylor}, independent exposure $\EPEFVAIndep{t}{u}$ from Equation~\eqref{eq:epe2} is written as
\begin{align} \displaystyle
  \EPEFVAIndep{t}{u}
    &= \zcb_{I}(t,u) \zcb_{C}(t,u)\mu_{S}(t, u)\condExpSmall{\expPower{-\int_{t}^{u} \shortRate(v)\dv}\maxOperator{\tradeVal(u)}}{t}  \nonumber \\
    &\quad + \LGD_I\condExpSmall{\expPower{-\int_{t}^{u} \intensity_I(v) + \intensity_C(v)\dv} y_I(t,u)}{t} \condExpSmall{\expPower{-\int_{t}^{u} \shortRate(v)\dv}\maxOperator{\tradeVal(u)}}{t},\label{eq:epeIndepCredita}
\end{align}
where the second term captures the dependence between the credit randomness driving the borrowing spread and the default probabilities.
Using the Taylor expansion from Equation~\eqref{eq:taylor4} with $n_{\intensity}=1$, which is a first-order expansion, $\EPEFVAIndep{t}{u}$ gives us:
\\[1.0ex]
\noindent\fbox{\parbox[][40pt][c]{0.98\textwidth}{
\begin{align} \displaystyle
  \EPEFVAIndep{t}{u}
    &= \zcb_{I}(t,u) \zcb_{C}(t,u)\mu_{S}(t, u)\condExpSmall{\expPower{-\int_{t}^{u} \shortRate(v)\dv}\maxOperator{\tradeVal(u)}}{t}  \nonumber \\
    &\quad - \LGD_I H_{I,C}(t,u)\condExpSmall{ Y_I(t,u) y_I(t,u)}{t} \condExpSmall{\expPower{-\int_{t}^{u} \shortRate(v)\dv}\maxOperator{\tradeVal(u)}}{t} + \errorIndep, \label{eq:epeIndepCreditb}
\end{align}
}}\\[1.0ex]
where $\errorIndep$ is the truncation error from cutting off the Taylor series expansion:
\begin{align} \displaystyle
  \errorIndep
    &\ldef \LGD_I H_{I,C}(t,u)\condExpSmall{\taylorTrunc{2}{\infty}(Y_I(t,u) + Y_C(t,u)) y_I(t,u)}{t} \condExpSmall{\expPower{-\int_{t}^{u} \shortRate(v)\dv}\maxOperator{\tradeVal(u)}}{t} . \label{eq:errorIndep}
\end{align}
An approximation for $\EPEFVAIndep{t}{u}$ is obtained by omitting the $\errorIndep$ term in Equation~\eqref{eq:epeIndepCreditb}.
The corresponding approximation error results from truncating the Taylor series expansion, by discarding the product $\taylorTrunc{2}{\infty}(Y_I(t,u) + Y_C(t,u)) y_I(t,u)$.
The choice of $n_{\intensity}=1$ means that no calibration needs to be performed for this approximation parameter.

\subsubsection{WWR exposure} \label{sec:fvaEquationFundingSpreadWithWWR}
In a similar fashion as the independent exposure, WWR exposure $\EPEFVAWWR{t}{u}$ from Equation~\eqref{eq:epe2} can be rewritten:
\begin{align} \displaystyle
  \EPEFVAWWR{t}{u}
    &= \condExpSmall{\left(\expPower{-\int_{t}^{u} \shortRate(v)\dv}\maxOperator{\tradeVal(u)} - \condExpSmall{\expPower{-\int_{t}^{u} \shortRate(v)\dv}\maxOperator{\tradeVal(u)}}{t} \right)\expPower{-\int_{t}^{u} \intensity_I(v) + \intensity_C(v)\dv}\borrowingSpread(u) }{t}. \label{eq:epeWWRCredit1}
\end{align}
If IR and credit are independent, then clearly $\EPEFVAWWR{t}{u}=0$.

We take the WWR exposure from Equation~\eqref{eq:epeWWRCredit1} and apply the modelling choices from Section~\ref{sec:SDE}, the funding spread from Equation~\eqref{eq:fundingSpreadCredit} together with the two Taylor expansions from Equations~\eqref{eq:taylor3} and~\eqref{eq:taylor4} with $n_{\intensity}=1$, resulting in:
\\[1.0ex]
\noindent\fbox{\parbox[][94pt][c]{0.98\textwidth}{
\begin{align} \displaystyle
  \EPEFVAWWR{t}{u}
    &=  H_{\shortRate,I,C}(t,u) \mu_{S}(t, u)  \condExpSmall{\taylorTrunc{0}{n_{\shortRate}}(Y_{\shortRate}(t,u)) \left(-Y_I(t,u) - Y_C(t,u)\right) \maxOperator{\tradeVal(u)}}{t} \nonumber \\
    &\quad +  \LGD_I  H_{\shortRate,I,C}(t,u) \condExpSmall{\taylorTrunc{0}{n_{\shortRate}}(Y_{\shortRate}(t,u)) y_I(t,u)\left( 1 - Y_I(t,u)- Y_C(t,u) \right)\maxOperator{\tradeVal(u)}  }{t}  \nonumber \\
    &\quad  +  \LGD_I H_{I,C}(t,u) \condExpSmall{Y_I(t,u)y_I(t,u)}{t}  \condExpSmall{\expPower{-\int_{t}^{u} \shortRate(v)\dv}\maxOperator{\tradeVal(u)}}{t}    \nonumber \\
    &\quad  + \errorWWRPartOne, \label{eq:epeWWRCredit2} \\
  \errorWWRPartOne
    &\ldef \mu_{S}(t, u) \left[\error{1}(1) + \error{2}(1)\right] + \LGD_I \left[\error{1}(y_I) + \error{2}(y_I) + \error{3}(y_I)\right], \label{eq:errorWWR1}
\end{align}
}}\\[1.0ex]
where $H_{\shortRate,I,C}(t,u)$ and $H_{I,C}(t,u)$ are as in Equation~\eqref{eq:HzMultiple}.
In Equation~\eqref{eq:errorWWR1}, we have defined the generic truncation error $\errorWWRPartOne$, which is the result of truncating multiple Taylor series.
An approximation for $\EPEFVAWWR{t}{u}$ is obtained by omitting $\errorWWRPartOne$ in Equation~\eqref{eq:epeWWRCredit2}.
See~\ref{app:fvaExposureDerivationCredit} for a derivation of the WWR exposure.

First, $\errorWWRPartOne$ contains the scaled truncation errors, $\mu_{S}(t, u) \error{1}(1) + \LGD_I \error{1}(y_I)$, where
\begin{align} \displaystyle
  \error{1}(x)
    &\ldef H_{I,C}(t,u)\cdot \nonumber \\
    &\qquad \condExpSmall{\left(\expPower{-\int_{t}^{u} \shortRate(v)\dv}\maxOperator{\tradeVal(u)} - \condExpSmall{\expPower{-\int_{t}^{u} \shortRate(v)\dv}\maxOperator{\tradeVal(u)}}{t}\right)\taylorTrunc{2}{\infty}(Y_I(t,u) + Y_C(t,u)) \cdot x}{t}. \label{eq:error1}
\end{align}
Error $\mu_{S}(t, u) \error{1}(1)$ results from removing the terms $\taylorTrunc{2}{\infty}(Y_I(t,u) + Y_C(t,u))$, which corresponds to the Taylor series expansion of  $\expPower{-Y_I(t,u)-Y_C(t,u)}$, combined with the deterministic part of the funding spread, i.e., $\mu_{S}(t, u)$.
Error $\LGD_I \error{1}(y_I)$ is a consequence of omitting the product $y_I(t,u) \taylorTrunc{2}{\infty}(Y_I(t,u) + Y_C(t,u))$, which corresponds to the Taylor series expansion of  $\expPower{-Y_I(t,u)-Y_C(t,u)}$, combined with the stochastic part of the funding spread, i.e., $\LGD_I y_I(t,u)$.
The choice of $n_{\intensity}=1$ means that no calibration needs to be performed for this approximation parameter.

Furthermore, $\errorWWRPartOne$ contains scaled truncation errors, $\mu_{S}(t, u) \error{2}(1) + \LGD_I \error{2}(y_I)$, where
\begin{align} \displaystyle
  \error{2}(x)
    &\ldef H_{\shortRate,I,C}(t,u)  \condExpSmall{\taylorTrunc{n_{\shortRate} + 1}{\infty}(Y_{\shortRate}(t,u)) \cdot x \cdot \left(-Y_I(t,u) - Y_C(t,u)\right) \maxOperator{\tradeVal(u)}}{t}. \label{eq:error2}
\end{align}
Similarly, $\errorWWRPartOne$ contains the scaled truncation errors $\LGD_I \error{3}(y_I)$, with
\begin{align} \displaystyle
  \error{3}(x)
    &\ldef H_{\shortRate,I,C}(t,u) \condExpSmall{\taylorTrunc{n_{\shortRate} + 1}{\infty}(Y_{\shortRate}(t,u)) \cdot x \cdot \maxOperator{\tradeVal(u)}  }{t}. \label{eq:error3}
\end{align}
Both errors $\error{2}$ and $\error{3}$ are from the truncation of the infinite summations to $n_{\shortRate}$, and thus discard the terms $\taylorTrunc{n_{\shortRate} + 1}{\infty}(Y_{\shortRate}(t,u))$ of the corresponding full Taylor series.
The choice of an appropriate value for $n_{\shortRate}$ is discussed in Section~\ref{sec:numericalResultsError}.
So far, all the truncation errors are at the exposure level and are generic, regardless of the choice of portfolio $\tradeVal$.

Equation~\eqref{eq:epeWWRCredit2} forms the starting point of the WWR approximation proposed in this paper.
The third term in Equation~\eqref{eq:epeWWRCredit2} will cancel out the second term in Equation~\eqref{eq:epeIndepCreditb}.
This is a direct result of the split of the two types of exposure.

\section{Gaussian approximation for WWR exposure}  \label{sec:approximation}
In Section~\ref{sec:wwr}, we have obtained an expression for the WWR exposure using the affine dynamics, correlation assumptions, funding spread and Taylor series expansions.
Next, we propose an efficient and generic approximation of the expectations in Equation~\eqref{eq:epeWWRCredit2}.
The distributions of the various processes are approximated by normal distributions.
This significantly simplifies the equations while preserving the dependence structure in an approximate fashion, allowing for a useful approximation of the WWR exposure.
Furthermore, the approximation preserves intuition on WWR from a modelling perspective.

We choose dynamics that fit the framework from Section~\ref{sec:SDE}.
The Hull-White dynamics (HW1F) with constant volatility are used for IR process $\shortRate(t)$.
For Bilateral Credit Valuation Adjustment (BCVA) WWR modelling, a CIR model is often used for the credit processes~\cite{BrigoPallavicini201405,BrigoPallaviciniPapatheodorou201107}. 
Hence, the CIR++ dynamics~\cite{BrigoMercurio2006} with constant volatility are used for the credit process $\intensity_z$ of each counterparty $z \in \{I,\ C\}$.
See~\ref{app:dynamics} for the full specification of the model dynamics.
We are interested in the integrated model dynamics, such that both drift and diffusion appear in integrated form.

The stochastic processes for IR, defined in Equation~\eqref{eq:shortRateFramework}, follow normal distributions, i.e., $y_{\shortRate}\sim \N\left(0,\condVarSmall{y_{\shortRate}}{t}\right)$ and $Y_{\shortRate}\sim \N\left(0,\condVarSmall{Y_{\shortRate}}{t}\right)$.
Hence, the use of a Gaussian approximation is justified for the IR processes, as both processes are already Gaussian.

The CIR++ model has a scaled non-central chi-square distribution, with fatter tails than a normal distribution.
When the Feller condition is satisfied, both tails of the density will decay fast~\cite{OosterleeGrzelak201911}, such that it can be well approximated by a Gaussian.~\footnote{Under some market conditions, the Feller condition can be too restrictive in the sense that the model cannot generate enough volatility to fit the market~\cite{LichtersStammGallagher2015}.
In this case, jumps could be added to the model to increase the generated volatility.}
Still, information from the tails is lost in the approximation.
However, it is challenging to calibrate these tails properly due to insufficient market data.
Therefore, the Gaussian approximation is also justified for the credit processes.
The Gaussian approximation might lead to a small probability of negative values of the credit distribution.
As the approximation is only applied to the joint distributions required for WWR, and not to the marginals distributions used to compute the independent part of the exposure, this is an acceptable consequence of the Gaussian approximation.

\begin{rem}[Black-Karasinski for credit]
    Alternatively to the CIR++ dynamics, the Black-Karasinski (BK) dynamics can be used to model the default intensity, where positive rates are guaranteed due to the lognormal distribution.
    The main drawback of these dynamics is that there is no analytic ZCB formula, and the dynamics do not belong to the affine class.
    Turfus proposes an analytic approximation of the ZCB price by truncating an infinite power series of the rate, which preserves positivity~\cite{TurfusShubert202006}.
    This approximation is done under the assumption of a small deviation of rates from the forward curve.
    Due to the lack of affinity of the BK dynamics, and the need for an approximation to obtain an analytic ZCB price, these dynamics are not considered in this paper.
\end{rem}

\subsection{Gaussian approximation for stochastic processes} \label{sec:stochProcessApprox}
The essence of the proposed WWR approximation is to approximate all relevant stochastic processes by a single normally distributed variable.
Currently, $y_{\shortRate}$ follows a normal distribution.
Hence, we approximate all stochastic processes in terms of $y_{\shortRate}$, without loss of generality.

As $y_{\shortRate}(t,u)$ and $Y_{\shortRate}(t,u)$ are both normal with zero mean, $Y_{\shortRate}(t,u)$ can be expressed as $y_{\shortRate}(t,u)$ times a scaling factor:
\\[1.0ex]
\noindent\fbox{\parbox[][30pt][c]{0.98\textwidth}{
\begin{align} \displaystyle
  Y_{\shortRate}(t,u)
    &\equalDistr \sqrt{\frac{\condVarSmall{Y_{\shortRate}(t,u)}{t}}{\condVarSmall{y_{\shortRate}(t,u)}{t}}}   y_{\shortRate}(t,u)
    \rdef \Sigma(Y_\shortRate(t,u)) y_{\shortRate}(t,u).  \label{eq:YrApprox}
\end{align}
}}\\[1.0ex]
Here, we have introduced the notation $\Sigma(x) \ldef \sqrt{\frac{\condVarSmall{x}{t}}{\condVarSmall{y_{\shortRate}(t,u)}{t}}} > 0$ for the ratio of variances.

The CIR++ model has dynamics which follow a scaled non-central chi-square distribution.
Thus, $y_z$ and $Y_z$ are not Gaussian for $z \in \{I,\ C\}$.
We propose to approximate $Y_z(t,u)$ by a normally distributed variable, and then rewrite this in terms of $y_{\shortRate}(t,u)$:
\\[1.0ex]
\noindent\fbox{\parbox[][30pt][c]{0.98\textwidth}{
\begin{align} \displaystyle
  Y_z(t,u)
    &\approx \N\left(0, \condVarSmall{Y_z(t,u)}{t}\right)
    \equalDistr \sqrt{\frac{\condVarSmall{Y_z(t,u)}{t}}{\condVarSmall{y_{\shortRate}(t,u)}{t}}}   y_{\shortRate}(t,u)
    \rdef \Sigma(Y_z(t,u)) y_{\shortRate}(t,u),  \label{eq:YzApprox}
\end{align}
}}\\[1.0ex]
where for the WWR exposures, the correlations $\corr_{\shortRate,z}$ enter the approximation.

\begin{rem}[Correlation]
    As written in Equation~\eqref{eq:YzApprox}, the Gaussian approximation is applied to the marginal distributions.
    However, for WWR exposures, not the marginal distributions but the joint distributions are relevant.
    These define the WWR exposures in the form of cross-moments.
    Therefore, correlation $\corr_{\shortRate,z}$ should enter the equations, because the Brownian motions are correlated, i.e., $W_{\shortRate}(t)W_z(t) = \corr_{\shortRate,z}t$.
    For example, when the marginal of $Y_z(t,u)$ is approximated using Equation~\eqref{eq:YzApprox}, the cross-moment with $y_{\shortRate}(t,u)$ will be approximated as
    \begin{align} \displaystyle
      \condExpSmall{Y_z(t,u)y_{\shortRate}(t,u)}{t}
        &\approx \corr_{\shortRate,z} \Sigma(Y_z(t,u)) \condExpSmall{y_{\shortRate}^2(t,u)}{t},  \nonumber
    \end{align}
    where the dependence structure is preserved in an approximate fashion.
    The marginals are unaffected, as $\EPEFVAIndep{t}{u}$ is computed in the existing $\xva$ calculation, where WWR is not included.
    The error size mainly depends on how close $Y_z(t,u)$ is to a normal distribution.
\end{rem}

\subsection{WWR exposure approximation} \label{sec:approxWWR}
The WWR exposure in Equation~\eqref{eq:epeWWRCredit2} forms the starting point of the proposed WWR approximation.
The individual terms in the Taylor summation yield expressions of the form $\condExpSmall{ Y_{\shortRate}^j(t,u) \cdot x \cdot  \maxOperator{\tradeVal(u)}}{t}$.
The essence of the WWR exposure approximation is to apply the stochastic process approximations from Section~\ref{sec:stochProcessApprox}, which results in:
\\[1.0ex]
\noindent\fbox{\parbox[][102pt][c]{0.98\textwidth}{
\begin{align} \displaystyle
  \EPEFVAWWR{t}{u}
    &=  H_{\shortRate,I,C}(t,u)\left(\mu_{S}(t, u) \alpha(t,u) + \LGD_I \gamma(t,u)\right)\sum_{j=0}^{n_{\shortRate}} \beta_j(t,u)  \condExpSmall{ y_{\shortRate}^{j+1}(t,u)  \maxOperator{\tradeVal(u)} }{t}\nonumber \\
    &\quad +  \LGD_I H_{\shortRate,I,C}(t,u) \nu(t,u)
     \sum_{j=0}^{n_{\shortRate}} \beta_j(t,u)    \condExpSmall{ y_{\shortRate}^{j+2}(t,u)  \maxOperator{\tradeVal(u)} }{t}  \nonumber \\
    &\quad  + \LGD_I H_{I,C}(t,u) \condExpSmall{Y_I(t,u)y_I(t,u)}{t}  \condExpSmall{\expPower{-\int_{t}^{u} \shortRate(v)\dv}\maxOperator{\tradeVal(u)}}{t}    \nonumber \\
    &\quad  + \errorWWRPartOne + \errorWWRPartTwo, \label{eq:epeWWRCredit3}
\end{align}
}}\\[1.0ex]
with $y_{\shortRate}(t,u)$ from Equation~\eqref{eq:shortRateFramework}.
The first term in Equation~\eqref{eq:epeWWRCredit3} accounts for the linear correlation effects, while the second term covers the non-linear correlation effects.
We introduced the following notation for simplicity and to help identify the WWR exposure drivers:
\begin{align} \displaystyle
  \gamma(t,u)
    &\ldef \corr_{\shortRate,I} \Sigma(y_I(t,u)), \label{eq:epeWWRCredit3Gamma}  \\
  \alpha(t,u)
    &\ldef -\left[\corr_{\shortRate,I} \Sigma(Y_I(t,u))  + \corr_{\shortRate,C} \Sigma(Y_C(t,u))\right], \label{eq:epeWWRCredit3Alpha} \\
  \nu(t,u)
    &\ldef -\left[ \corr_{\shortRate,I}^2 \Sigma(Y_I(t,u)) + \corr_{\shortRate,I}\corr_{\shortRate,C} \Sigma(Y_C(t,u))\right] \Sigma(y_I(t,u)) , \label{eq:epeWWRCredit3Nu}  \\
  \beta_j(t,u)
    &\ldef \frac{(-\Sigma(Y_\shortRate(t,u)))^j}{j!}, \label{eq:epeWWRCredit3Beta}
\end{align}
where $\gamma(t,u)$ represents the main WWR effect resulting from the stochastic component of the funding spread.
The factor $\alpha(t,u)$ corresponds to the WWR from the credit adjustment effects, where $\nu(t,u)$ is a second-order cross-effect between the funding spread and the credit adjustment factors.
Both $\gamma(t,u)$ and $\alpha(t,u)$ are linear in correlations, while $\nu(t,u)$ is non-linear in the correlations.
This is in line with~\cite{ZwaardGrzelakOosterlee202210}, where linear correlation effects were identified as the main WWR drivers, while the non-linear correlation effects were of second order.

Furthermore, $\errorWWRPartTwo$ represents the overall Gaussian approximation error:
\begin{align} \displaystyle
  &\errorWWRPartTwo \nonumber \\
    &= H_{\shortRate,I,C}(\cdot)\sum_{j=0}^{n_{\shortRate}} \frac{(-1)^j}{j!} \Bigg\{\mu_{S}(\cdot) \condExpSmall{ \left[Y_{\shortRate}^{j}(\cdot) \left(-Y_I(\cdot) - Y_C(\cdot)\right)  - \alpha(\cdot) (\Sigma(Y_\shortRate(\cdot)))^j y_{\shortRate}^{j+1}(\cdot) \right]\maxOperator{\tradeVal(u)} }{t}\nonumber \\
    &\quad \quad + \LGD_I \condExpSmall{ \left[Y_{\shortRate}^{j}(\cdot) y_I(\cdot)  - \gamma(\cdot) (\Sigma(Y_\shortRate(\cdot)))^j y_{\shortRate}^{j+1}(\cdot) \right]\maxOperator{\tradeVal(u)} }{t} \nonumber \\
    &\quad \quad + \LGD_I \E_t \Big[ \left[Y_{\shortRate}^{j}(\cdot) \left(-Y_I(\cdot) - Y_C(\cdot)\right) y_I(\cdot)  - \nu(\cdot) (\Sigma(Y_\shortRate(\cdot)))^j y_{\shortRate}^{j+2}(\cdot) \right]\maxOperator{\tradeVal(u)} \Big]
    \Bigg\}, \label{eq:errorWWR2}
\end{align}
and $\errorWWRPartOne$ is given in Equation~\eqref{eq:errorWWR1}.
An approximation of the WWR exposure~\eqref{eq:epeWWRCredit3} is obtained by omitting $\errorWWRPartOne$ and $\errorWWRPartTwo$.

In summary, Equation~\eqref{eq:epeWWRCredit3} provides a generic expression for the WWR exposure of portfolio $\tradeVal$.
Once $\tradeVal$ is specified, two approaches can be taken.
The first is a generic one, where we use the fact that $y_{\shortRate}(t,u)$ and $\maxOperator{\tradeVal(u)}$ are available on a path level from the existing $\xVA$ calculation (without any WWR).
The expectations $\condExpSmall{ y_{\shortRate}^{l}(t,u)  \maxOperator{\tradeVal(u)} }{t}$, $l \in \{1,2,\ldots\}$, can then be computed as an average over these existing paths.
Alternatively, in some specific cases, $\maxOperator{\tradeVal(u)}$ can be written as a function of $y_{\shortRate}(t,u)$, such that $\condExpSmall{ y_{\shortRate}^{l}(t,u)  \maxOperator{\tradeVal(u)} }{t}$ can be approximated analytically.
An example of the latter approach for an IR swap is given in Section~\ref{sec:exampleIRSwap}.

The WWR approximation is flexible and allows one to quickly compute the WWR exposure for various counterparties.
This property can also be exploited when one wants to see how the WWR numbers would change if the credit quality of one of the relevant parties changed.
For example, consider the hypothetical case of an identical trade with a new counterparty $D$.
The first step for any approach would be to calibrate the model dynamics $\intensity_D(t)$ and correlation $\corr_{\shortRate,D}$ to the (historical) market data.
For the WWR approximation, only $H_D(t,u)$ and $\Sigma(Y_D(t,u))$ are required, which can be computed analytically, and no simulation is required.
These results can directly be substituted into Equation~\eqref{eq:epeWWRCredit3}, where all other components can be reused from the simulations which have already been performed.

\subsection{Understanding WWR effects using the approximation} \label{sec:approxWWRunderstanding}

The representation of the WWR exposure in Equation~\eqref{eq:epeWWRCredit3} helps us identify under what conditions the various model components generate either WWR or RWR.
As noted in Section~\ref{sec:stochProcessApprox}, by definition, $\Sigma(Y_z(t,u)) > 0$.
Recall that $H_z(t,u)>0$ can be interpreted as a discount factor.
By construction, $\mu_S(t,u) >0$, and by definition, $\LGD_I > 0$.

Consider the example where $\tradeVal$ is an uncollateralized IR receiver swap, and $\corr_{\shortRate,I},\corr_{\shortRate,C} < 0$.
Then, $\gamma(t,u) < 0$, $\alpha(t,u) > 0$, $\nu(t,u) < 0$, which implies
\begin{align} \displaystyle
  H_{\shortRate,I,C}(t,u)\LGD_I \gamma(t,u) &< 0, \nonumber \\
  H_{\shortRate,I,C}(t,u)\mu_{S}(t, u) \alpha(t,u) &> 0, \nonumber \\
  \LGD_I H_{\shortRate,I,C}(t,u) \nu(t,u) &< 0.\nonumber
\end{align}
For ease of notation, we define
\begin{align} \displaystyle
  \psi_m(t,u)
    &\ldef \sum_{j=0}^{n_{\shortRate}} \beta_j(t,u) \condExpSmall{ y_{\shortRate}^{j+m}(t,u)  \maxOperator{\tradeVal(u)}}{t}. \nonumber
\end{align}
If $\tradeVal$ is a receiver swap, then~\footnote{The $j=0$'th term determines the sign, the rest of the terms are decaying in magnitude due to the factor $\frac{1}{j!}$ in $\beta_j(t,u)$, see Equation~\eqref{eq:epeWWRCredit3Beta}. One can examine the $0$'th term in the sum, decompose the payoff and eventually come to the conclusion that $\beta_0(t,u) \condExpSmall{ y_{\shortRate}(t,u)  \maxOperator{\tradeVal(u)} }{t} < 0$.} $\psi_1(t,u) < 0$, and $\psi_2(t,u) > 0$.
Hence, in this example, the $\gamma(t,u)$ term in Equation~\eqref{eq:epeWWRCredit3} gives rise to WWR, i.e.,
\begin{align} \displaystyle
  H_{\shortRate,I,C}(t,u)\LGD_I \gamma(t,u) \psi_1(t,u) &> 0. \nonumber
\end{align}
The $\alpha(t,u)$ and $\nu(t,u)$ terms however give rise to RWR, i.e.,
\begin{align} \displaystyle
  H_{\shortRate,I,C}(t,u)\mu_{S}(t, u) \alpha(t,u) \psi_1(t,u) &< 0, \nonumber \\
  \LGD_I H_{\shortRate,I,C}(t,u) \nu(t,u) \psi_2(t,u)  &< 0.\nonumber
\end{align}
So, the stochastic component of the funding spread generates WWR, while the credit adjustment effects (direct and cross) generate RWR in this example of a receiver swap with negative IR-credit correlations.
This is in line with the findings from~\cite{ZwaardGrzelakOosterlee202210}.

Whether the net result is WWR or RWR depends on the correlations, credit parameters, IR parameters and product type~\cite{ZwaardGrzelakOosterlee202210}.
The proposed approximation helps us to understand this in both a qualitative and quantitative sense.
As $y_{\shortRate}(t,u)\sim \N\left(0,\condVarSmall{y_{\shortRate}(t,u)}{t}\right)$, we have that $\condProbSmall{y_r(t,u) > 1}{t} = 1 - \normCDF\left( \frac{1}{\sqrt{\condVarSmall{y_{\shortRate}(t,u)}{t}}} \right) $ with $\condVarSmall{y_{\shortRate}(t,u)}{t} = \frac{\vol_{\shortRate}^2}{2a_{\shortRate}} \left( 1 - \expPower{-2a_{\shortRate} (u-t)}\right)$, see~\ref{app:dynamicsIR}. Given model parameters $a_{\shortRate}$ and $\vol_{\shortRate}$, the variance will be largest at the portfolio's maturity $u=T$. One can then verify that $\condProbSmall{y_r(t,u) > 1}{t}$ will be small enough to claim that generally $y_{\shortRate}(t,u)<1$.
In this case, $\left|\psi_2(t,u)\right| \ll \left|\psi_1(t,u)\right|$.

Furthermore, the third term in Equation~\eqref{eq:epeWWRCredit3} is of a lower order than the first two terms, as $\condExpSmall{Y_I(t,u)y_I(t,u)}{t}$ is small.~\footnote{See the analytic expression for $\condExpSmall{Y_I(t,u)y_I(t,u)}{t}$ in Equation~\eqref{eq:expYzyz} in~\ref{app:dynamicsCredit}.}
Hence, the first term in Equation~\eqref{eq:epeWWRCredit3} is dominant in determining whether there is net WWR or RWR.
This first term accounts for the linear correlation effects, which we already identified as the main WWR driver.
In particular, the net WWR will depend on the sign of $\mu_{S}(t, u) \alpha(t,u) + \LGD_I \gamma(t,u)$, which depends on correlations, credit parameters and IR parameters.
This quantity can be computed analytically given a calibrated set of model dynamics and is payoff-independent.
The type of payoff will determine the sign of $\psi_1(t,u)$, which, combined with the sign of $\mu_{S}(t, u) \alpha(t,u) + \LGD_I \gamma(t,u)$, will indicate whether there is net WWR or RWR.
The proposed WWR approximation thus helps to identify the WWR/RWR drivers of exposures, as only a set of calibrated model dynamics and a few straightforward calculations are required.
Hence, we can conclude that the approximation preserves some form of WWR intuition.

\subsection{Analytic WWR exposure approximation for an IR swap} \label{sec:exampleIRSwap}
Here, a detailed example of the WWR exposure approximation is presented for the case where $\tradeVal$ is a single uncollateralized IR swap.
In this case, $\condExpSmall{ y_{\shortRate}^{l}(t,u)  \maxOperator{\tradeVal(u)} }{t}$ can be expressed in terms of $y_{\shortRate}(t,u)$, such that an analytic approximation of the WWR exposure is obtained.
In Section~\ref{sec:IRSwap}, $\maxOperator{\tradeVal(u)}$ is written as a function of stochastic process $y_{\shortRate}(t,u)$.
Next, in Section~\ref{sec:approxWWRIRSwap}, the results are combined to simplify the expectations of the form $\condExpSmall{ y_{\shortRate}^{l}(t,u)  \maxOperator{\tradeVal(u)} }{t}$, which appear in Equation~\eqref{eq:epeWWRCredit3}.

\subsubsection{IR swap formula} \label{sec:IRSwap}
Consider an IR swap under a single yield curve setup starting at $T_0$, maturing at $T_m$ with intermediate payment dates $T_1 < T_2 < \ldots < T_{m-1} < T_m$, $\dct_k = T_k - T_{k-1}$.
To apply the Gaussian approximation from Section~\ref{sec:approximation}, the IR swap value needs to be written differently than the conventional formulations.
In~\ref{app:swapPayoff}, the value of an IR swap $\tradeVal$ with strike $\strike$ and notional $\notional$ is expressed in terms of $y_{\shortRate}$:
\begin{align} \displaystyle
  \tradeVal(u)
    &= \swapType\notional\left(-\indicator{u > T_0} + \sum_{k=\beta(u)}^m \overline{w}_k  \expPower{-y_{\shortRate}(t,u)B_{\shortRate}(u,T_k)}\right), \label{eq:swap} \\
  \swapType
    &= \left\{
       \begin{array}{ll}
    	-1, & \text{payer swap}, \\
    	1, & \text{receiver swap},
       \end{array}
       \right.
  \beta(u)
    = \left\{
    \begin{array}{ll}
      0, & u \in [t_0, T_0],\\
      j+1, & u \in (T_j, T_{j+1}],
    \end{array}
    \right. \nonumber\\
  \overline{w}_k
    &= w_k \expPower{\overline{A}_{{\shortRate}}(u,T_k) - \mu_{\shortRate}(t,u) B_{\shortRate}(u,T_k)}, \
  w_k
    = \left\{
    \begin{array}{ll}
      -1, & k=0, \\
      K\dct_k, & k=1,\ldots,m-1, \\
      1 + K \dct_k, & k=m.
    \end{array}
    \right.  \nonumber
\end{align}

Recall that the term $\maxOperator{\tradeVal(u)}$ appears in the approximation~\eqref{eq:epeWWRCredit3}, where $\maxOperator{\tradeVal(u)} = \tradeVal(u) \indicator{\tradeVal(u)\geq 0}$.
In~\ref{app:derivationsMaxProb}, we simplify the indicator term $\indicator{\tradeVal(u)\geq 0}$.
These results are combined with a Taylor expansion of $\expPower{-y_{\shortRate}(t,u)B_{\shortRate}(u,T_k)}$, yielding the following expression for $\maxOperator{\tradeVal(u)}$:
\begin{align} \displaystyle
  \maxOperator{\tradeVal(u)}
    &= \swapType\notional \left(-\indicator{u > T_0} + \sum_{k=\beta(u)}^m \overline{w}_k  \taylor(y_{\shortRate}(t,u)B_{\shortRate}(u,T_k)) \right) \left( \indicator{\swapType = -1} + \swapType \cdot \indicator{y_{\shortRate}(t,u) \leq y_{\shortRate}^{*}(t,u)}  \right), \label{eq:swaptionPayoff}
\end{align}
where $y_{\shortRate}^*(t,u)$ is computed numerically, independent of $y_{\shortRate}(t,u)$, see~\ref{app:derivationsMaxProb} for details.

\subsubsection{WWR approximation for an IR swap} \label{sec:approxWWRIRSwap}
In Equation~\eqref{eq:swaptionPayoff}, $\maxOperator{\tradeVal(u)}$ is expressed as a function of $y_{\shortRate}(t,u)$.
In~\ref{sec:moments}, the moments of a normal and generic truncated normal random variable are given in Results~\ref{res:normMoment} and~\ref{res:truncNormMoment}, respectively.
Using these analytic results, we write $\condExpSmall{ y_{\shortRate}^{l}(t,u)}{t} \rdef \moment_l$ and $\condExpSmall{ y_{\shortRate}^{l}(t,u)\indicator{y_{\shortRate}(t,u) \leq y_{\shortRate}^{*}(t,u)} }{t} \rdef \momentTruncNorm_{l}   F_{y_{\shortRate}}\left(y_{\shortRate}^{*}(t,u)\right)$, where $F_{y_{\shortRate}}\left(\cdot\right)$ is the CDF of $y_{\shortRate}(t,u)$.
Combining these moments with the results from Equation~\eqref{eq:swaptionPayoff}, the expectations of the form $\condExpSmall{ y_{\shortRate}^{l}(t,u)  \maxOperator{\tradeVal(u)} }{t}$ in Equation~\eqref{eq:epeWWRCredit3} can be written in terms of these analytic moments:
\\[1.0ex]
\noindent\fbox{\parbox[][100pt][c]{0.98\textwidth}{
\begin{align} \displaystyle
  &\condExpSmall{ y_{\shortRate}^{l}(t,u) \maxOperator{\tradeVal(u)}}{t} \nonumber \\
    & = \swapType\notional \Bigg(-\indicator{u > T_0} \left( \indicator{\swapType = -1}\moment_l + \swapType \cdot \momentTruncNorm_{l}   F_{y_{\shortRate}}\left(y_{\shortRate}^{*}(t,u)\right) \right) \nonumber \\
    &\qquad \quad + \sum_{k=\beta(u)}^m \overline{w}_k  \sum_{a=0}^{n_a}\frac{\left(-B_{\shortRate}(u,T_k)\right)^a}{a!} \left( \indicator{\swapType = -1}\moment_{a+l} + \swapType \cdot \momentTruncNorm_{a+l}   F_{y_{\shortRate}}\left(y_{\shortRate}^{*}(t,u)\right)\right)\Bigg) + \error{4}(l). \label{eq:approxExp3}
\end{align}
}}\\[1.0ex]
This expression can be computed fully analytically, provided that $y_{\shortRate}^*(t,u)$ is known.

Product-level truncation error $\error{4}(l)$ manifests itself after the application of the Gaussian approximation.
It is the error due to the truncation of the infinite summations of $\taylor(y_{\shortRate}(t,u)B_{\shortRate}(u,T_k))$ in $\condExpSmall{ y_{\shortRate}^{l}(t,u) \maxOperator{\tradeVal(u)}}{t}$ to $n_a$, see Equation~\eqref{eq:approxExp3}.
The choice of an appropriate value for $n_a$ is discussed in Section~\ref{sec:numericalResultsError}.

The final step is to substitute the payoff-dependent expectation from Eq.~\eqref{eq:approxExp3}, i.e.,  $\condExpSmall{ y_{\shortRate}^{l}(t,u) \maxOperator{\tradeVal(u)}}{t}$, into the WWR exposure approximation~\eqref{eq:epeWWRCredit3}.
Then, $\errorWWRPartThree$ is the overall product-level truncation error after the Gaussian approximation is performed:
\begin{align} \displaystyle
  \errorWWRPartThree
    &= H_{\shortRate,I,C}(t,u)\left(\mu_{S}(t, u) \alpha(t,u) + \LGD_I \gamma(t,u)\right)\sum_{j=0}^{n_{\shortRate}} \beta_j(t,u)  \error{4}(j+1) \nonumber \\
    &\quad +  \LGD_I H_{\shortRate,I,C}(t,u) \nu(t,u)
     \sum_{j=0}^{n_{\shortRate}} \beta_j(t,u) \error{4}(j+2). \label{eq:errorWWR3}
\end{align}
In Section~\ref{sec:numericalResults}, we will demonstrate that for $n_a$ sufficiently large, the error $\errorWWRPartThree$ is negligible compared to $\errorWWRPartOne$ and $\errorWWRPartTwo$.

\subsection{Scope of products and asset classes in the generic case} \label{sec:approxGeneralization}
The Gaussian approximation method presented in Section~\ref{sec:stochProcessApprox} is based on a HW1F dynamics for IR and CIR++ dynamics for credit.
The approximation in Equation~\eqref{eq:epeWWRCredit3} is generic and is applicable to any portfolio of IR derivatives.
Here, we rely on the availability of $\condExpSmall{ y_{\shortRate}^{l}(t,u)  \maxOperator{\tradeVal(u)} }{t}$ from the $\xva$ calculation without WWR.
Therefore, any derivative that can be priced in the original $\xva$ calculation can also be in the portfolio when approximating the WWR exposure.

The FX extension is trivial as this happens in exactly the same way as the non-WWR FX extension of $\xva$.
The only thing one must keep in mind is that only the IR-credit relationship is captured in the approximation.
Whether or not this is acceptable depends on the business model of a financial institution.
Furthermore, extending to other asset classes, e.g., inflation, proceeds in a similar fashion.
Also, the extension to a multi-curve framework is not a problem, and the HW1F model for IR can easily be replaced by a multi-factor IR model.

\subsubsection{Special cases of fully analytical approximations} \label{sec:approxGeneralizationFully Analytical}
As seen in Section~\ref{sec:exampleIRSwap} for the IR swap, for some cases it is possible to find an analytical WWR approximation~\footnote{Up to the numerical computation of $y_{\shortRate}^*(t,u)$.}.
If this is desired, a portfolio of derivatives $\tradeVal$ allows for an analytic WWR approximation if it can be written in terms of $y_{\shortRate}(t,u)$, such that $\condExpSmall{ y_{\shortRate}^{l}(t,u)  \maxOperator{\tradeVal(u)}}{t}$ allows for an analytic approximation.
Linear IR derivatives which can be written as a monotonically decreasing function of $y_{\shortRate}(t,u)$ are already covered.
This can be extended to all linear IR derivatives that are not of this particular form at the cost of some additional computations.
Vanilla options on linear IR derivatives are also covered.
For example, consider a swaption payoff exposure: $\left(\tradeVal^{\text{swaption}}\right)^+ = \tradeVal^{\text{swaption}} = \left(\tradeVal^{\text{swap}}\right)^+$.
Hence, the exposure of vanilla options on linear derivatives collapses to the exposure of the underlying derivative itself, which is covered in our framework.
For exotic derivatives, like non-linear derivatives, which are not vanilla options on linear derivatives, it is less trivial to see if they allow for an analytic WWR approximation.
Fortunately, there is always the generic approach which covers all derivatives that can be priced in the original $\xva$ calculation.

To extend to FX derivatives, one would need two additional SDEs: a foreign short-rate process $\shortRate_f(t)$ for IR and an FX process $\FX_f^d(t)$ to convert currency $f$ into $d$.
The original IR process is then denoted as $\shortRate_d(t)$.
Furthermore, correlations need to be imposed between all the processes, again with the assumption of independent counterparties.
For both IR processes, we use the HW1F model as before.
The $\shortRate_f(t)$ dynamics will be given under the $\Q^f$ risk-neutral measure.
When setting up a Monte Carlo simulation, all processes must be under the same measure~\cite{Green201511}.
Hence, we use a change of measure $\d \brownian_{\shortRate_f}^{f}(t) = \d \brownian_{\shortRate_f}^d(t) - \rho_{\shortRate_f, \FX} \cdot \vol_{\FX} \d t$ and write the $\shortRate_f(t)$ dynamics under the $\Q^d$ risk-neutral measure.
This will lead to an additional term in the drift, which is a quanto correction term, see~\ref{app:dynamicsIR}.
For the FX process, $\FX_f^d$, we choose log-normal dynamics, see~\ref{app:dynamicsFX}.
Next, we write this in integrated form, and express it as $\FX_f^d(u) = \expPower{\mu_{\FX}(t,u) - \overline{y}_{\FX}(t,u)}$, where $\overline{y}_{\FX}(t,u) = -\left(Y_{\shortRate_d}(t,u) - Y_{\shortRate_f}(t,u) + y_{\FX}(t,u)\right)$, with $y_{\FX}(t,u) = \vol_{\FX} \int_t^u \d\brownian_{\FX}^d(v)$.
The three stochastic processes $Y_{\shortRate_d}(t,u)$, $Y_{\shortRate_f}(t,u)$, $y_{\FX}(t,u)$ are correlated normals with zero mean.
Then, $\overline{y}_{\FX}(t,u)$ is normal with zero mean and a variance where the correlations are included.

Next, we examine an FX derivative, for example, an FX forward contract with maturity $T$.
This payoff can be written in terms of the quantities $\zcb_d(t,T)$, $\zcb_f(t,T)$ and $\FX_f^d(t)$.
In turn, these quantities can be expressed in terms of the corresponding integrated drift and integrated stochastic processes.
Using the Gaussian approximation for the stochastic processes from Section~\ref{sec:stochProcessApprox}, $\tradeVal(u)$ can be approximated in terms of $y_{\shortRate_d}(t,u)$.
Further manipulations result in an expression of $\left(\tradeVal(u)\right)^+$ in terms of $y_{\shortRate_d}(t,u)$.
See~\ref{app:fxExtension} for a detailed derivation of this example.
Using this expression, we decompose the expressions $\condExpSmall{y_{\shortRate}^{l}(t,u) \maxOperator{\tradeVal(u)}}{t}$  in terms of moments of a normal and truncated normal distribution.
This approach can easily be extended to all linear FX derivatives.

\section{Error analysis} \label{sec:approxError}

In the proposed WWR approximation, various errors appear.
First, $\errorWWRPartOne$ from Equation~\eqref{eq:errorWWR1} is a truncation error from the truncated Taylor series expansions.
Next, $\errorWWRPartTwo$ from Equation~\eqref{eq:errorWWR2} is the Gaussian approximation error.
Finally, for specific derivatives $\tradeVal$, a product-level truncation error $\errorWWRPartThree$ is introduced in Equation~\eqref{eq:errorWWR3}.
In general, there are two types of errors: truncation errors and Gaussian approximation errors.

Here, we dive further into the two error types.
The goal is to find out when the approximations are working well and when not; and to find error bounds where possible.
In particular, we look at the impact of various model parameters on the quality of the approximations, including extreme conditions.
In this way, it becomes apparent when the approximation can safely be used and when some more caution is required.

\subsection{Truncation error} \label{sec:approxErrorTrunc}
First, we examine $\errorWWRPartOne$ from Equation~\eqref{eq:errorWWR1}, where the contributing terms all contain Taylor sums of the form $\taylorTrunc{n+1}{\infty}(Y_z)$.
Here, $Y_{z}$ are stochastic integrals with $\condExpSmall{Y_{z}(t,u)}{t} = 0$ for all $u$.
The variance grows through time, as $Y_{z}(t,u)$ is the integrated stochastic part of the dynamics $x_z(u)$.
Consider a portfolio with maturity $T$.
If the error for $Y_{z}(t,T)$ is sufficiently small, then the same holds for $Y_{z}(t,u)$ with $u\in[t,T]$.
Therefore, only $Y_{z}(t,T)$ is considered in the error analysis.

The terms of $\errorWWRPartOne$ in Equation~\eqref{eq:errorWWR1} all contain an expectation of the form \\
$\condExpSmall{\taylorTrunc{n+1}{\infty}(Y_z(t,u)) \cdot \overline{x} \cdot \maxOperator{\tradeVal(u)}}{t}$, which we aim to bound in the $L^2$ norm.
The Cauchy-Schwarz inequality gives
\begin{align}\displaystyle
  \left( \condExpSmall{\taylorTrunc{n+1}{\infty}(Y_z(t,u)) \cdot \overline{x} \cdot \maxOperator{\tradeVal(u)}}{t}\right)^2
    &\leq \condExpSmall{\left(\taylorTrunc{n+1}{\infty}(Y_z(t,u)) \cdot \overline{x}\right)^2}{t} \condExpSmall{ \left(\maxOperator{\tradeVal(u)}\right)^2}{t}. \label{eq:genericErrorBoundB1}
\end{align}

\paragraph{First term}
For the first term in the RHS of Equation~\eqref{eq:genericErrorBoundB1}, we use that $\taylorTrunc{n+1}{\infty}(x) = \bigOh\left(\taylorTrunc{n+1}{n+1}(x)\right)$, i.e., $\exists \ C_{T,n+1} > 0 $ s.t. $\left|\taylorTrunc{n+1}{\infty}(x)\right| \leq \frac{C_{T,n+1}}{(n+1)!} \cdot x^{n+1}$.
Hence,
\begin{align}\displaystyle
  \condExpSmall{\left(\taylorTrunc{n+1}{\infty}(Y_z(t,u)) \cdot \overline{x}\right)^2}{t}
  &\leq \frac{C_{T,n+1}^2}{((n+1)!)^2}\condExpSmall{\left(Y_z(t,u)\right)^{2(n+1)} \cdot \overline{x}^2}{t}. \label{eq:genericErrorBoundB3}
\end{align}

Next, we use the correlation definition for two correlated random variables $X$ and $Y$ to obtain the following inequality:
\begin{align} \displaystyle
  \condExpSmall{XY}{t}
    &= \corr_{X,Y}\sigma_X \sigma_Y + \condExpSmall{X}{t}\condExpSmall{Y}{t}
    \leq \sigma_X \sigma_Y + \condExpSmall{X}{t}\condExpSmall{Y}{t}, \label{eq:corrDefBound}
\end{align}
with correlation $\corr_{X,Y} \in [-1,1]$ and $\sigma_X, \sigma_Y > 0$.

Using this result, we can bound $\condExpSmall{\left(Y_z(t,u)\right)^{2(n+1)} \cdot \overline{x}^2}{t}$ in Equation~\eqref{eq:genericErrorBoundB3}, as follows:
\begin{align}\displaystyle
  \condExpSmall{\left(Y_z(t,u)\right)^{2(n+1)} \cdot \overline{x}^2}{t}
    &\leq \sqrt{\condExpSmall{\left(Y_z(t,u)\right)^{4(n+1)}}{t} - \left(\condExpSmall{\left(Y_z(t,u)\right)^{2(n+1)}}{t}\right)^2} \sqrt{\condExpSmall{\overline{x}^4}{t} - \left(\condExpSmall{\overline{x}^2}{t}\right)^2} \nonumber \\
    &\quad + \condExpSmall{\left(Y_z(t,u)\right)^{2(n+1)}}{t}\condExpSmall{\overline{x}^2}{t}.  \label{eq:genericErrorBoundC1}
\end{align}

\paragraph{Second term}
Since the approximation is generic, the second term at the RHS in Equation~\eqref{eq:genericErrorBoundB1}, which is the product-specific component of the error, can be computed straightaway from the available grid of simulated future portfolio values.
Hence, we write
\begin{align}\displaystyle
  \condExpSmall{ \left(\maxOperator{\tradeVal(u)}\right)^2}{t}
    &\rdef C_{\tradeVal}. \label{eq:swapSquaredBound}
\end{align}
In the example of an IR swap, see Section~\ref{sec:exampleIRSwap}, an analytic bound $C_{\tradeVal}$ can be derived, such that $\condExpSmall{ \left(\maxOperator{\tradeVal(u)}\right)^2}{t} \leq C_{\tradeVal}$.
See~\ref{app:derivationsSwapErrorBound} for a derivation of $C_{\tradeVal}$ in this case.

\paragraph{Overall error bound}
Plugging Equations~\eqref{eq:genericErrorBoundB3} and~\eqref{eq:swapSquaredBound} into Equation~\eqref{eq:genericErrorBoundB1} yields the following generic bound
\\[1.0ex]
\noindent\fbox{\parbox[][35pt][c]{0.98\textwidth}{
\begin{align}\displaystyle
  \left| \condExpSmall{\taylorTrunc{n+1}{\infty}(Y_z(t,u)) \cdot \overline{x} \cdot \maxOperator{\tradeVal(u)}}{t}\right|
    &\leq  \frac{ \sqrt{C_{\tradeVal}} \cdot C_{T,n+1}}{(n+1)!} \sqrt{\condExpSmall{\left(Y_z(t,u)\right)^{2(n+1)} \cdot \overline{x}^2}{t}}, \label{eq:genericErrorBoundC3a}
\end{align}
}}\\[1.0ex]
where the remaining expectation depends on the specific error under consideration, e.g., $\error{1}$, and is bounded in Equation~\eqref{eq:genericErrorBoundC1}.
The error bound in Equation~\eqref{eq:genericErrorBoundC3a} clearly shows an exponential decay of the error, so only a few terms $n$ will be needed for a high-precision approximation.
This is only relevant if the higher order (cross-)moments in Equation~\eqref{eq:genericErrorBoundC1} are finite.
In~\ref{app:approxErrorTruncSpecificForm}, the obtained error bound is applied to $\error{1}(x)$ from Equation~\eqref{eq:error1}, to obtain an explicit form of the error bound.

Looking at Equation~\eqref{eq:genericErrorBoundC3a}, given a portfolio and a number of terms $n$ in the Taylor approximation, the error will be driven by $\condExpSmall{\left(Y_z(t,u)\right)^{2(n+1)} \cdot \overline{x}^2}{t}$, which is bounded in Equation~\eqref{eq:genericErrorBoundC1} based on higher-order moments of $Y_z(t,u)$ and $\overline{x}$.
Since both $Y_z(t,u)$ and $\overline{x}$ are in several cases composed of multiple stochastic integrals~\footnote{For $\error{1}(x)$, $Y_z(t,u) = Y_I(t,u) + Y_C(t,u)$, $\overline{x} = x \cdot \expPower{-Y_{\shortRate}(t,u)}$, $ x \in \{1,y_I(t,u)\}$.
For $\error{2}(x)$, $Y_z(t,u) = Y_{\shortRate}(t,u)$, $\overline{x} = x \left(-Y_I(t,u) - Y_C(t,u)\right)$, $ x \in \{1,y_I(t,u)\}$.
For $\error{3}(x)$, $Y_z(t,u) = Y_{\shortRate}(t,u)$, $\overline{x} = x = y_I(t,u)$.}, the leading order of the error will be the higher-order (cross-)moments of these stochastic integrals.
For simplicity of the argument, we focus on $Y_z(t,u)$.

As observed before, as $u$ increases, the distributions of $Y_z(t,u)$ will increase in magnitude; a pattern which the error will follow, depending on the shape of the exposure $\left(\tradeVal(u)\right)^+$.
For example, for an IR swap, the exposure decays towards maturity, which dampens the effect of an increased truncation error for larger $u$.
Furthermore, given a fixed value of mean reversion parameter $a_z$ in the model dynamics, increased volatility $\vol_z$ will result in an increase in error.
Alternatively, for a fixed value of $\vol_z$, a lower magnitude of $a_z$ will increase the error, where the effect is larger for larger values of $\vol_z$.

More details on the impact of the model parameters will be given in Section~\ref{sec:numericalResultsError}, where also the interplay with the Gaussian approximation error $\errorWWRPartTwo$ is discussed.

\subsection{Gaussian approximation error} \label{sec:approxErrorDistr}
In Section~\ref{sec:stochProcessApprox}, the Gaussian approximations of the stochastic processes that appear in the exposure formula are introduced.
When computing cross-moments, stochastic processes $Y_z(t,u)$ are approximated in terms of $y_{\shortRate}(t,u)$ times a scaling factor, which depends on the ratio of the variances and the correlation between the two variables.
For the integrated stochastic processes, recall that $\condExpSmall{Y_z(t,u)}{t}=0$.
When approximating WWR exposure, this results in Gaussian approximation error $\errorWWRPartTwo$ as defined in Equation~\eqref{eq:errorWWR2}.

The Gaussian approximation is done after truncating the Taylor series, so it is only applied to the first $n$ terms in the Taylor series.
Furthermore, it is applied in the context of cross-moments, i.e., the marginal distributions are left untouched.
Covariance, being a linear measure of joint variability, looks at the first-order relationship between two variables.
Since it is at first order, the dependence can be summarized through a set of correlated Gaussians, replacing the original joint distributions.
The information that is lost in this step mainly relates to the tails.
However, as we are looking at first-order dependence, it is key to get the bulk of the distribution correct, and it is justified to lose some information from the tails.
Further justification of the loss of tail information is already given at the start of Section~\ref{sec:approximation}.
This will provide practitioners with an intuitive and fast way to get further intuition on WWR exposures.

Naturally, the further away the distribution of $Y_z(t,u)$ is from a Gaussian, the worse the approximation will be.
Here, the (a)symmetry of the original distribution plays a role, as well as the fatness of the tails.
Though it is difficult to control this error at a portfolio level, at a lower level, it can be quantified.
The error may be measured by comparing the CDF of $Y_z(t,u)$ to a normal distribution with equivalent variance, i.e., $\N\left(0, \condVarSmall{Y_z(t,u)}{t}\right)$.
A way to measure the distance between the two CDFs is the summed squared distance between the two, which is an $L^2$-norm.
If an empirical CDF of $Y_z(t,u)$ is available, this is equivalent to computing the Cramer-von Mises test statistic.
Alternatively, a Wasserstein distance can be used.

Numerical examples of the approximation errors are presented in Section~\ref{sec:numericalResultsError}.
In particular, we examine how the Gaussian approximation error behaves under different circumstances, and also assess the interplay with the truncation error $\errorWWRPartOne$ is discussed.

\section{Numerical results}  \label{sec:numericalResults}

To demonstrate the effectiveness of the proposed WWR approximation, various examples with different portfolio compositions $\tradeVal$ are considered.
A Monte Carlo simulation is used for the benchmark $\FVA$ numbers, with $10^5$ paths and $10$ dates per year.
Algorithm~\ref{algo:fvaCalculationProcedure} summarizes the $\FVA$ calculation procedure.

\begin{algorithm}[h!]
    \footnotesize
    \SetAlgoLined 
    \DontPrintSemicolon 


    \KwIn{Portfolio $\tradeVal$, monitoring dates ($t=t_0<t_1<t_2<\ldots<t_N=T$), market data (or synthetic model parameters), modelling choices (including/excluding $\default_I$ and $\default_C$, funding spread type), Monte Carlo parameters (number of paths), approximation parameters ($n_{\shortRate}$, $n_{\intensity}$, $n_a$).}
    \KwOut{$\FVA(t)$ and $\FVAWWR(t)$ for the chosen method (approximation or Monte Carlo).}
    \BlankLine

    Model calibration to market data and/or set synthetic model parameters. \;

    Simulating the market scenarios, and computing future portfolio values. \;

    \For{$u \leftarrow t_0$ \KwTo $t_N$}{
        Compute $\EPEFVAIndep{t}{u}$ using Equation~\eqref{eq:epeIndepCreditb}. \;
        \Switch{Method}{
            \uCase{Approximation}{Compute $\EPEFVAWWR{t}{u}$ using Equation~\eqref{eq:epeWWRCredit3}, where
            \begin{itemize}
              \item for a generic portfolio, $\condExpSmall{ y_{\shortRate}^{l}(t,u)  \maxOperator{\tradeVal(u)}}{t}$ is computed through Monte Carlo;
              \item for an IR swap, $\condExpSmall{ y_{\shortRate}^{l}(t,u)  \maxOperator{\tradeVal(u)}}{t}$ is approximated using Equation~\eqref{eq:approxExp3}.
            \end{itemize}}
            \Case{Monte Carlo}{Compute $\EPEFVAWWR{t}{u}$ using Equation~\eqref{eq:epeWWRCredit1}.\;}
        }
        Compute $\EPEFVA{t}{u} = \EPEFVAIndep{t}{u} + \EPEFVAWWR{t}{u}$. \;
    }

    Numerical integration of $\FVAIndep(t)$ and $\FVAWWR(t)$ from Equation~\eqref{eq:fca2}:
    \begin{align} \displaystyle
      \FVA(t)
        &= \sum_{i=0}^{N-1} \int_{t_i}^{t_{i+1}}  \EPEFVA{t}{u} \du
        \approx  \sum_{i=0}^{N-1}  \left(t_{i+1} - t_i\right) \EPEFVA{t}{t_{i+1}}. \label{eq:fca6}
    \end{align}\;
    \caption{$\FVA$ calculation procedure.}
    \label{algo:fvaCalculationProcedure}
\end{algorithm}

We use market data that is representative of the market situation in April 2020, following the market stress in March 2020~\cite{ZwaardGrzelakOosterlee202210}.
At that time, interest rates were negative and continued to drop, and credit quality was deteriorating for all ratings.
In particular, we use yield curves with low or negative interest rates, and credit curves corresponding AAA and BBB ratings.
The detailed set of model inputs can be found in~\ref{app:params}.

The IR-credit correlations $\corr_{\shortRate,z}$ are chosen manually, but we focus on negative correlations to match the WWR scenario for receiver swaps in case of a negative correlation between IR and funding spreads.
For the stochastic funding spread assumption, this implies negative IR-credit correlations.
This adverse relationship between IR and credit is also understood through the empirical evidence that in case of an IR drop, credit spreads will widen in response to a flight-to-quality by investors.
These market moves are typically observed during a bear market (financial downturn).
In general, the lower the credit quality of a party, the higher the IR-credit correlation is in the absolute sense.
The reverse of this argument holds for financially good times: if the default frequency goes down, credit spreads will go down, while at the same time, the central bank might raise IR to counter inflation.
In practical settings, one cannot handpick correlation values, but a historical calibration can be performed based on the correlation between time series.
For example, the correlation can be calibrated by taking the terminal correlation between the swap rate and CDS par rate as generated by the model setup, and match that to the historical correlation between those quantities.

When presenting $\FVA$ and WWR numbers, the percentage of $\FVAWWR(t)$ w.r.t. $\FVAIndep(t)$, i.e., $\frac{\FVAWWR(t)}{\FVAIndep(t)} \cdot 100\%$, is denoted by `$\WWR \%$'.
Furthermore, `$\WWR$ RD' refers to the relative difference w.r.t. the Monte Carlo result, i.e., $\frac{\FVA(t) - \FVA^{\text{MC}}(t)}{\FVA^{\text{MC}}(t)}$.
This puts the WWR approximation error in the perspective of the overall exposure size, summarized in the $\FVA$ numbers.
The runtime (in seconds) refers to the extra computing time to compute the $\FVA$ WWR according to the specified method.

In Section~\ref{sec:numericalResultsIRSwap}, the performance of the approximation is assessed in the case of a single IR swap.
Then, in Section~\ref{sec:numericalResultsPortfolio}, we examine the practically relevant case of a multi-currency portfolio of IR swaps.
The approximation errors are analyzed in Section~\ref{sec:numericalResultsError}.
Finally, the approximation of WWR sensitivities is discussed in Section~\ref{sec:numericalResultSensi}.

\subsection{Single IR swap}  \label{sec:numericalResultsIRSwap}
The first example is that of a single receiver IR swap $\tradeVal$, expiring in $1$ year and maturing in $30$ years, paying yearly coupons on both legs, with zero margin on the floating leg, based on a $10000$ notional such that the value can be interpreted as a basis point value.
The model parameters used in this example are given in~\ref{app:paramsSingleIRS}.
For a receiver swap, the aforementioned market conditions are an example of WWR, with an adverse relationship between IR and credit.
For a single IR swap, the fully analytic WWR exposure approximation from Section~\ref{sec:exampleIRSwap} can be used.

\begin{figure}[h]
  \centering
  \begin{subfigure}[b]{\resultFigureSize}
    \includegraphics[width=\linewidth]{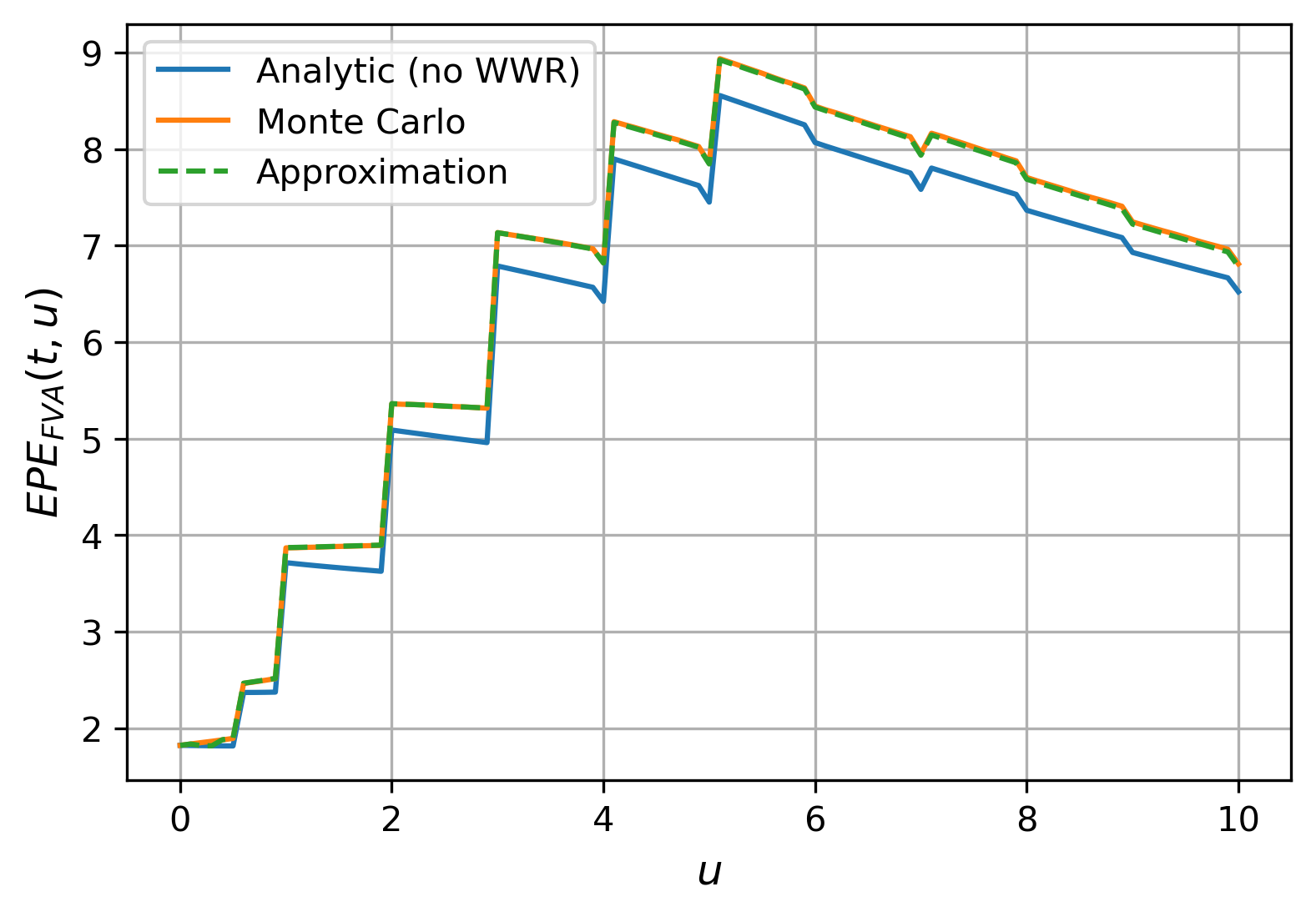}
    \caption{Zoomed exposures.}
    \label{fig:singleSwap31ITMInclInclExpZoom}
  \end{subfigure}
  \begin{subfigure}[b]{\resultFigureSize}
    \includegraphics[width=\linewidth]{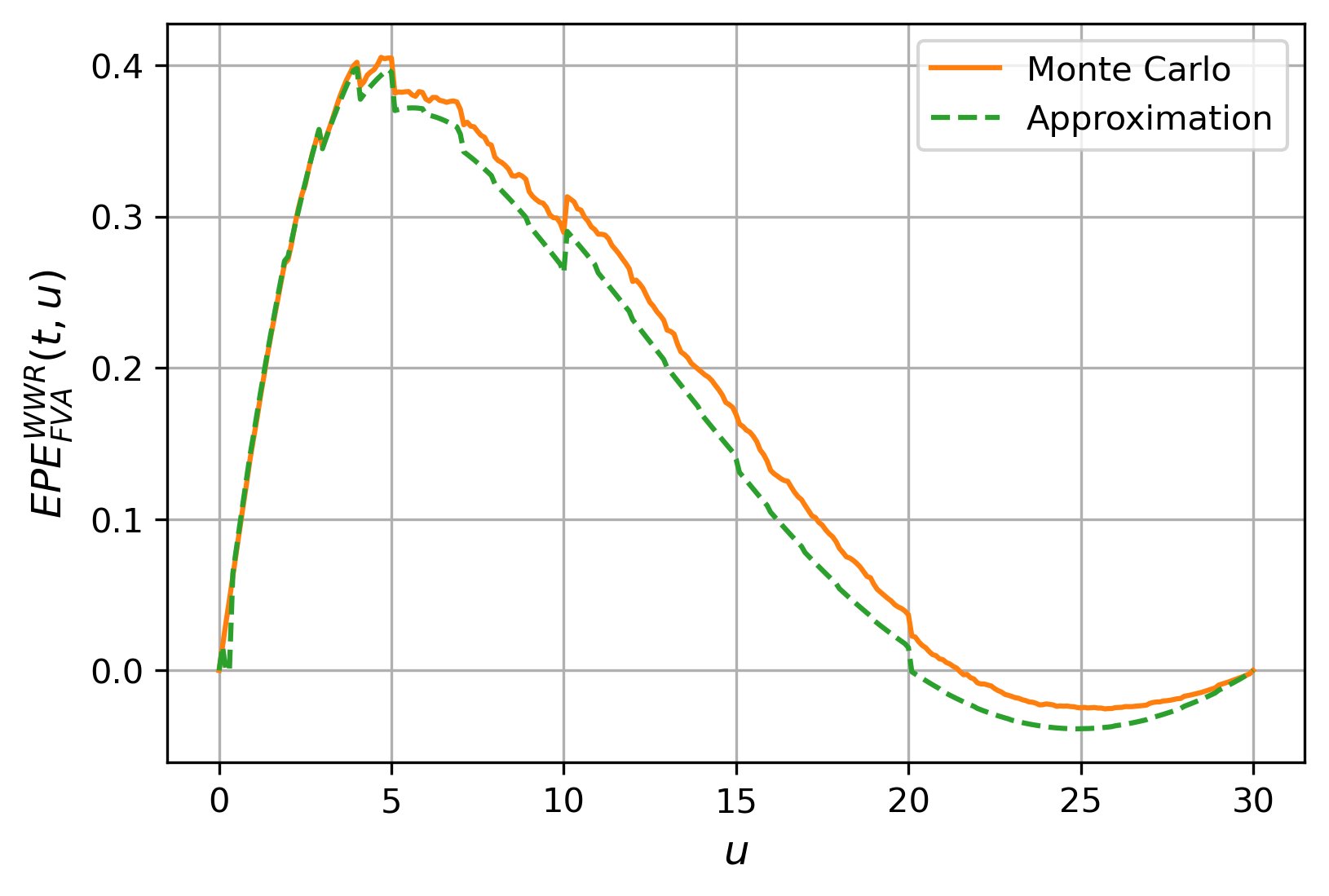}
    \caption{WWR exposure.}
    \label{fig:singleSwap31ITMInclInclWWR}
  \end{subfigure}
  \begin{subfigure}[b]{\resultFigureSize}
    \includegraphics[width=\linewidth]{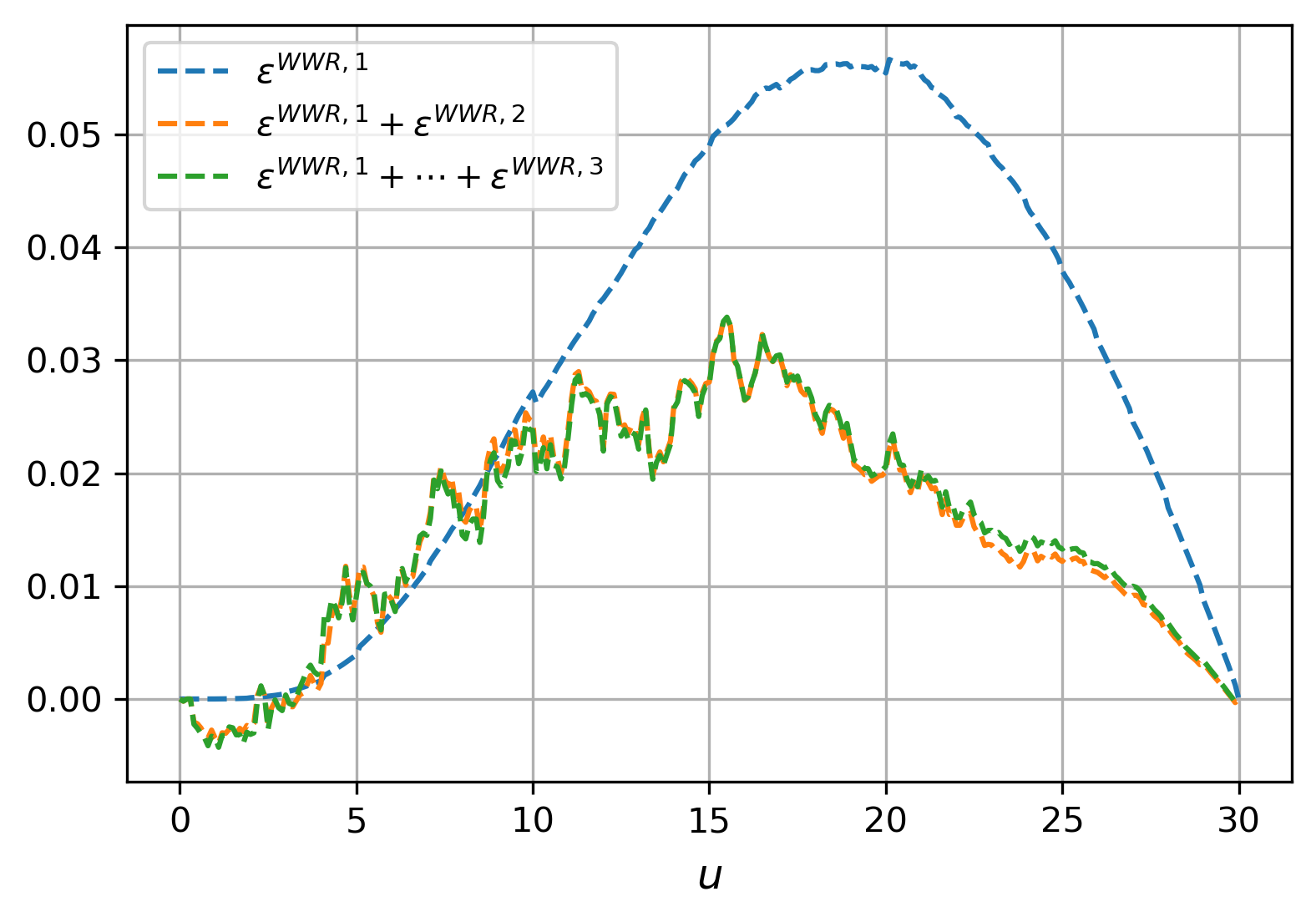}
    \caption{Approximation error.}
    \label{fig:singleSwap31ITMInclInclErrorCum}
  \end{subfigure}
  \caption{Exposures for a single ITM receiver swap.}
  \label{fig:singleSwap31ITMInclIncl}
\end{figure}

The various exposures (for the first $10$ out of $30$ years) and approximation errors are presented in Figure~\ref{fig:singleSwap31ITMInclIncl}.
From Figure~\ref{fig:singleSwap31ITMInclInclExpZoom}, one can immediately conclude that the WWR component of the exposure needs to be included in the $\FVA$ computation.
The approximation captures the same WWR pattern as the Monte Carlo benchmark, albeit at a slightly different scale, which is also visible from Figure~\ref{fig:singleSwap31ITMInclInclWWR}.
This is the result of the overall approximation error $\errorWWRPartOne + \errorWWRPartTwo + \errorWWRPartThree$ as plotted in Figure~\ref{fig:singleSwap31ITMInclInclErrorCum}.
This figure also shows that the Taylor series approximation $\errorWWRPartOne$ and the Gaussian approximation $\errorWWRPartTwo$cancel out to some degree.
Furthermore, the product-specific Taylor approximation $\errorWWRPartThree$ from Section~\ref{sec:exampleIRSwap}, which allows WWR exposures to be computed analytically, is of high quality.
All in all, in the presented example, the WWR exposure approximation is of high quality.

Table~\ref{tab:CreditSpreadParamSet31ReceiverITMDefaultIinclDefaultCinclFvaResults} reports the $\FVA$ numbers and WWR runtimes corresponding to the exposures from Figure~\ref{fig:singleSwap31ITMInclIncl}.
Naturally, the high approximation quality is preserved when looking at $\FVA$ numbers.
The WWR contribution, being roughly $3.9\%$ of the overall exposure, is non-negligible.
Also, looking at the runtimes, the approximation yields a significant speedup of more than 20 times compared to the benchmark methodology to compute the WWR component.
Hence, next to being of high quality, the approximation is also fast, and the approximation runtime does not scale in the number of derivatives.
In this particular example, the analytic approximation from Section~\ref{sec:exampleIRSwap} is used.
Hence, the computation of $\condExpSmall{ y_{\shortRate}^{l}(t,u)  \maxOperator{\tradeVal(u)} }{t}$ is analytical, and therefore slightly faster as the generic approach where this expectation is computed as a Monte Carlo average over the existing set of simulated paths.
\begin{table}[h]
  \renewcommand{\arraystretch}{1.1}
  \centering
  \footnotesize
  \begin{tabular}{l|rrrrr}
                          & $\FVA(t)$ & $\FVAWWR(t)$ & $\WWR \%$ & $\WWR$ RD & Runtime (sec) \\ \hline
     Analytic (no WWR)    & 122.3386 & 0.0000 & 0.0000 & 0.0000 & 0.00 \\
     Monte Carlo          & 127.0546 & 4.7160 & 3.8549 & 0.0000 & 6.48 \\
     Approximation        & 126.5516 & 4.2130 & 3.4437 & -0.0040 & 0.27 \\
  \end{tabular}
  \caption{$\FVA$ values for a single ITM receiver swap.}
  \label{tab:CreditSpreadParamSet31ReceiverITMDefaultIinclDefaultCinclFvaResults}
\end{table}

The behaviour of the approximation errors for the single IR swap portfolio will be analyzed in Section~\ref{sec:numericalResultsError}.

\subsection{Multi-currency portfolio of IR swaps}  \label{sec:numericalResultsPortfolio}
The encouraging results for the single IR swap motivate us to look at a practically relevant example.
Therefore, we consider a portfolio of swaps in multiple currencies, similarly as in~\cite[Section 3.2]{Grzelak202212}, i.e., portfolio $\tradeVal(t)$ in domestic currency $d$, with also trades in foreign currencies $f_1, \ldots, f_{N_f}$.
In this case, dynamics are required for $\shortRate_d$, and for every foreign currency $f_i$ two processes are needed: $\shortRate_{f_i}$ and $\FX_{f_i}^d$.
Naturally, all these processes are correlated with each other, as well as with the credit processes.
The portfolio contains $M_d$ swaps in domestic currency $d$, and $M_{f_k}$ swaps in foreign currency $f_k$ for each of the $N_f$ foreign currencies, and is then denoted as:
\begin{align}
  \tradeVal(t)
    &= \sum_{i=1}^{M_d} \tradeVal_i^d\left(t, \shortRate_d(t)\right)
     + \sum_{k=1}^{N_f} \FX_{f_k}^d(t) \sum_{i=1}^{M_{f_k}} \tradeVal_i^{f_k}\left(t, \shortRate_{f_k}(t)\right). \label{eq:portfolio}
\end{align}
Here, swap $\tradeVal_i^{f_k}(t)$ is in foreign currency $f_k$ at time $t$, and is then converted to domestic currency $d$ using the FX rate at that date, i.e., $\FX_{f_k}^d(t)$.

For the modelling, we use HW1F for the IR processes, and a GBM process for the FX rates.
See~\ref{app:dynamics} for the model dynamics in the form introduced in Section~\ref{sec:setupSummary}.

As an example, consider the currencies (EUR, USD, GBP) $= (d, f_1, f_2)$, such that the following processes are involved: $\shortRate_d$, $\shortRate_{f_1}$, $\shortRate_{f_2}$, $\FX_{f_1}^d$, $\FX_{f_2}^d$, $\intensity_I$, $\intensity_C$.
The example portfolio consists of two swaps in each currency, i.e., $M_{d} = M_{f_1} = M_{f_2} = 2$.
The notional is 10k for all swaps, such that the $\FVA$ results can be expressed in basis points, similarly to the single IR swap case.
The model parameters used in this example are given in~\ref{app:paramsPortfolio}.
All swaps are receiver swaps, have a maturity of 30 years, with a yearly frequency, and either an expiry 1 year or 10 years from now.
In this case, the generic approximation from Section~\ref{sec:approxWWR} is used, where expectations $\condExpSmall{ y_{\shortRate}^{l}(t,u)  \maxOperator{\tradeVal(u)} }{t}$, $l \in \{1,2,\ldots\}$, are computed using Monte Carlo simulation, since everything has been pre-computed in the $\xva$ calculation where no WWR is present.

\begin{figure}[h]
  \centering
  \begin{subfigure}[b]{\resultFigureSize}
    \includegraphics[width=\linewidth]{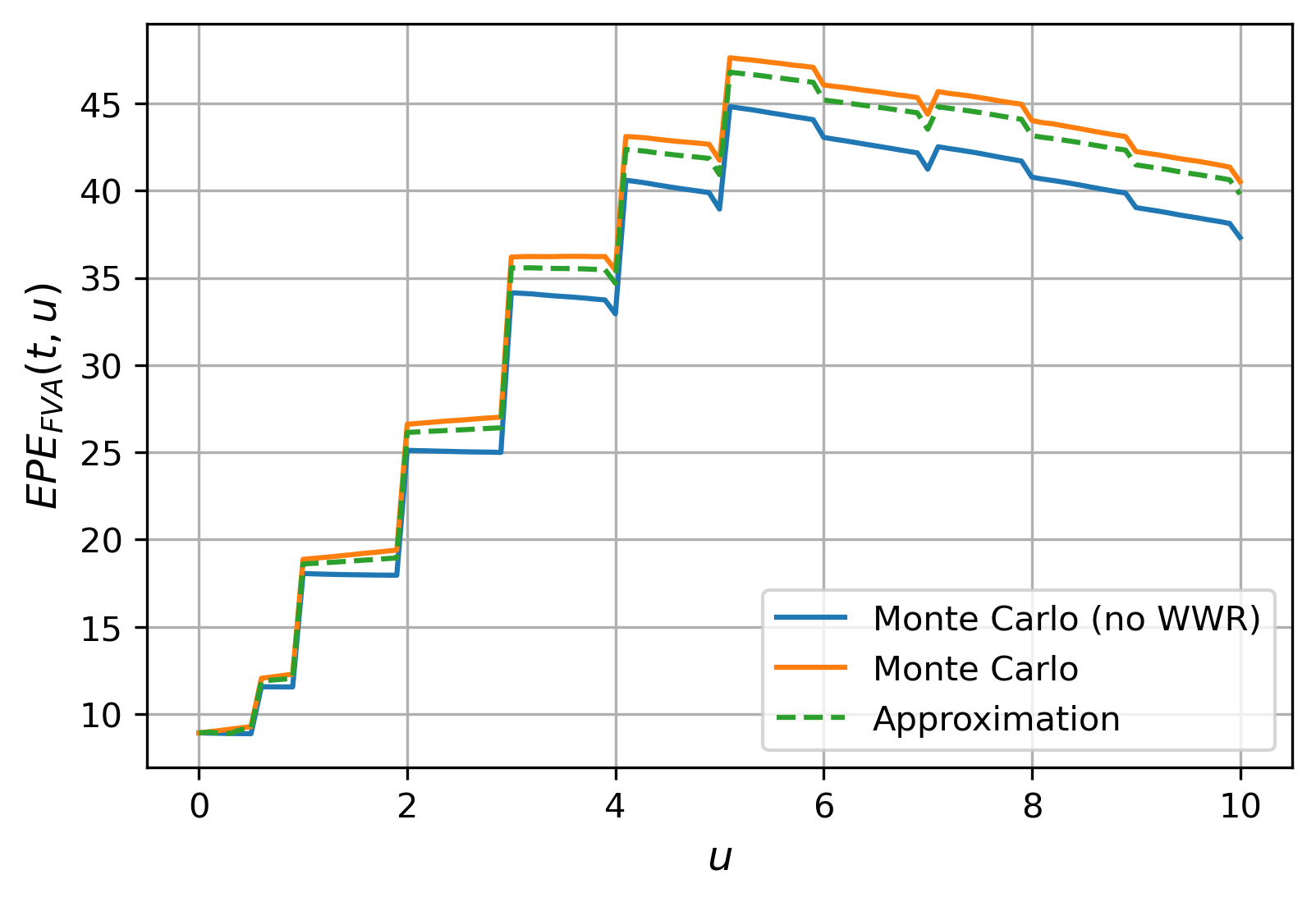}
    \caption{Zoomed exposures.}
    \label{fig:portfolio38ITMInclInclExpZoom}
  \end{subfigure}
  \begin{subfigure}[b]{\resultFigureSize}
    \includegraphics[width=\linewidth]{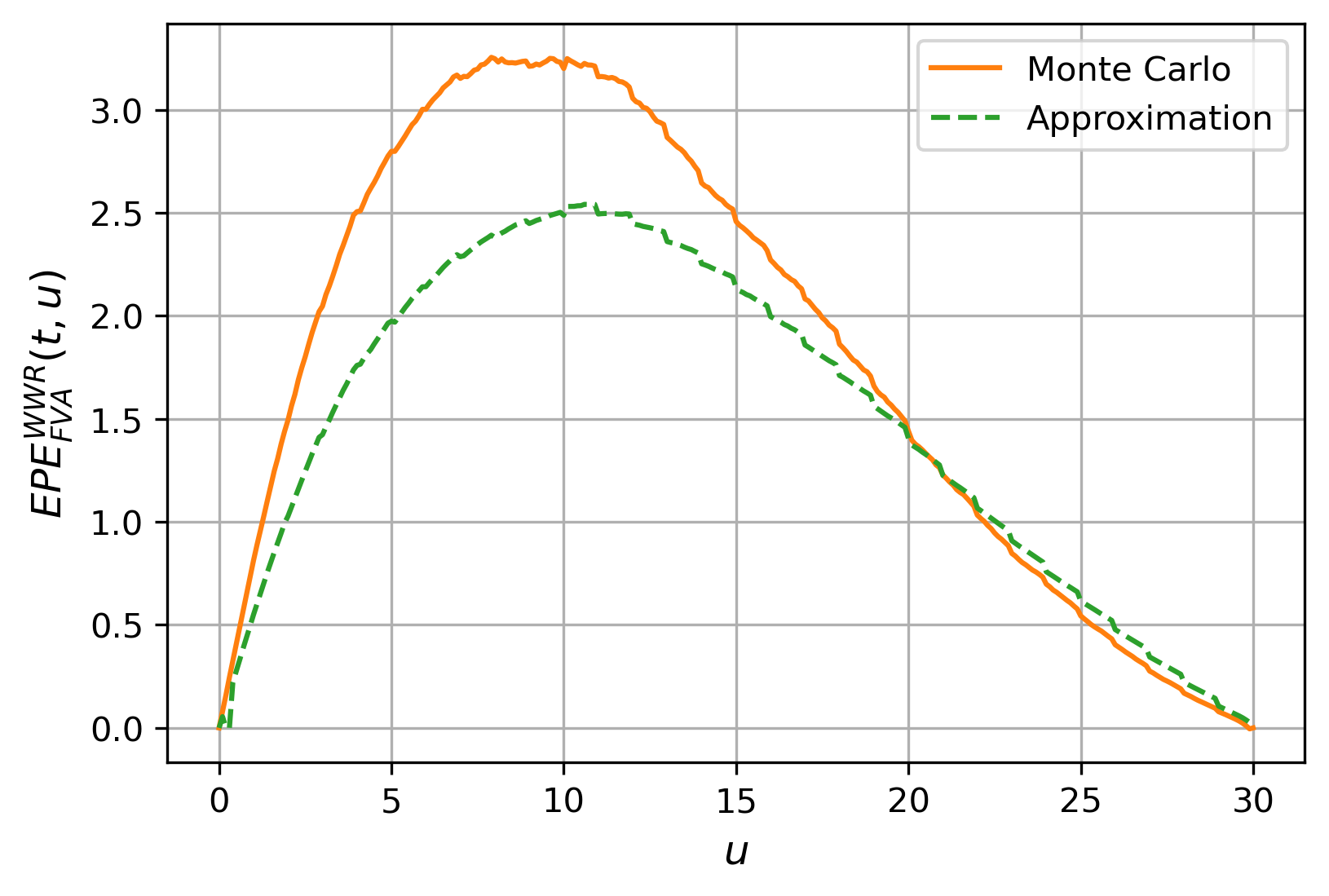}
    \caption{WWR exposure.}
    \label{fig:portfolio38ITMInclInclWWR}
  \end{subfigure}
  \begin{subfigure}[b]{\resultFigureSize}
    \includegraphics[width=\linewidth]{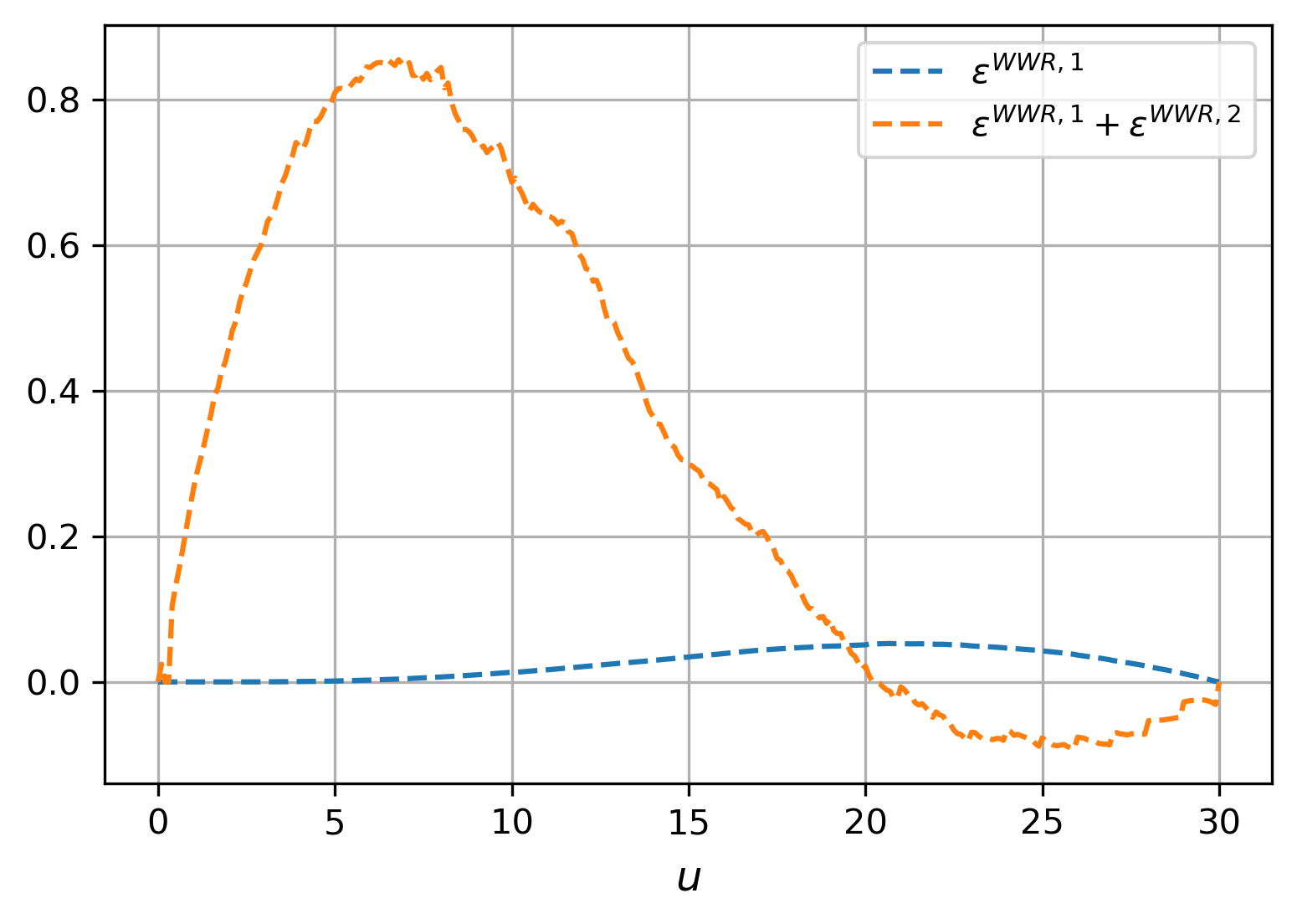}
    \caption{Approximation error.}
    \label{fig:portfolio38ITMInclInclErrorCum}
  \end{subfigure}
  \caption{Exposures for a portfolio of ITM receiver swaps in multiple currencies.}
  \label{fig:portfolio38ITMInclIncl}
\end{figure}

The various exposures and approximation errors are presented in Figure~\ref{fig:portfolio38ITMInclIncl}.
Again, the WWR is non-negligible, and the approximation captures the global pattern of the WWR, despite a larger difference in WWR exposure compared to the single IR swap case, see Figure~\ref{fig:portfolio38ITMInclInclWWR}.
The majority of the increase in overall approximation error can be attributed to the Gaussian approximation error $\errorWWRPartTwo$, see Figure~\ref{fig:portfolio38ITMInclInclErrorCum}.

When looking at the approximation error in terms of the $\FVA$ difference, Table~\ref{tab:PortfolioCreditSpreadParamSet38ITMDefaultIinclDefaultCinclFvaResults} shows that the relative difference is roughly $1.4\%$.
We are approximating the WWR that the full model generates based on the chosen model dynamics.
It is key to understand that this reference WWR is not a `true' market value to begin with but a consequence of model choices.
Despite not capturing all the WWR from the reference, the approximation gives a useful and practical indication of the amount of WWR in a portfolio.
Especially due to the fast computation when compared with the Monte Carlo benchmark.
\begin{table}[h]
  \renewcommand{\arraystretch}{1.1}
  \centering
  \footnotesize
  \begin{tabular}{l|rrrrr}
                          & $\FVA(t)$ & $\FVAWWR(t)$ & $\WWR \%$ & $\WWR$ RD & Runtime (sec) \\ \hline
     Monte Carlo (no WWR) & 658.4338 & 0.0000 & 0.0000 & 0.0000 & 0.00 \\
     Monte Carlo          & 713.0611 & 54.6272 & 8.2965 & 0.0000 & 7.59 \\
     Approximation        & 703.2882 & 44.8544 & 6.8123 & -0.0137 & 0.47 \\
  \end{tabular}
  \caption{$\FVA$ values for a portfolio of ITM receiver swaps in multiple currencies.}
  \label{tab:PortfolioCreditSpreadParamSet38ITMDefaultIinclDefaultCinclFvaResults}
\end{table}

The approximation errors presented for this directional portfolio of linear derivatives can be considered as a lower bound for portfolios with more complicated derivatives.
Given that $\xva$ makes all derivatives non-linear, the presented portfolio is a simple but relevant one, as many of these portfolios exist in practice.
The behaviour of the approximation errors for the multi-currency portfolio of IR swaps will be analyzed further in Section~\ref{sec:numericalResultsError}.

\subsection{Error analysis}  \label{sec:numericalResultsError}

In the various Taylor approximations, a truncation point need to be chosen, i.e., $n_{\intensity}$ and $n_{\shortRate}$ in Eq.~\eqref{eq:errorWWR1}, and $n_a$ in Eq.~\eqref{eq:approxExp3} in the case of a fully analytic WWR approximation for an IR swap.
We assess appropriate choices of these truncation points, as well as their interplay with the Gaussian approximation error, which is particularly relevant for the choice of $n_{\intensity}$.
Furthermore, we examine how the Gaussian approximation error behaves, particularly in the context of various portfolios with different shapes of exposure profiles, as well as for cases where  the distribution of $Y_z(t,u)$ deviates from a Gaussian. 
The latter is also illustrated through an additional numerical example of a stressed market scenario.

\paragraph{Choice of $n_{\intensity}$} The use of $n_{\intensity}=1$ is hard-coded in Section~\ref{sec:fvaEquationFundingSpreadWithWWR}, and results in a first-order approximation of the credit adjustment factors, i.e., $\taylorTrunc{0}{n_{\intensity}}\left(Y_I(t,u) + Y_C(t,u)\right) = 1 - Y_I(t,u) - Y_C(t,u)$.
Consecutively, the Gaussian approximation is applied to these terms of the Taylor approximation.
The Gaussian approximation will be worse for higher-order Taylor terms if the distribution of the approximated variable is far from normal.
This implies that the approximation quality cannot be improved by increasing $n_{\intensity}$.
By $n_{\intensity}=1$, the Gaussian approximation is limited to first-order terms, limiting the Gaussian approximation error $\errorWWRPartTwo$.
At the same time, the Taylor truncation error $\errorWWRPartOne$ will be larger in absolute sense for lower $n_{\intensity}$.
Therefore, there is a tradeoff between $\errorWWRPartOne$ and $\errorWWRPartTwo$ when choosing $n_{\intensity}$.
In some cases, for example, when the IR volatility dominates, and the credit volatilities are low, $n_{\intensity}=2$ yields a slightly improved overall approximation than $n_{\intensity}=1$.
However, since the credit distributions are typically not that close to Gaussian, the choice of $n_{\intensity}=1$, leading to a first-order approximation, is appropriate and no calibration of this approximation parameter is required.

\begin{figure}[h]
  \centering
  \begin{subfigure}[b]{\resultFigureSize}
    \includegraphics[width=\linewidth]{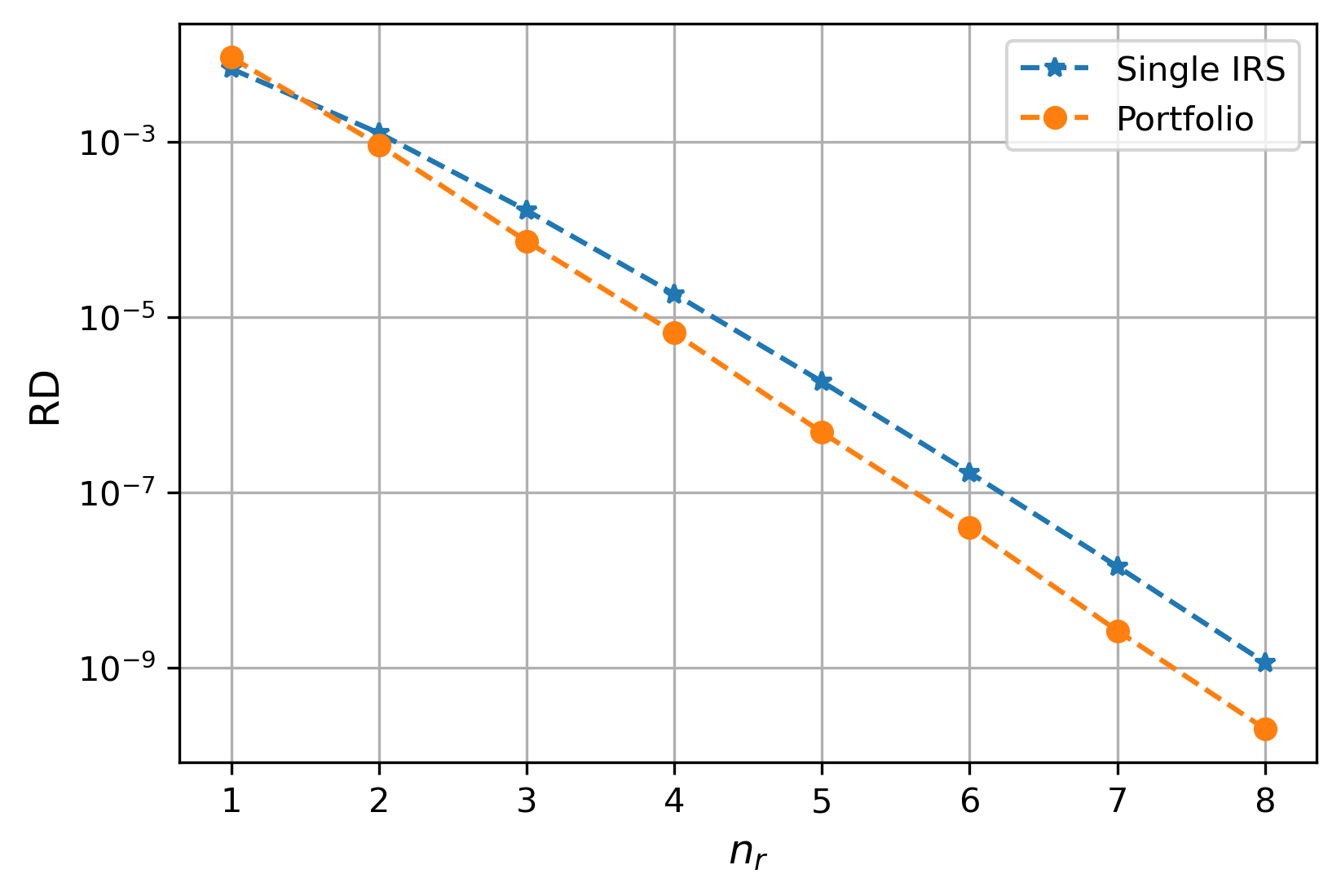}
    \caption{RD indicates the relative difference in $\FVA$ w.r.t. $\FVA$ for $n_{\shortRate} = 20$ terms.}
    \label{fig:NumberTermsDFTaylorError}
  \end{subfigure}
  \begin{subfigure}[b]{\resultFigureSize}
    \includegraphics[width=\linewidth]{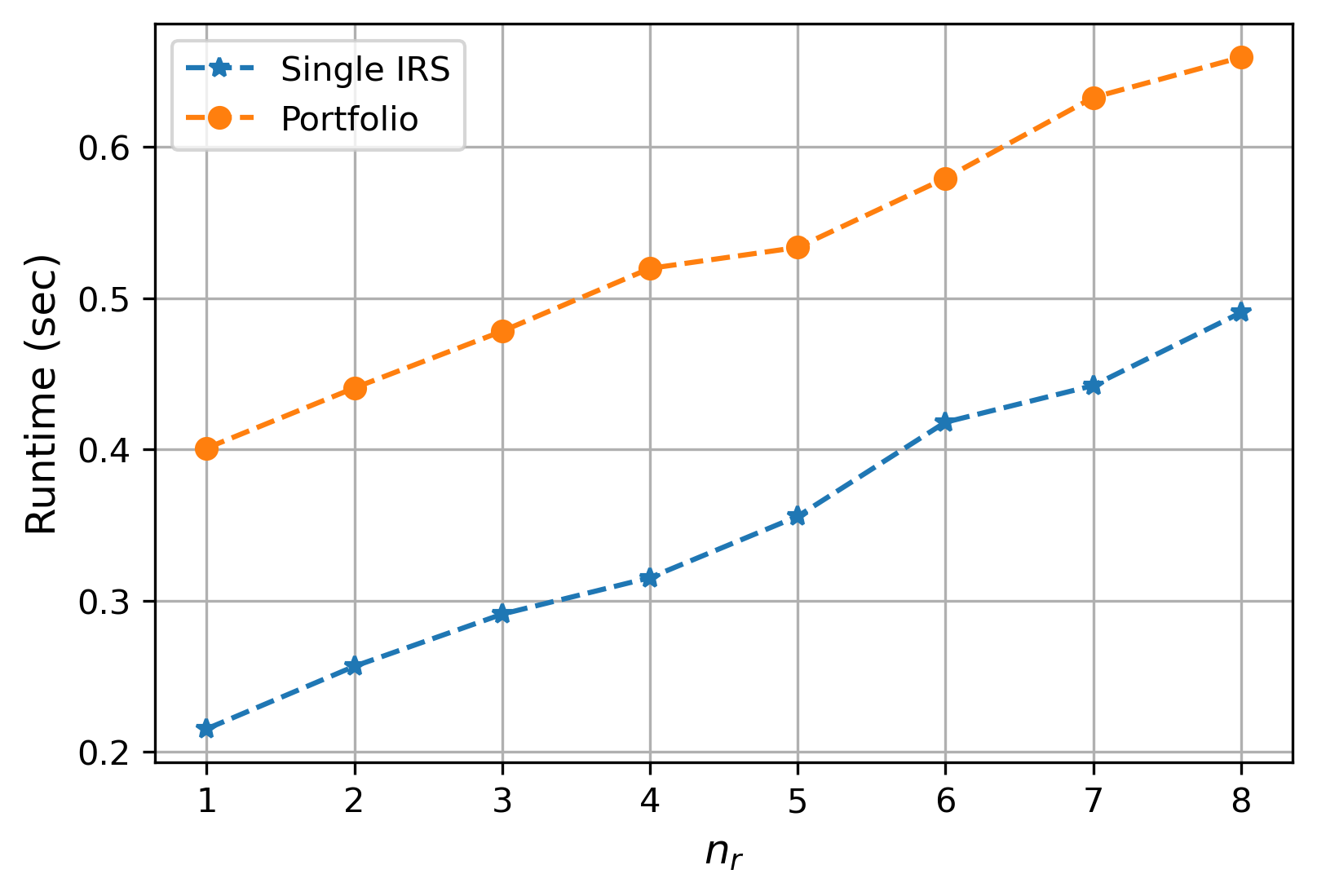}
    \caption{Approximation runtime in seconds.}
    \label{fig:NumberTermsDFTaylorRuntime}
  \end{subfigure}
  \caption{Impact of $n_\shortRate$ on the approximation error and runtime.}
  \label{fig:NumberTermsDFTaylor}
\end{figure}

\paragraph{Choice of $n_{\shortRate}$} Given $n_{\intensity}=1$, $n_{\shortRate}$ from Equation~\eqref{eq:epeWWRCredit2} is the next truncation point that needs to be chosen.
Since $n_{\shortRate}$ represents the truncation of the Taylor approximation involving $Y_{\shortRate}(t,u)$, and given that $Y_{\shortRate}(t,u)$ is normally distributed, no tradeoff with the Gaussian approximation needs to be considered.
In the results presented so far, $n_{\shortRate} = 5$ has been used.
From Figure~\ref{fig:NumberTermsDFTaylor} it is clear that the choice of $n_{\shortRate}=5$ is suitable, given the tradeoff between accuracy and speed, and no further significant improvement in accuracy can be achieved.
Alternatively, the number of terms $n_{\shortRate}$ can be made adaptive by imposing a stopping criterion on the contribution of the $n_{\shortRate}$'th term, but this will only lead to marginal improvement of accuracy.
In both cases, no calibration of the approximation parameter is required.

\begin{figure}[h]
  \centering
  \begin{subfigure}[b]{\resultFigureSize}
    \includegraphics[width=\linewidth]{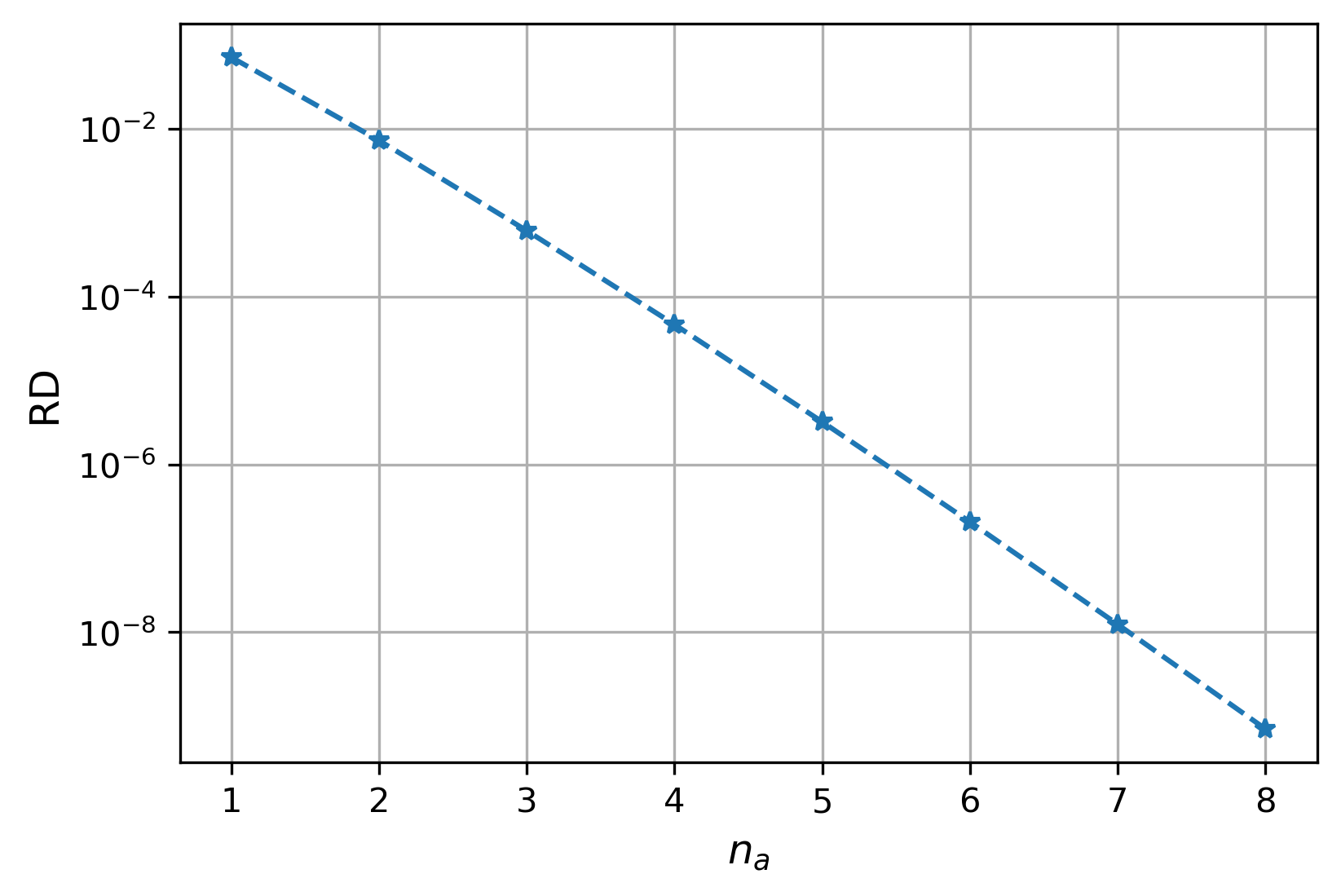}
    \caption{RD indicates the relative difference in $\FVA$ w.r.t. $\FVA$ for $n_a = 20$ terms.}
    \label{fig:NumberTermsProductTaylorError}
  \end{subfigure}
  \begin{subfigure}[b]{\resultFigureSize}
    \includegraphics[width=\linewidth]{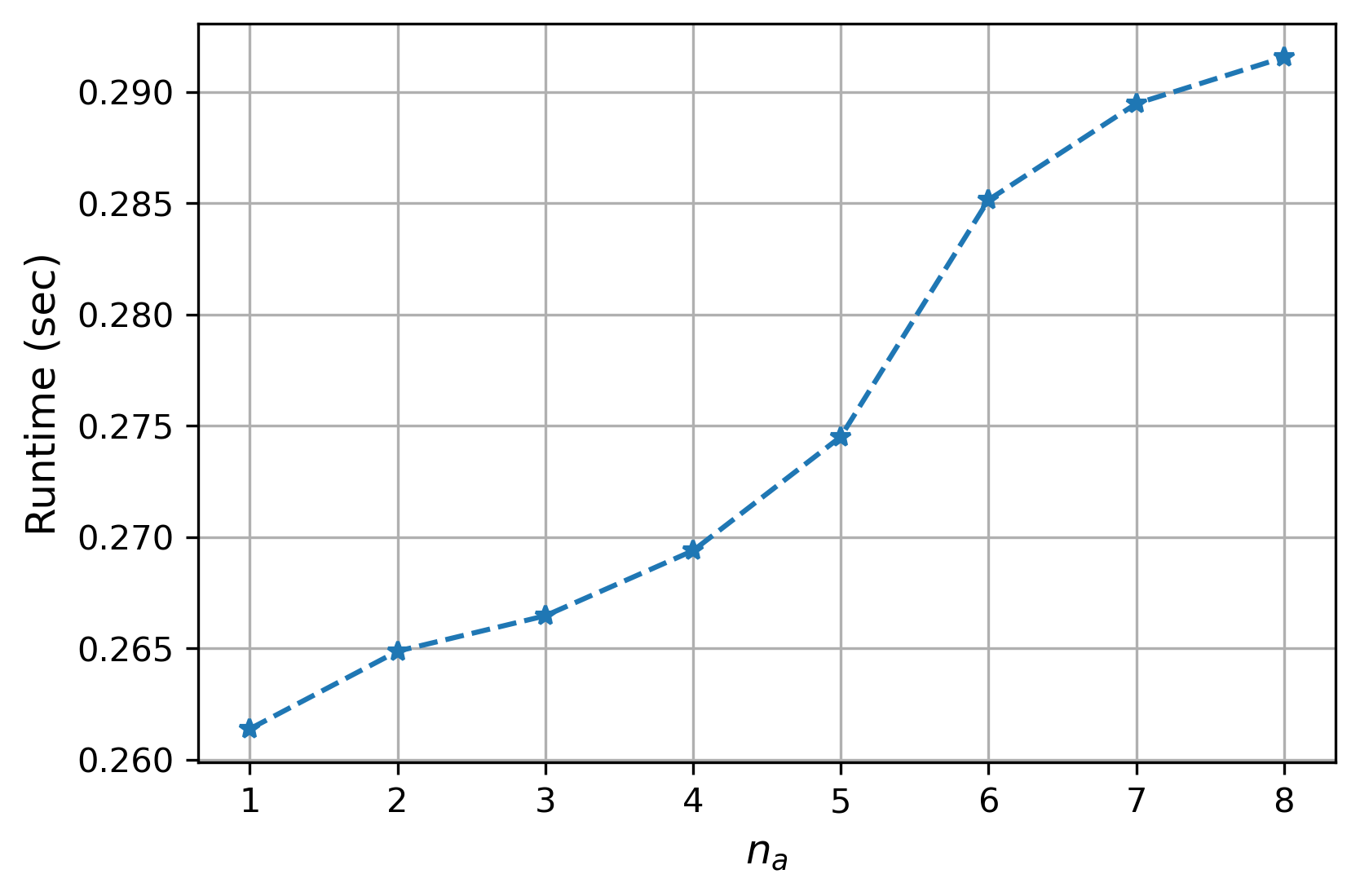}
    \caption{Approximation runtime in seconds.}
    \label{fig:NumberTermsProductTaylorRuntime}
  \end{subfigure}
  \caption{Impact of $n_a$ on the approximation error and runtime.}
  \label{fig:NumberTermsProductTaylor}
\end{figure}

\paragraph{Choice of $n_a$} For the fully analytic approximation for an IR swap from Section~\ref{sec:exampleIRSwap}, an additional truncation of a Taylor series is done in Equation~\eqref{eq:approxExp3} at the $n_a$'th term.
As this involves the truncation of $\taylor(y_{\shortRate}(t,u)B_{\shortRate}(u,T_k))$, and since $y_{\shortRate}(t,u)$ is Gaussian, no tradeoff with the Gaussian approximation needs to be considered.
In the results presented so far, $n_a = 5$ has been used.
From Figure~\ref{fig:NumberTermsProductTaylor} it is clear that the choice of $n_a=5$ is suitable.
Compared to $n_{\shortRate}$ from Figure~\ref{fig:NumberTermsDFTaylor}, the impact on the approximation runtime is much smaller.
Recall from Figure~\ref{fig:singleSwap31ITMInclInclErrorCum} that $\errorWWRPartThree$ has negligible impact on the overall approximation error.
Alternatively to $n_a = 5$, the number of terms $n_a$ can be made adaptive in a similar fashion as $n_{\shortRate}$.

\paragraph{Gaussian approximation error} As mentioned before, there is a tradeoff between the choice of $n_{\intensity}$ and the magnitude of the Gaussian approximation error.
Naturally, the Gaussian approximation error also depends on the distribution $Y_z(t,u)$.
The distribution of $Y_z(t,u)$ will increase in magnitude for later exposure dates $u$, so this pattern is also expected for the Gaussian approximation error: the error is expected to increase for larger $u$.
However, the portfolio decomposition may dampen this error, for example, due to a decaying exposure profile for later exposure dates.
This is precisely the case for the single IR swap in Figure~\ref{fig:singleSwap31ITMInclInclErrorCum} and for the multi-currency portfolio of IR swaps in Figure~\ref{fig:portfolio38ITMInclInclErrorCum}.

\begin{figure}[h]
  \centering
  \begin{subfigure}[b]{\resultFigureSize}
    \includegraphics[width=\linewidth]{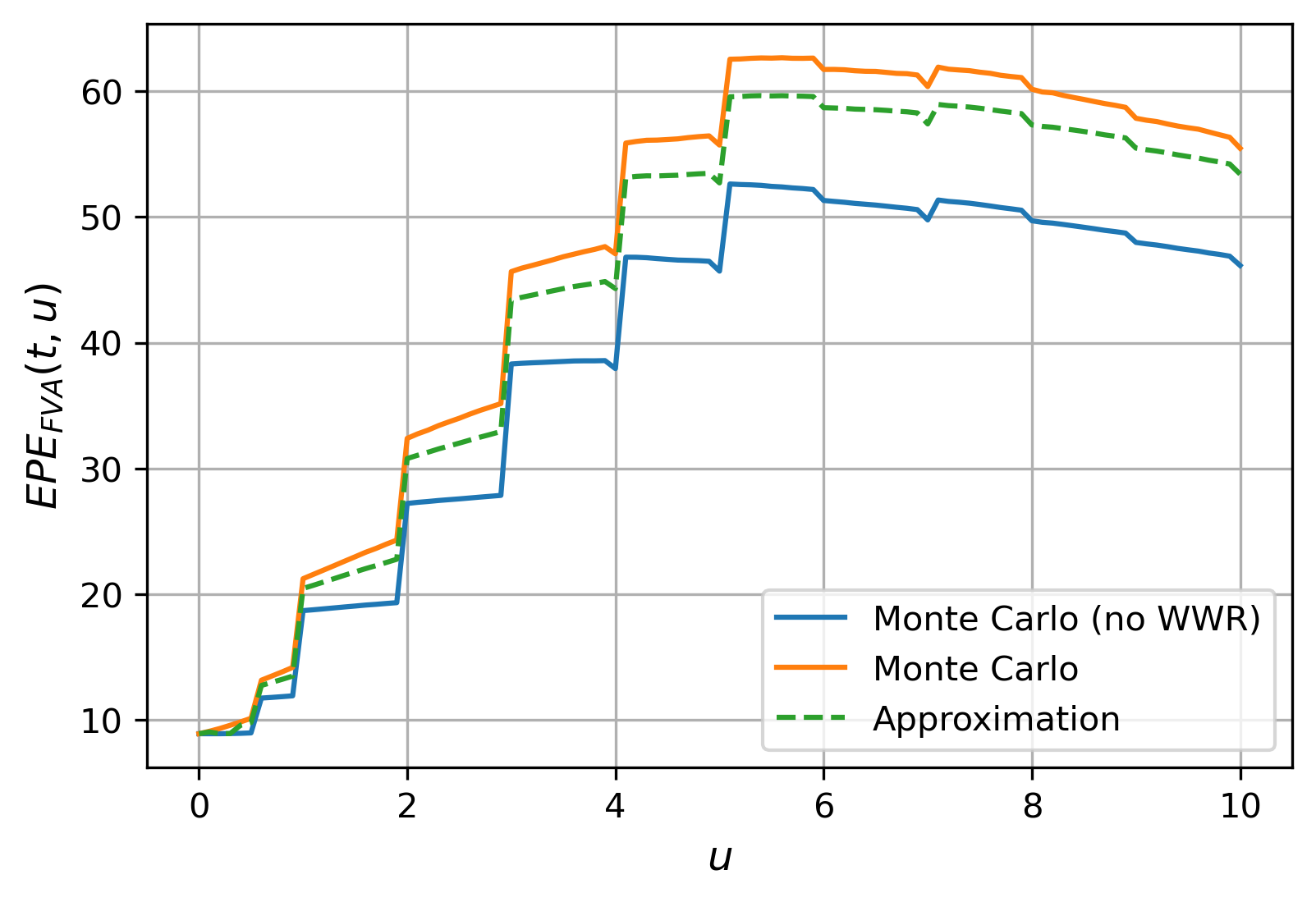}
    \caption{Zoomed exposures.}
    \label{fig:portfolio38StressedITMInclInclExpZoom}
  \end{subfigure}
  \begin{subfigure}[b]{\resultFigureSize}
    \includegraphics[width=\linewidth]{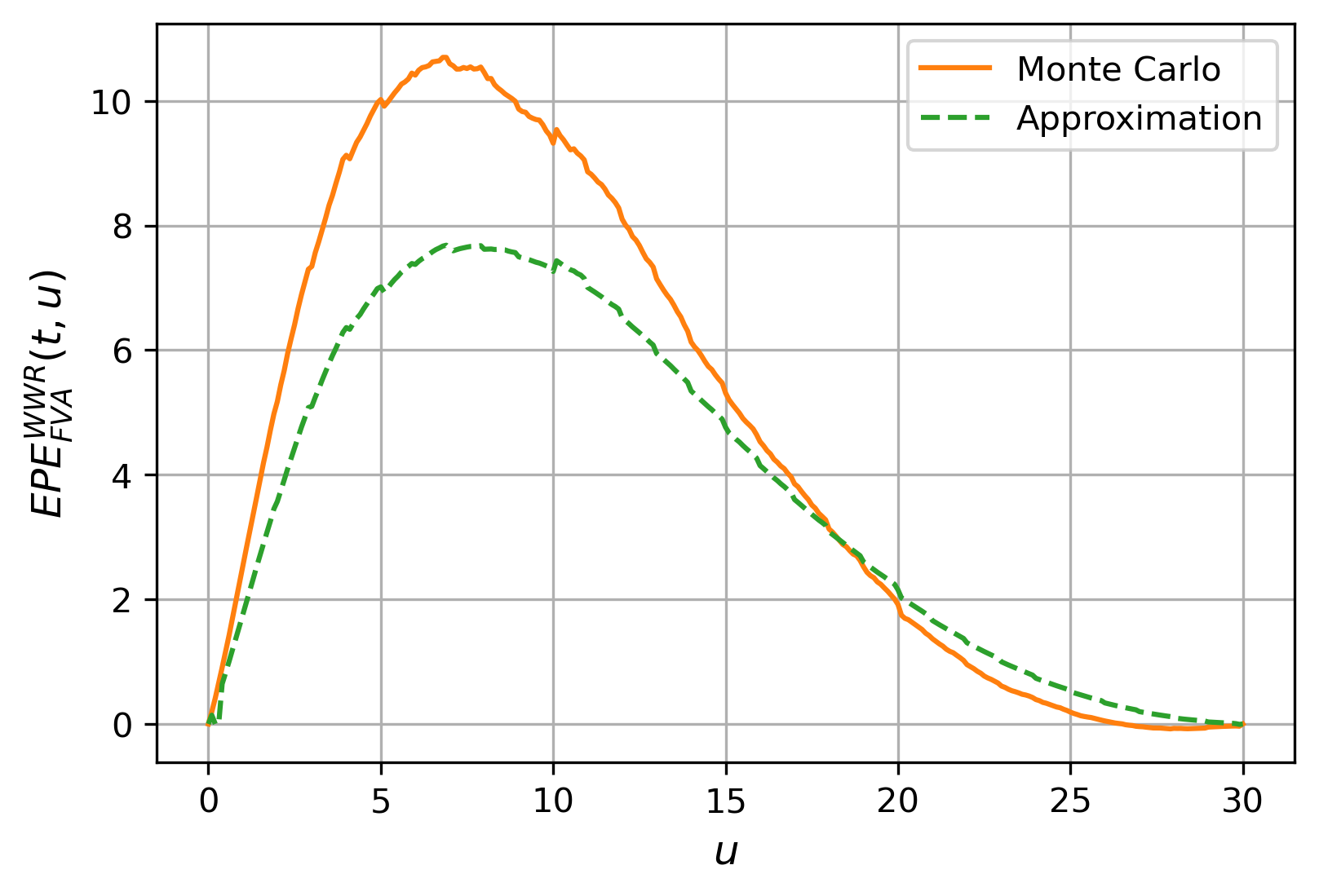}
    \caption{WWR exposure.}
    \label{fig:portfolio38StressedITMInclInclWWR}
  \end{subfigure}
  \begin{subfigure}[b]{\resultFigureSize}
    \includegraphics[width=\linewidth]{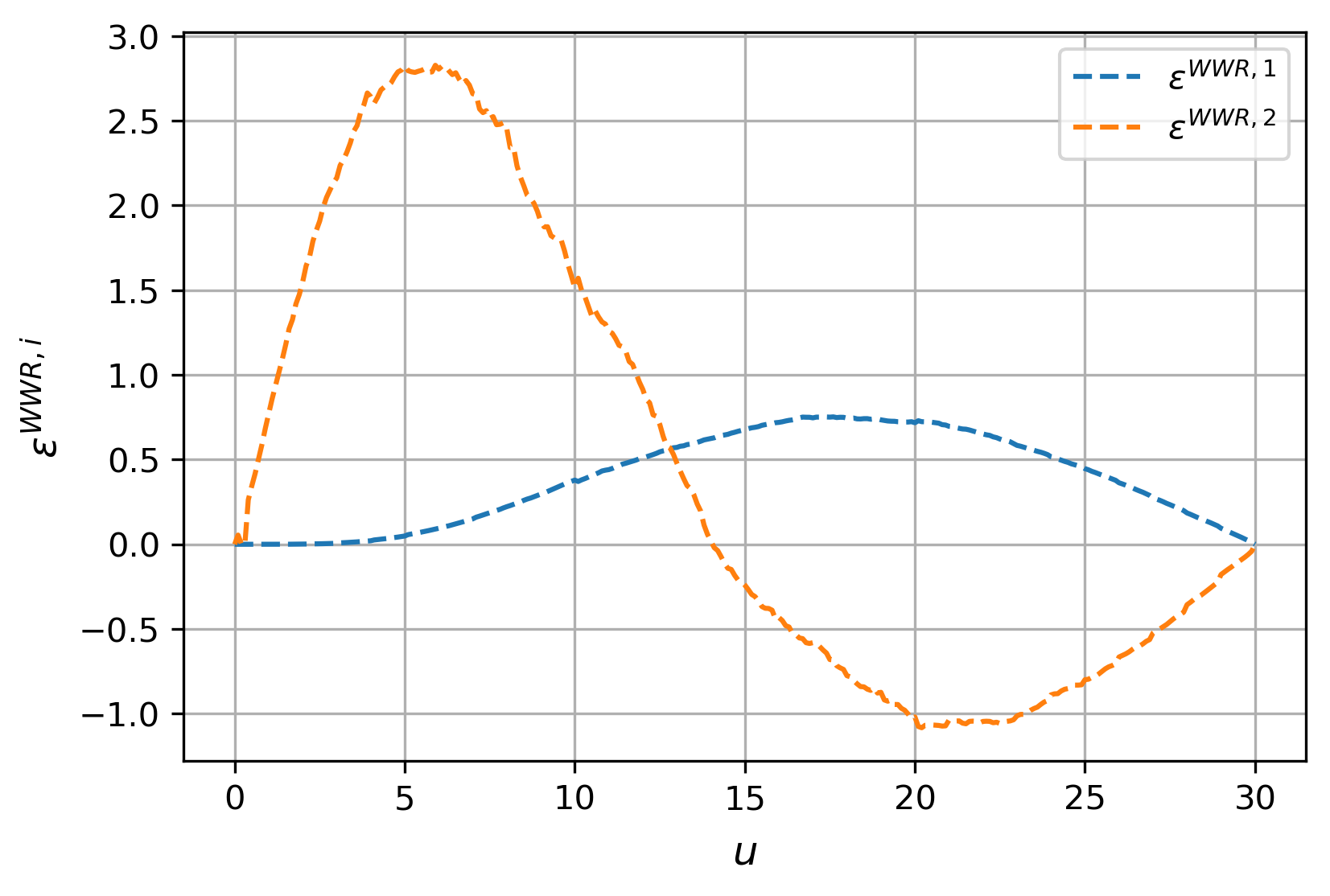}
    \caption{Incremental approximation error.}
    \label{fig:portfolio38StressedITMInclInclErrorIncr}
  \end{subfigure}
  \begin{subfigure}[b]{\resultFigureSize}
    \includegraphics[width=\linewidth]{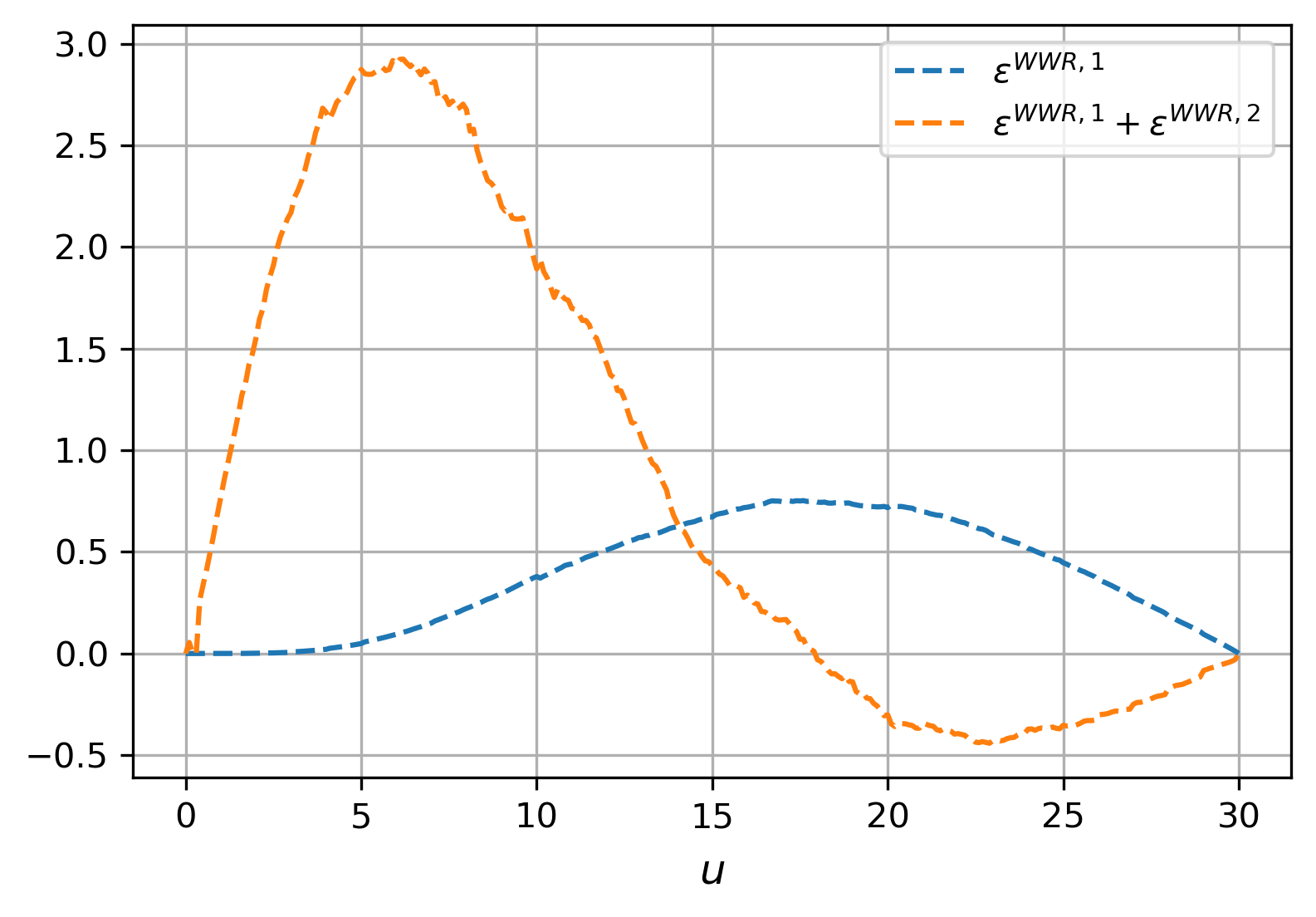}
    \caption{Cumulative approximation error.}
    \label{fig:portfolio38StressedITMInclInclErrorCum}
  \end{subfigure}
  \caption{Exposures for a portfolio of ITM receiver swaps in multiple currencies in a stressed market scenario.}
  \label{fig:portfolio38StressedITMInclIncl}
\end{figure}

On the other hand, the Gaussian approximation error is larger when the distribution of $Y_z(t,u)$ deviates from a Gaussian, where the distribution asymmetry and fatness of the distribution tails play a role.
A numerical example of a stressed scenario for the multi-currency portfolio of IR swaps is presented in Figure~\ref{fig:portfolio38StressedITMInclIncl} and Table~\ref{tab:PortfolioCreditSpreadParamSet38StressedITMDefaultIinclDefaultCinclFvaResults}, where especially in Figure~\ref{fig:portfolio38StressedITMInclInclErrorIncr} we observe the same portfolio dampening effect on the Gaussian approximation error.

\paragraph{Impact of model parameters} From Section~\ref{sec:approxError} it is clear that in some situations the WWR approximation performs less well.
Hence, we illustrate the impact of the model parameters on the various approximation errors.
Looking solely at the truncation error, increased credit volatility $\vol_I$ and $\vol_C$ will result in an increased error, and lower mean-reversion $a_I$ and $a_C$ will also increase the truncation error.
For the Gaussian approximation, as opposed to the truncation error, larger values of $a_I$ and $a_C$ lead to less normality in the credit distributions, increasing the error.
Like for the truncation error, for increased credit volatilities $\vol_I$ and $\vol_C$, the credit distributions deviate from Gaussian, such that the Gaussian approximation error increases.
This effect is especially visible for higher values of $\vol_I$ combined with low $\vol_C$, since the credit of institution $I$ enters through both the funding spread and the credit adjustment factor, whereas $C$'s credit enters only through the latter.
Furthermore, for larger IR volatility, the amount of WWR increases, and the overall approximation error increases along with it.
Finally, increased IR-credit correlation magnitudes also result in larger approximation errors.

In stressed market scenarios, the WWR approximation has its limitations, but as the analysis above illustrates, there is a sufficient level of understanding what the impact would be of various market stresses.
In Figure~\ref{fig:portfolio38StressedITMInclIncl} a stressed scenario is considered for the same portfolio as in Section~\ref{sec:numericalResultsPortfolio}.
The model parameters used in this example are given in~\ref{app:paramsPortfolioStressed}.
In comparison with Figure~\ref{fig:portfolio38ITMInclIncl}, the error pattern is comparable, but with a different scaling.
A part of this scaling is explained by the larger degree of WWR that is present in this example.

In terms of $\FVA$, see Table~\ref{tab:PortfolioCreditSpreadParamSet38StressedITMDefaultIinclDefaultCinclFvaResults}, the approximation error has indeed also increased, but is still at an acceptable level for the approximation to be practically relevant, especially given the large degree of WWR that is captured.
\begin{table}[h]
  \renewcommand{\arraystretch}{1.1}
  \centering
  \footnotesize
  \begin{tabular}{l|rrrrr}
                          & $\FVA(t)$ & $\FVAWWR(t)$ & $\WWR \%$ & $\WWR$ RD & Runtime (sec) \\ \hline
     Monte Carlo (no WWR) & 795.4959 & 0.0000 & 0.0000 & 0.0000 & 0.00 \\
     Monte Carlo          & 935.8899 & 140.3940 & 17.6486 & 0.0000 & 7.83 \\
     Approximation        & 908.3324 & 112.8365 & 14.1844 & -0.0294 & 0.53 \\
  \end{tabular}
  \caption{$\FVA$ values for a portfolio of ITM receiver swaps in multiple currencies in a stressed scenario.}
  \label{tab:PortfolioCreditSpreadParamSet38StressedITMDefaultIinclDefaultCinclFvaResults}
\end{table}

\subsection{Sensitivities} \label{sec:numericalResultSensi}
Next to the $\FVA$ values, the relevant hedge ratios are required from a risk-management point of view.
When computing these sensitivities w.r.t. market inputs, the $\FVA$ modelling assumptions play a significant role, impacting first-order delta and vega risks, and adding cross-gamma risks between the risk factors.

Using a Monte Carlo simulation to calculate $\FVA$, the corresponding sensitivities w.r.t. the various risk drivers can be computed efficiently using Algorithmic Differentiation (AD) techniques.
However, for a transparent implementation, some institutions resort to Finite Difference (FD) approximations, which require at least one recalculation of the entire Monte Carlo calculation chain for each sensitivity.
We approximate the sensitivities using FD, both for the Monte Carlo benchmark and the Gaussian approximation.
The latter also allows for several semi-analytic sensitivities, but for the proof of concept, we employ the generic FD method.

Following from Equation~\eqref{eq:fca2}, the sensitivity of $\FVA$ w.r.t. risk factor $\theta$ is given by
\begin{align} \displaystyle
  \pderiv{\FVA(t)}{\theta}
    &= \pderiv{\FVAIndep(t)}{\theta} + \pderiv{\FVAWWR(t)}{\theta}.  \label{eq:fca2Sensi}
\end{align}
With the WWR approximation, like for value, only the WWR component of the sensitivities, i.e., $\pderiv{}{\theta}\FVAWWR(t)$, is approximated using the Gaussian approximation.
The other part of the required sensitivity, $\pderiv{}{\theta} \FVAIndep(t)$, is taken from the current $\xva$ calculation, where independence between the funding spread and the market risks is assumed.
In this way, the WWR approximation error will not affect the `independent part' of the sensitivities.

From a risk-management and hedging perspective, the most relevant first-order risks are the deltas and vegas.
The IR and FX risks are the most relevant, as $\xva$ desks at financial institutions are likely to hedge these.

In addition to the first-order risks, the cross-gamma risks are insightful, especially for WWR effects.
These cross-gamma effects are always present, even if correlations are zero, but with non-zero correlations, they become more relevant.
Especially the cross-gamma risks with funding risk and market risk are essential: the funding risk increases if the market risk (exposure) increases.
Cross-gamma risks are difficult to hedge, but computing them will at least give us insight into this risk for a portfolio under consideration.

\begin{table}[h]
  \renewcommand{\arraystretch}{1.1}
  \centering
  \footnotesize
  \begin{tabular}{l|rrrr}
                                    &                  & $\FVAWWR(t)$   & $\FVAWWR(t)$       & \\
                                    & $\FVAIndep(t)$   & Monte Carlo    & Approximation      & $\WWR$ RD \\ \hline
    Value                           & 658.43           & 54.63          & 44.85              & -0.0137\\ \hline
    IR delta EUR                    & -32680.74        & -1629.76       & -1627.51           & 0.0001 \\
    IR delta USD                    & -25311.11        & -925.64        & -801.76            & 0.0047 \\
    IR delta GBP                    & -34115.96        & -1154.91       & -1042.12           & 0.0032 \\ \hline
    FX delta EUR/USD                & 202.14           & 16.70          & 10.22              & -0.0296\\
    FX delta EUR/GBP                & 221.77           & 15.83          & 10.18              & -0.0238\\ \hline
    Credit delta I                  & -4360.76         & 80.68          & 65.84              & -0.0035\\
    Credit delta C                  & 932.22           & 79.04          & 67.02              & -0.0119\\ \hline
    IR vega EUR                     & 729.18           & 170.16         & 212.04             & 0.0466 \\
    IR vega USD                     & 793.98           & 146.77         & 76.29              & -0.0749\\
    IR vega GBP                     & 1014.62          & 166.41         & 87.63              & -0.0667\\ \hline
    FX vega EUR/USD                 & 21.10            & -29.75         & -12.12             & 2.0380 \\
    FX vega EUR/GBP                 & 19.56            & -39.78         & -16.16             & 1.1682 \\ \hline
    Credit vega I                   & -4.55            & 872.67         & 800.98             & -0.0826\\
    Credit vega C                   & 0.02             & -215.31        & -207.46            & 0.0365 \\ \hline
    IR EUR / credit I cross-gamma   & 328215.00        & -2410.00       & -2372.00           & 0.0001 \\
    IR EUR / credit C cross-gamma   & -44484.00        & -2358.00       & -2430.00           & -0.0015\\
  \end{tabular}
  \caption{Sensitivities for a portfolio of ITM receiver swaps in multiple currencies.}
  \label{tab:FvaSensisParamSet38}
\end{table}

To measure the performance of the approximated WWR sensitivities, the relative difference of the complete sensitivity $\pderiv{\FVA(t)}{\theta}$ from the approximation is computed w.r.t. the Monte Carlo benchmark result.
In Table~\ref{tab:FvaSensisParamSet38} the sensitivities corresponding to the multi-currency portfolio example from Section~\ref{sec:numericalResultsPortfolio} are presented.
The Gaussian approximation correctly determines the sign of the WWR part of the sensitivity in all cases.
The IR deltas, most important for hedging, are of high quality.
This is expected, as in the approximation from Equation~\eqref{eq:epeWWRCredit3}, $H_{\shortRate, I, C}(t,u)$ is taken outside the expectation in the exposure formula, and no Gaussian approximation is applied to this factor.
Only $\left(\tradeVal(u)\right)^+$ is affected when computing an IR delta, but the effect is limited in this example.
The IR vega approximation errors go up to 8\%, but the approximation captures the tendency of the WWR sensitivities well.
Also, the cross-gamma sensitivities between IR and credit are remarkably close.

Furthermore, the FX deltas are approximated at an acceptable level.
The FX vegas are close to zero, as the portfolio does not include FX derivatives.
Therefore, the high values of the RD metric are not an issue.

Like the IR deltas, the credit deltas are well approximated in this case.
The credit vegas have similar error magnitudes as the IR vegas.

As expected, under stressed market conditions, like in Section~\ref{sec:numericalResultsError}, the sensitivity approximation error increases.
More caution must thus be taken in this case, yet the tendency of the WWR sensitivity is well captured by the approximation.
Especially the IR deltas are performing well, with maximum errors of around 2.5\%.
The IR vega error increases to 10\%.
For the credit sensitivities and the cross-gammas, the sensitivity approximation errors increase.
As credit risks are more challenging to hedge due to limited liquidity in some of the credit markets, this is less of an issue.

The approximation allows for efficient sensitivity calculations.
For example, for the IR deltas this efficiency becomes apparent.
For the benchmark method, for each pillar on the yield curve, a new Monte Carlo simulation would need to be undertaken.
Due to the semi-analytic nature of the Gaussian approximation, these extra simulations are not required.
This speed-up in computing the WWR sensitivities can give an institution an edge over the rest of the market when setting up hedging positions.

\section{Conclusion}  \label{sec:conclusion}

In summary, the proposed WWR approximation provides a practical, robust and efficient method to compute $\FVA$ WWR when the existing $\xva$ calculation cannot simulate stochastic credit and funding processes which are correlated to the market risk factors.
The approximation avoids simulation of these additional processes, and is therefore also faster than the Monte Carlo benchmark method.
Furthermore, the approximation runtime does not scale in the number of derivatives in the portfolio.
Within the affine setting, the approximation is generic, does not require any calibration of approximation parameters, and is applicable to a wide variety of derivatives and asset classes.
Only the WWR exposure is approximated, so it is an add-on to the $\xva$ calculation in place at financial institutions, where the original exposures without WWR are left untouched.
In specific cases, the approximation gives rise to analytical expressions at the cost of one additional Taylor series approximation.
This has been demonstrated for the example of a single IR swap.
The approximation has been demonstrated for the practically relevant example of a multi-currency portfolio of multiple IR swaps.
This approximation error can be considered a lower bound for portfolios with more complicated derivatives.
After analyzing the various approximation errors, a good understanding is established of their behaviour and interplay.
The proposed approximation has its limitations under stressed market conditions, but at the same time the approximation allows for insights in the effect of different market stresses on the approximation error.
Recall that the quantity that is being approximated is not a `true market quantity', but the WWR resulting from a series of modelling choices.
Therefore, the WWR approximation is relevant from a practical perspective, giving a fast method to assess the impact of portfolio-level $\FVA$ WWR.
While the approach is applied to the calculation of $\FVA$, it is equally applicable to other metrics such as $\CVA$.

\section*{Acknowledgements} This work has been financially supported by Rabobank.

{\footnotesize
\bibliographystyle{abbrv}
\bibliography{bib/MacroStrings,bib/Articles,bib/Books,bib/Misc,bib/Regulation} 
}

\appendix
\small
\setlength\parindent{0pt}

\section{FVA exposure derivation} \label{app:fvaExposureDerivationCredit}
Using the modelling choices from Section~\ref{sec:SDE} together with the funding spread from Equation~\eqref{eq:fundingSpreadCredit}, we write:
\begin{align} \displaystyle
  \expPower{-\int_{t}^{u} \intensity_I(v) + \intensity_C(v)\dv}\borrowingSpread(u)
    & = H_{I,C}(t,u) \expPower{-Y_I(t,u) - Y_C(t,u)}  \left(\mu_{S}(t, u) + \LGD_I y_I(t,u)\right). \nonumber
\end{align}
Applying this to the WWR exposure from Equation~\eqref{eq:epeWWRCredit1} yields:
\begin{align} \displaystyle
  &\EPEFVAWWR{t}{u}\nonumber \\
    &\qquad=  H_{I,C}(t,u)\mu_{S}(t, u)\condExpSmall{\left(\expPower{-\int_{t}^{u} \shortRate(v)\dv}\maxOperator{\tradeVal(u)} - \condExpSmall{\expPower{-\int_{t}^{u} \shortRate(v)\dv}\maxOperator{\tradeVal(u)}}{t} \right)\expPower{-Y_I(t,u) - Y_C(t,u)}}{t} \nonumber \\
    &\qquad\quad +   H_{I,C}(t,u)\LGD_I\condExpSmall{\left(\expPower{-\int_{t}^{u} \shortRate(v)\dv}\maxOperator{\tradeVal(u)} - \condExpSmall{\expPower{-\int_{t}^{u} \shortRate(v)\dv}\maxOperator{\tradeVal(u)}}{t} \right)\expPower{-Y_I(t,u) - Y_C(t,u)}y_I(t,u)}{t}. \label{eq:epeWWRCredit1aApp}
\end{align}

The first term can be rewritten using Taylor expansions~(\ref{eq:taylor3}--\ref{eq:taylor4}) with $n_{\intensity}=1$:
\begin{align} \displaystyle
  &H_{I,C}(t,u) \mu_{S}(t, u)\condExpSmall{\left(\expPower{-\int_{t}^{u} \shortRate(v)\dv}\maxOperator{\tradeVal(u)} - \condExpSmall{\expPower{-\int_{t}^{u} \shortRate(v)\dv}\maxOperator{\tradeVal(u)}}{t} \right)\expPower{-Y_I(t,u) - Y_C(t,u)}}{t} \nonumber \\
    &=  H_{I,C}(t,u) \mu_{S}(t, u) \cdot \nonumber \\
    & \qquad \condExpSmall{\left(\expPower{-\int_{t}^{u} \shortRate(v)\dv}\maxOperator{\tradeVal(u)} - \condExpSmall{\expPower{-\int_{t}^{u} \shortRate(v)\dv}\maxOperator{\tradeVal(u)}}{t} \right) \left[\taylorTrunc{0}{1}(Y_I(t,u) + Y_C(t,u)) + \taylorTrunc{2}{\infty}(Y_I(t,u) + Y_C(t,u))\right] }{t} \nonumber \\
    &= H_{I,C}(t,u)\mu_{S}(t, u)\cdot \condExpSmall{\left(\expPower{-\int_{t}^{u} \shortRate(v)\dv}\maxOperator{\tradeVal(u)} - \condExpSmall{\expPower{-\int_{t}^{u} \shortRate(v)\dv}\maxOperator{\tradeVal(u)}}{t} \right)\left( - Y_I(t,u) - Y_C(t,u) \right)}{t} \nonumber \\
    & \quad + \mu_{S}(t, u)\error{1}(1) \nonumber \\
    &= H_{I,C}(t,u)\mu_{S}(t, u)\condExpSmall{\expPower{-\int_{t}^{u} \shortRate(v)\dv} \left(-Y_I(t,u) - Y_C(t,u)\right) \maxOperator{\tradeVal(u)}}{t} + \mu_{S}(t, u)\error{1}(1)\nonumber \\
    &= H_{\shortRate,I,C}(t,u)\mu_{S}(t, u)\condExpSmall{ \taylor(Y_{\shortRate}(t,u)) \left(-Y_I(t,u) - Y_C(t,u)\right) \maxOperator{\tradeVal(u)}}{t} + \mu_{S}(t, u)\error{1}(1), \nonumber
\end{align}
where
\begin{align} \displaystyle
  \error{1}(x)
    &\ldef H_{I,C}(t,u)\condExpSmall{\left(\expPower{-\int_{t}^{u} \shortRate(v)\dv}\maxOperator{\tradeVal(u)} - \condExpSmall{\expPower{-\int_{t}^{u} \shortRate(v)\dv}\maxOperator{\tradeVal(u)}}{t}\right)\taylorTrunc{2}{\infty}(Y_I(t,u) + Y_C(t,u)) \cdot x}{t}. \label{eq:error1App}
\end{align}

The remaining unknown term in Equation~\eqref{eq:epeWWRCredit1aApp} is rewritten in a similar fashion:
\begin{align} \displaystyle
  &H_{I,C}(t,u) \LGD_I \condExpSmall{\left(\expPower{-\int_{t}^{u} \shortRate(v)\dv}\maxOperator{\tradeVal(u)} - \condExpSmall{\expPower{-\int_{t}^{u} \shortRate(v)\dv}\maxOperator{\tradeVal(u)}}{t} \right)\expPower{-Y_I(t,u) - Y_C(t,u)}y_I(t,u)}{t} \nonumber \\
    & = H_{I,C}(t,u)\LGD_I \condExpSmall{\left(\expPower{-\int_{t}^{u} \shortRate(v)\dv}\maxOperator{\tradeVal(u)} - \condExpSmall{\expPower{-\int_{t}^{u} \shortRate(v)\dv}\maxOperator{\tradeVal(u)}}{t} \right)y_I(t,u)\left( 1 - Y_I(t,u) - Y_C(t,u) \right)}{t} \nonumber \\
    &\quad +\LGD_I \error{1}(y_I) \nonumber \\
    &= H_{I,C}(t,u)\LGD_I \condExpSmall{\expPower{-\int_{t}^{u} \shortRate(v)\dv} y_I(t,u)\left( 1 - Y_I(t,u) - Y_C(t,u) \right)\maxOperator{\tradeVal(u)}}{t} \nonumber \\
    & \quad - H_{I,C}(t,u)\LGD_I \condExpSmall{\expPower{-\int_{t}^{u} \shortRate(v)\dv}\maxOperator{\tradeVal(u)}}{t} \Big(\condExpSmall{y_I(t,u)}{t}  - \condExpSmall{Y_I(t,u)y_I(t,u)}{t}  - \condExpSmall{Y_C(t,u)y_I(t,u)}{t} \Big) \nonumber \\
    & \quad +\LGD_I \error{1}(y_I)\nonumber \\
    &= H_{\shortRate,I,C}(t,u)\LGD_I \condExpSmall{\taylor(Y_{\shortRate}(t,u))y_I(t,u)\left( 1 - Y_I(t,u) - Y_C(t,u) \right)\maxOperator{\tradeVal(u)}  }{t} \nonumber \\
    & \quad + H_{I,C}(t,u)\LGD_I  \condExpSmall{Y_I(t,u)y_I(t,u)}{t} \condExpSmall{\expPower{-\int_{t}^{u} \shortRate(v)\dv}\maxOperator{\tradeVal(u)}}{t}   +\LGD_I \error{1}(y_I). \nonumber
\end{align}

With these results, WWR exposure from Equation~\eqref{eq:epeWWRCredit1aApp} changes into:
\begin{align} \displaystyle
  \EPEFVAWWR{t}{u}
    &=  H_{\shortRate,I,C}(t,u) \mu_{S}(t, u)  \condExpSmall{\taylor(Y_{\shortRate}(t,u)) \left(-Y_I(t,u) - Y_C(t,u)\right) \maxOperator{\tradeVal(u)}}{t} \nonumber \\
    &\quad +  \LGD_I  H_{\shortRate,I,C}(t,u) \condExpSmall{\taylor(Y_{\shortRate}(t,u)) y_I(t,u)\left( 1 - Y_I(t,u)- Y_C(t,u) \right)\maxOperator{\tradeVal(u)}  }{t}  \nonumber \\
    &\quad  +  \LGD_I H_{I,C}(t,u) \condExpSmall{Y_I(t,u)y_I(t,u)}{t}  \condExpSmall{\expPower{-\int_{t}^{u} \shortRate(v)\dv}\maxOperator{\tradeVal(u)}}{t}    \nonumber \\
    &\quad  + \mu_{S}(t, u) \error{1}(1) + \LGD_I\error{1}(y_I). \nonumber
\end{align}

Finally, we apply the second part of Taylor expansion~\eqref{eq:taylor3}, which means we truncate the summation at the $n_{\shortRate}$-th term:
\begin{align} \displaystyle
  \EPEFVAWWR{t}{u}
    &=  H_{\shortRate,I,C}(t,u) \mu_{S}(t, u)  \condExpSmall{\taylorTrunc{0}{n_{\shortRate}}(Y_{\shortRate}(t,u)) \left(-Y_I(t,u) - Y_C(t,u)\right) \maxOperator{\tradeVal(u)}}{t} \nonumber \\
    &\quad +  \LGD_I  H_{\shortRate,I,C}(t,u) \condExpSmall{\taylorTrunc{0}{n_{\shortRate}}(Y_{\shortRate}(t,u)) y_I(t,u)\left( 1 - Y_I(t,u)- Y_C(t,u) \right)\maxOperator{\tradeVal(u)}  }{t}  \nonumber \\
    &\quad  +  \LGD_I H_{I,C}(t,u) \condExpSmall{Y_I(t,u)y_I(t,u)}{t}  \condExpSmall{\expPower{-\int_{t}^{u} \shortRate(v)\dv}\maxOperator{\tradeVal(u)}}{t}    \nonumber \\
    &\quad  + \mu_{S}(t, u) \left[\error{1}(1) + \error{2}(1)\right] + \LGD_I \left[\error{1}(y_I) + \error{2}(y_I) + \error{3}(y_I)\right] , \label{eq:epeWWRCredit2App}
\end{align}
where:
\begin{align} \displaystyle
  \error{2}(x)
    &\ldef H_{\shortRate,I,C}(t,u)  \condExpSmall{\taylorTrunc{n_{\shortRate} + 1}{\infty}(Y_{\shortRate}(t,u)) \cdot x \cdot \left(-Y_I(t,u) - Y_C(t,u)\right) \maxOperator{\tradeVal(u)}}{t}, \label{eq:error2App} \\
  \error{3}(x)
    &\ldef H_{\shortRate,I,C}(t,u) \condExpSmall{\taylorTrunc{n_{\shortRate} + 1}{\infty}(Y_{\shortRate}(t,u)) \cdot x \cdot \maxOperator{\tradeVal(u)}  }{t}. \label{eq:error3App}
\end{align}

\section{IR and credit dynamics} \label{app:dynamics}

\subsection{IR dynamics} \label{app:dynamicsIR}
For the IR process $\shortRate(t)$ we choose the HW1F (G1++) dynamics with constant volatility:
\begin{align} \displaystyle
  \shortRate(t)
    &= x_{\shortRate}(t) + b_{\shortRate}(t), \
  \dx_{\shortRate}(t)
    = -a_{\shortRate}x_{\shortRate}(t)\dt + \vol_{\shortRate} \d\brownian_{\shortRate}(t). \nonumber
\end{align}
This fits the framework we introduced in Section~\ref{sec:SDE}, where
{ \allowdisplaybreaks
\begin{align} \displaystyle
  \mu_{\shortRate}(t,u)
    &= x_{\shortRate}(t)\expPower{-a_{\shortRate} (u-t)}, \
  y_{\shortRate}(t,u)
    = \vol_{\shortRate} \int_{t}^{u}  \expPower{-a_{\shortRate} (u-v)} \d\brownian_{\shortRate}(v), \nonumber \\
  M_{\shortRate}(t,u)
    &= x_{\shortRate}(t) B_{\shortRate}(t,u), \
  Y_{\shortRate}(t,u)
    = \vol_{\shortRate} \int_{t}^{u} B_{\shortRate}(v,u)\d\brownian_{\shortRate}(v), \nonumber \\
  A_{\shortRate}(t,u)
    &= \frac{\vol_{\shortRate}^2}{2a_{\shortRate}^2} \left[ (u-t) - 2 B_{\shortRate}(t,u)  + \frac{ 1}{2a_{\shortRate}}\left( 1 - \expPower{-2a_{\shortRate} (u-t)} \right) \right],
  B_{\shortRate}(t,u)
    = \frac{ 1}{a_{\shortRate}}\left( 1 - \expPower{-a_{\shortRate} (u-t)} \right), \nonumber \\
  b_{\shortRate}(t)
    &= f^{\text{M}}_{\shortRate}(0,t) + \half \vol_{\shortRate}^2 B_{\shortRate}^2(0,t) -  x_{\shortRate}(0)  \expPower{-a_{\shortRate}t}, \nonumber \\
  \expPower{- \int_{t}^{T} b_{\shortRate}(u) \du}
    &= \frac{\zcb^{\text{M}}_{\shortRate}(0,T)}{\zcb^{\text{M}}_{\shortRate}(0,t)} \expBrace{ A_{\shortRate}(0,t) - A_{\shortRate}(0,T) + x_{\shortRate}(0) \left[ B_{\shortRate}(0,T) - B_{\shortRate}(0,t) \right]}. \nonumber
\end{align}
}
$x_{\shortRate}(t)$ and $\int_{t}^{u} x_{\shortRate}(v) \dv$  both follow a normal distribution with mean $\mu_{\shortRate}(t,u)$ and $M_{\shortRate}(t,u)$ respectively, and variances
\begin{align} \displaystyle
  \condVarSmall{x_{\shortRate}(u)}{t}
    &= \frac{\vol_{\shortRate}^2}{2a_{\shortRate}} \left( 1 - \expPower{-2a_{\shortRate} (u-t)}\right),  \
  \condVarSmall{\int_{t}^{u} x_{\shortRate}(v) \dv}{t}
    = 2 A_{\shortRate}(t,u). \nonumber
\end{align}

In the case of a multi-currency portfolio, the domestic IR is denoted as $\shortRate_d(t)$, which adheres to the equations above.
For foreign IR, $\shortRate_f(t)$, use a similar model:
\begin{align} \displaystyle
  \shortRate_f(t)
    &= x_{\shortRate_f}(t) + b_{\shortRate_f}(t), \
  \dx_{\shortRate_f}(t)
    = -a_{\shortRate_f}x_{\shortRate_f}(t)\dt + \vol_{\shortRate_f} \d\brownian_{\shortRate_f}^f(t). \nonumber
\end{align}
Here, $\d\brownian_{\shortRate_f}^f(t)$ denotes the Brownian motion of process $\shortRate_f$ under the $f$ risk-neutral measure.

When setting up a Monte Carlo simulation, all processes must be under the same measure~\cite{Green201511}.
Hence, we use a change of measure $\d \brownian_{\shortRate_f}^f(t) = \d \brownian_{\shortRate_f}^d(t) - \rho_{\shortRate_f, \FX} \cdot \vol_{\FX} \d t$ and write the $\shortRate_f(t)$ dynamics under the $\Q^d$ risk-neutral measure.
This will lead to an additional term in the drift, which is a quanto correction term:
\begin{align} \displaystyle
  \dx_{\shortRate_f}(t)
    &= -a_{\shortRate_f}x_{\shortRate_f}(t)\dt - \rho_{\shortRate_f, \FX} \cdot \vol_{\shortRate_f} \vol_{\FX} \dt + \vol_{\shortRate_f} \d\brownian_{\shortRate_f}^d(t). \nonumber
\end{align}
In this case,
\begin{align} \displaystyle
  \mu_{\shortRate_f}(t,u)
    &= x_{\shortRate_f}(t)\expPower{-a_{\shortRate_f} (u-t)} - \rho_{\shortRate_f, \FX} \cdot \vol_{\shortRate_f} \vol_{\FX} B_{\shortRate_f}(t,u), \
  y_{\shortRate_f}(t,u)
    = \vol_{\shortRate_f} \int_{t}^{u}  \expPower{-a_{\shortRate_f} (u-v)} \d\brownian_{\shortRate_f}^d(v), \nonumber \\
  M_{\shortRate_f}(t,u)
    &= x_{\shortRate_f}(t) B_{\shortRate_f}(t,u) - \frac{\rho_{\shortRate_f, \FX} \cdot \vol_{\shortRate_f} \vol_{\FX}}{a_{\shortRate_f}} \left[ (u-t) -  B_{\shortRate_f}(t,u) \right], \
  Y_{\shortRate_f}(t,u)
    = \vol_{\shortRate_f} \int_{t}^{u} B_{\shortRate_f}(v,u)\d\brownian_{\shortRate_f}^d(v), \nonumber
\end{align}
where further information on the FX process is given in~\ref{app:dynamicsFX}.

\subsection{Credit dynamics} \label{app:dynamicsCredit}
In literature on WWR modelling for BCVA purposes, a CIR type model is commonly used~\cite{BrigoPallavicini201405,BrigoPallaviciniPapatheodorou201107}. 
Hence, for the credit process we choose a CIR++ credit dynamics with constant volatility for each counterparty $z \in \{I,\ C\}$:
\begin{align} \displaystyle
  \intensity_z(t)
    &= x_z(t) + b_z(t), \
  \dx_z(t)
    = a_z\left( \theta_z -  x_z(t)\right) \dt + \vol_z \sqrt{x_z(t)}\d\brownian_z(t). \nonumber
\end{align}
When the Feller condition $2a_z\theta_z > \vol_z^2$ is satisfied, we are not affected by potential issues around the origin, see~\cite{OosterleeGrzelak201911} for more information.
This model dynamics fits the framework introduced in Section~\ref{sec:SDE}, where
{  \allowdisplaybreaks
\begin{align} \displaystyle
  \mu_{z}(t,u)
    &= x_z(t)\expPower{-a_z (u-t)} + \theta_z  \left( 1 - \expPower{-a_z (u-t)} \right) , \
  y_{z}(t,u)
    = \vol_z \int_{t}^{u}  \expPower{-a_z (u-v)} \sqrt{x_z(v)}\d\brownian_z(v), \nonumber \\
  M_{z}(t,u)
    &= \frac{ x_z(t)}{a_z}\left( 1 - \expPower{-a_z (u-t)} \right) + \theta_z \left[ (u-t) - \frac{ 1}{a_z}\left( 1 - \expPower{-a_z (u-t)} \right) \right], \nonumber \\
  Y_{z}(t,u)
    &= \frac{\vol_z}{a_z} \int_{t}^{u} \left( 1 - \expPower{-a_z (u-v)} \right)\sqrt{x_z(v)}\d\brownian_z(v), \
  B_{z}(t,u)
    = \frac{2\left( \expPower{h(u-t)} - 1 \right)}{2h + (a_z + h)\left( \expPower{h(u-t)} - 1 \right)}, \nonumber \\
  A_{z}(t,u)
    &= \frac{2a_z \theta_z}{\vol_z^2} \ln \left[ \frac{2h\expPower{\half (a_z + h)(u-t)} }{2h + (a_z + h)\left( \expPower{h(u-t)} - 1 \right)} \right], \
  b_{z}(t)
    = f^{\text{M}}_z(0,t) - f^{CIR++}_z(0,t), \nonumber \\
  f^{CIR++}_z(0,t)
    &= \frac{2 a_z \theta_z \left( \expPower{ht} - 1 \right)}{2h + (a_z + h)\left( \expPower{ht} - 1 \right)} +  x_z(0) \frac{4h^2\expPower{ht}}{\left[2h + (a_z + h)\left( \expPower{ht} - 1 \right)\right]^2}, \nonumber \\
  \expPower{- \int_{t}^{T} b_z(u) \du}
    &= \frac{\zcb^{\text{M}}_z(0,T)}{\zcb^{\text{M}}_z(0,t)} \expBrace{ A_z(0,t) - A_z(0,T) + x_z(0) \left[ B_z(0,T) - B_z(0,t) \right]}, \
  h
    = \sqrt{a_z^2 + 2\vol_z^2}. \nonumber
\end{align}
}
Conditional on $x_z(t)$, $x_z(u)$ follows a scaled non-central chi-square distribution~\cite{OosterleeGrzelak201911}:
\begin{align} \displaystyle
  x_z(u)|x_z(t)
    &\sim c(u,t) \chi^2(\delta, k(u,t)), \nonumber \\
  c(u,t)
    &= \frac{\vol_z^2}{4 a_z}  \left( 1 - \expPower{-a_z (u-t)}\right), \
  \delta
    = \frac{4 a_z \theta_z}{\vol_z^2}, \
  k(u,t)
    = \frac{4 a_z x_z(s) \expPower{-a_z (u-t)}}{\vol_z^2 \left( 1 - \expPower{-a_z (u-t)}\right)}. \nonumber
\end{align}
Processes $x_z(t)$ and $\int_{t}^{u}x_z(v)\dv$ have mean $\mu_{z}(t,u)$ and $M_{z}(t,u)$ respectively, and variance
\begin{align} \displaystyle
  &\condVarSmall{x_z(u)}{t}
    = \frac{\vol_z^2 }{a_z}\left( 1 - \expPower{-a_z (u-t)}\right)\left( \mu_z(t,u) -  \frac{\theta_z}{2} \left( 1 - \expPower{-a_z (u-t)} \right)\right). \nonumber \\
  &\condVarSmall{\int_{t}^{u}x_z(v)\dv}{t}
    = \frac{\vol_z^2 x_z(t)}{a_z^3} \left[ 1  - 2 a_z\expPower{-a_z (u-t)} (u-t) - \expPower{-2a_z (u-t)}  \right] \nonumber \\
    &\qquad \quad+ \frac{\vol_z^2 \theta_z}{a_z^3} \left[
    a_z(u-t) -   3 \left(1 -  \expPower{-a_z (u-t)}\right)
     +2 a_z\expPower{-a_z (u-t)} (u-t)
    +  \frac{1}{2}\left( 1 - \expPower{-a_z (u-t)} \right)^2
    \right]. \nonumber
\end{align}

Furthermore, the following analytic expression can be derived for the $\condExpSmall{ Y_z(t,u) y_z(t,u)}{t}$ term, which is needed when computing $\EPEFVAIndep{t}{u}$.
\begin{align} \displaystyle
  \condExpSmall{ Y_z(t,u) y_z(t,u)}{t}
    & = \frac{\vol_z^2 x_z(t)}{a_z^2} \expPower{-a_z (u-t)} \left( a_z(u-t) -  1 + \expPower{-a_z (u-t)} \right)  \nonumber \\
    &\quad   + \frac{\vol_z^2\theta_z }{a_z^2} \left( \frac{1}{2}\left( 1 - \expPower{-2a_z (u-t)} \right) - a_z \expPower{-a_z (u-t)} (u-t)\right).\label{eq:expYzyz}
\end{align}

\subsection{FX dynamics} \label{app:dynamicsFX}
For the FX process, $\FX_f^d$, which to converts currency $f$ into $d$, we use log-normal dynamics:
\begin{align} \displaystyle
  \d\left(\ln\FX_f^d(t)\right)
    &= \left[ \shortRate_d(t) - \shortRate_f(t)  - \half \vol_{\FX}^2 \right]\d t + \vol_{\FX} \d\brownian_{\FX}^d(t). \nonumber
\end{align}

The integrated form reads
\begin{align} \displaystyle
 \ln\FX_f^d(u)
   &= \mu_{\FX}(t,u) + Y_{\shortRate_d}(t,u) - Y_{\shortRate_f}(t,u) + \vol_{\FX} \int_t^u \d\brownian_{\FX}^d(v) , \nonumber \\
 \mu_{\FX}(t,u)
   &\ldef \ln\FX_f^d(t) + M_{\shortRate_d}(t,u) + \int_t^u b_{\shortRate_d}(v) \dv - M_{\shortRate_f}(t,u) - \int_t^u b_{\shortRate_f}(v) \dv -  \half \vol_{\FX}^2 (u-t). \nonumber
\end{align}
The three stochastic processes $Y_{\shortRate_d}(t,u)$, $Y_{\shortRate_f}(t,u)$, $y_{\FX}(t,u)$ are correlated normals with zero mean.
Here, $\overline{y}_{\FX}(t,u)$ is normal with zero mean and a variance where the correlations are included.

Conditional on time $t$, $\ln\FX_f^d(u)$ has mean $\condExpSmall{\ln\FX_f^d(u)}{t}= \mu_{\FX}(t,u)$ and variance
\begin{align}
  \condVarSmall{\ln\FX_f^d(u)}{t}
    &= \condVarSmall{\int_t^u x_{\shortRate_d}(v) \dv}{t} + \condVarSmall{\int_t^u x_{\shortRate_f}(v) \dv}{t} + \vol_{\FX}^2 (u-t) \nonumber \\
    &\quad - \frac{2 \corr_{\shortRate_d,\shortRate_f} \vol_{\shortRate_d}\vol_{\shortRate_f}}{a_{\shortRate_d}a_{\shortRate_f}}  \left[ (u-t) - B_{\shortRate_d}(t,u)  - B_{\shortRate_f}(t,u)  + \frac{1}{a_{\shortRate_d} + a_{\shortRate_f}}\left( 1 + \expPower{-\left(a_{\shortRate_d} + a_{\shortRate_f}\right)(u-t)} \right)\right]\nonumber \\
    &\quad + \frac{2 \corr_{\shortRate_d,\FX} \vol_{\shortRate_d}\vol_{\FX}}{a_{\shortRate_d}} \left[ (u-t) - B_{\shortRate_d}(t,u)\right]
    - \frac{2 \corr_{\shortRate_f,\FX} \vol_{\shortRate_f}\vol_{\FX}}{a_{\shortRate_f}} \left[ (u-t) - B_{\shortRate_f}(t,u)\right]. \nonumber
\end{align}

\subsection{Model parameters} \label{app:params}

\subsubsection{Single IR swap example} \label{app:paramsSingleIRS}
\begin{itemize}
  \item $\shortRate$: $x_{\shortRate}(0) = 0.0$, $a_{\shortRate} = 10^{-5}$, $\vol_{\shortRate} = 0.00284$, EUR1D yield curve;
  \item $\intensity_I$: $x_I(0) = 0.0016939$, $a_I = 0.05$, $\theta_I = 0.015390$, $\vol_I = 0.02$, $\LGD_I = 0.6$, AAA-rating curve;
  \item $\intensity_C$: $x_C(0) = 0.0063774$, $a_C = 0.2$, $\theta_C = 0.035447$, $\vol_C = 0.08$, $\LGD_C = 0.6$, BBB-rating curve;
  \item Correlation: $\corr_{\shortRate,I} = -0.35$, $\corr_{\shortRate,C} = -0.5$, $\corr_{I,C} = 0.0$.
\end{itemize}

\subsubsection{Multi-currency portfolio of IR swaps example} \label{app:paramsPortfolio}
\begin{itemize}
  \item Currencies: EUR, USD, GBP;
  \item $\shortRate$: $x_{\shortRate}(0) = 0.0$, $a_{\shortRate} = 10^{-5}$, $\vol_{\shortRate} = [0.00284, 0.00357, 0.00312]$, 1D yield curves;
  \item FX: $\FX_f^d(0) = [0.91802, 1.15069]$, $\vol_{\FX} = [0.15, 0.15]$;
  \item $\intensity_I$: $x_I(0) = 0.0016939$, $a_I = 0.05$, $\theta_I = 0.015390$, $\vol_I = 0.02$, $\LGD_I = 0.6$, AAA-rating curve;
  \item $\intensity_C$: $x_C(0) = 0.0098774$, $a_C = 0.05$, $\theta_C = 0.041033$, $\vol_C = 0.02$, $\LGD_C = 0.6$, BBB-rating curve;
  \item Correlation: $\corr_{\shortRate_i, \shortRate_{j}} = 0.5$, $\corr_{\shortRate_i, \FX_{j}^d} = 0.25$, $\corr_{\shortRate_i, \intensity_j} = -0.35$, $\corr_{\FX_{i}^d, \FX_{j}^d} = 0.5$, $\corr_{\FX_{i}^d, \intensity_j} = -0.2$, $\corr_{\intensity_I, \intensity_C} = 0.0$.
\end{itemize}

\subsubsection{Multi-currency portfolio of IR swaps stressed example} \label{app:paramsPortfolioStressed}
\begin{itemize}
  \item Currencies: EUR, USD, GBP;
  \item $\shortRate$: $x_{\shortRate}(0) = 0.0$, $a_{\shortRate} = 10^{-5}$, $\vol_{\shortRate} = [0.00426, 0.00536, 0.00469]$, 1D yield curves;
  \item FX: $\FX_f^d(0) = [0.91802, 1.15069]$, $\vol_{\FX} = [0.15, 0.15]$;
  \item $\intensity_I$: $x_I(0) = 0.00052392$, $a_I = 0.15$, $\theta_I = 0.012475$, $\vol_I = 0.04$, $\LGD_I = 0.6$, AAA-rating curve;
  \item $\intensity_C$: $x_C(0) = 0.0063774$, $a_C = 0.2$, $\theta_C = 0.035447$, $\vol_C = 0.0801$, $\LGD_C = 0.6$, BBB-rating curve;
  \item Correlation: $\corr_{\shortRate_i, \shortRate_{j}} = 0.5$, $\corr_{\shortRate_i, \FX_{j}^d} = 0.25$, $\corr_{\shortRate_i, \intensity_j} = -0.5$, $\corr_{\FX_{i}^d, \FX_{j}^d} = 0.5$, $\corr_{\FX_{i}^d, \intensity_j} = -0.2$, $\corr_{\intensity_I, \intensity_C} = 0.0$.
\end{itemize}

\section{IR swap} \label{app:swap}

\subsection{Payoff} \label{app:swapPayoff}
Consider an IR swap under a single curve setup starting at $T_0$, maturing at $T_m$ with intermediate payment dates $T_1 < T_2 < \ldots < T_{m-1} < T_m$, $\dct_k = T_k - T_{k-1}$.
The value of the swap with strike $\strike$ and notional $\notional$ is then:
\begin{align} \displaystyle
  \tradeVal(u)
    &= \swapType \notional \left(-\zcb_{\shortRate}(u,T_0) + \zcb_{\shortRate}(u,T_m) + \strike \cdot \sum_{k=1}^m \dct_k \zcb_{\shortRate}(u,T_k)\right), \label{eq:swap1App} \\
  \swapType
    &= \left\{
       \begin{array}{ll}
    	-1, & \text{payer swap}, \\
    	1, & \text{receiver swap}.
       \end{array}
       \right. \label{eq:swapTypeApp}
\end{align}
Using the assumptions from Section~\ref{sec:SDE}, we can express the swap value in terms of $y_{\shortRate}(t,u)$:
\begin{align} \displaystyle
  \tradeVal(u)
    &= \swapType\notional\sum_{k=0}^m w_k \zcb_{\shortRate}(u,T_k)
    = \swapType\notional\sum_{k=0}^m \overline{w}_k  \expPower{-y_{\shortRate}(t,u)B_{\shortRate}(u,T_k)}, \label{eq:swapValue1aApp}\\
  \overline{w}_k
    &= w_k \expPower{\overline{A}_{{\shortRate}}(u,T_k) - \mu_{\shortRate}(t,u) B_{\shortRate}(u,T_k)}, \
  w_k
    = \left\{
    \begin{array}{ll}
      -1, & k=0, \\
      K\dct_k, & k=1,\ldots,m-1, \\
      1 + K \dct_k, & k=m.
    \end{array}
    \right.  \label{eq:swapWeightsScaledApp}
\end{align}

Swap valuation formula~\eqref{eq:swapValue1aApp} only holds if $u \in [t_0, T_0]$, meaning before the first reset date of the swap, i.e., $T_0$.
As soon as $u > T_0$, the valuation changes, especially after several coupon payments have been made.
For $u\in(T_j, T_{j+1}]$ with $j\in\{0,1,\ldots,m-1\}$, the following valuation holds:
\begin{align} \displaystyle
  \tradeVal(u)
    &= \swapType\notional\left(-1 + \zcb_{\shortRate}(u,T_m) + \strike \cdot \sum_{k=j+1}^m \dct_k \zcb_{\shortRate}(u,T_k)\right)
    = \swapType\notional\left(-1 +  \sum_{k=j+1}^m \overline{w}_k  \expPower{-y_{\shortRate}(t,u)B_{\shortRate}(u,T_k)}\right) . \label{eq:swapValue1bApp}
\end{align}
We can combine Equations~(\ref{eq:swapValue1aApp}--\ref{eq:swapValue1bApp}) into the following generic result for the swap value:
\begin{align} \displaystyle
  \tradeVal(u)
    &= \swapType\notional\left(-\indicator{u > T_0} + \sum_{k=\beta(u)}^m \overline{w}_k  \expPower{-y_{\shortRate}(t,u)B_{\shortRate}(u,T_k)}\right), \label{eq:swapValue2} \\
  \beta(u)
    &= \left\{
    \begin{array}{ll}
      0, & u \in [t_0, T_0],\\
      j+1, & u \in (T_j, T_{j+1}].
    \end{array}
    \right. \label{eq:swapValue2BetaApp}
\end{align}

\subsection{Probability of positive value} \label{app:derivationsMaxProb}
We are interested in the case $\tradeVal(u) \geq 0$, with $\tradeVal(u)$ given in Equation~\eqref{eq:swapValue2}.
To find the values of $y_{\shortRate}(t,u)$ for which this holds, we rewrite Equation~\eqref{eq:swapValue2} as a monotonically decreasing function.
Consider the two different cases for $\beta(u)$ in~\eqref{eq:swapValue2BetaApp}:
\begin{itemize}
  \item $\beta(u) = 0$, hence $u \in [t_0, T_0]$, so $\indicator{u > T_0} = 0$.
  Then, starting from Equation~\eqref{eq:swapValue1aApp},
  \begin{align} \displaystyle
    \tradeVal(u)
      &= \swapType\notional\left( -1 + \sum_{k=1}^m \overline{w}_k  \expPower{-y_{\shortRate}(t,u)B_{\shortRate}(u,T_k)  -\overline{A}_{{\shortRate}}(u,T_0) + \mu_{\shortRate}(t,u)B_{\shortRate}(u,T_0) + y_{\shortRate}(t,u)B_{\shortRate}(u,T_0)}\right). \label{eq:swapValue2Case1App}
  \end{align}
  \item $\beta(u) = j+1$, hence $u \in (T_j, T_{j+1}]$, so $\indicator{u > T_0} = 1$.
  Then,
  \begin{align} \displaystyle
    \tradeVal(u)
      &= \swapType\notional\left(-1 + \sum_{k=j+1}^m \overline{w}_k  \expPower{-y_{\shortRate}(t,u)B_{\shortRate}(u,T_k)}\right). \label{eq:swapValue2Case2App}
  \end{align}
\end{itemize}
We combine Equations~(\ref{eq:swapValue2Case1App}--\ref{eq:swapValue2Case2App}) as follows:
\begin{align} \displaystyle
  \tradeVal(u; y_{\shortRate}(t,u))
    &= \swapType\notional\left( -1 + \sum_{k=\gamma(u)}^m \overline{w}_k  \expPower{-y_{\shortRate}(t,u)B_{\shortRate}(u,T_k)  + \indicator{u \leq T_0}  \left(-\overline{A}_{{\shortRate}}(u,T_0) + \mu_{\shortRate}(t,u)B_{\shortRate}(u,T_0) + y_{\shortRate}(t,u)B_{\shortRate}(u,T_0)\right)}\right), \label{eq:swapValue2CasesCombined} \\
  \gamma(u)
    &= \beta(u) + \indicator{u \leq T_0}
    = \left\{
    \begin{array}{ll}
      1, & u \in [t_0, T_0],\\
      j+1, & u\in(T_j, T_{j+1}],
    \end{array}
    \right. \label{eq:probSwapPosOptY2GammaApp}
\end{align}
where in Equation~\eqref{eq:swapValue2CasesCombined} we now explicitly write the dependence of $\tradeVal$ on $y_{\shortRate}(t,u)$.

Similarly to Jamshidian's trick for swaption pricing~\cite{Jamshidian198903}, we choose $y_{\shortRate}^{*}(t,u)$ numerically such that $\tradeVal(u; y_{\shortRate}^{*}(t,u)) = 0$.
With $\notional > 0$, $\tradeVal(u; y_{\shortRate}(t,u)) \geq  0$ is equivalent to $d\left(y_{\shortRate}(t,u)\right) \geq d\left(y_{\shortRate}^{*}(t,u)\right)$, where
\begin{align} \displaystyle
  d\left(y\right)
    &= \swapType\sum_{k=\gamma(u)}^m \overline{w}_k  \expPower{-y \left[ B_{\shortRate}(u,T_k) - \indicator{u \leq T_0} B_{\shortRate}(u,T_0)\right]  + \indicator{u \leq T_0}  \left(-\overline{A}_{{\shortRate}}(u,T_0) + \mu_{\shortRate}(t,u)B_{\shortRate}(u,T_0) \right)}. \nonumber
\end{align}

Recall $B_{\shortRate}(u,T_k) = \frac{1}{a_{\shortRate}}\left( 1 - \expPower{-a_{\shortRate} (T_k - u)} \right)$.
Then, for all $a_{\shortRate} \in \R$ the function $\expPower{-y \left[ B_{\shortRate}(u,T_k) - \indicator{u \leq T_0} B_{\shortRate}(u,T_0)\right]}$ is monotonically decreasing in $y$.
In Equation~\eqref{eq:swapWeightsScaledApp}, $\overline{w}_k  > 0$ for all $k\in\{\gamma(u), \ldots, m\}$.

For a payer swap (i.e., when $\swapType = -1$), $d(y)$ is a monotonically increasing function in $y$ as it is a sum of monotonically increasing functions.
Alternatively, for a receiver swap (i.e., when $\swapType = 1$), $d(y)$ is a monotonically decreasing function in $y$ as it is a sum of monotonically decreasing functions.

Using the monotonicity we can write the following for the indicator when the swap value is positive:
\begin{align} \displaystyle
    \indicator{\tradeVal(u) \geq 0}
        &= \indicator{f(y_{\shortRate}(t,u)) \geq f(y_{\shortRate}^{*}(t,u))}
        = \indicator{\swapType = -1} + \swapType \cdot \indicator{y_{\shortRate}(t,u) \leq y_{\shortRate}^{*}(t,u)}. \label{eq:indSwapPosApp}
\end{align}
From this, we have a straightforward expression of the positive swap value probability:
\begin{align} \displaystyle
    \condProbSmall{\tradeVal(u) \geq 0}{t}
        &= \indicator{\swapType = -1} + \swapType \cdot \condProbSmall{y_{\shortRate}(t,u) \leq y_{\shortRate}^{*}(t,u)}{t}
        = \indicator{\swapType = -1} + \swapType \cdot \normCDF\left( \frac{y_{\shortRate}^{*}(t,u)}{\sqrt{\condVarSmall{y_{\shortRate}(t,u)}{t}}}\right), \label{eq:probSwapPosApp}
\end{align}
where $\normCDF(\cdot)$ is the standard normal CDF.
In the last step, we have used the assumption that $y_{\shortRate}(t,u)$ is normally distributed $y_{\shortRate}(t,u)\sim \N\left(0,\condVarSmall{y_{\shortRate}(t,u)}{t}\right)$, which results from using the HW1F model.

\subsection{Error bound} \label{app:derivationsSwapErrorBound}
For the second term in Equation~\eqref{eq:genericErrorBoundB1}, which is the product-specific component of the error, we use
\begin{align}\displaystyle
  \left(\maxOperator{\tradeVal(u)}\right)^2
    &= \left(\tradeVal(u) \indicator{\tradeVal(u)\geq 0}\right)^2
    = \left(\tradeVal(u)\right)^2  \indicator{\tradeVal(u)\geq 0}
    \leq \left(\tradeVal(u)\right)^2, \label{eq:swapSquared0}
\end{align}
to obtain $\condExpSmall{\left(\maxOperator{\tradeVal(u)}\right)^2}{t} \leq \condExpSmall{\left(\tradeVal(u)\right)^2}{t}$.

Using the IR swap notation from Equation~\eqref{eq:swapValue2}, together with the Cauchy-Schwarz inequality for sums, i.e., $\left( \sum_{i=1}^{m} a_i \right)^2 \leq m \sum_{i=1}^{m} a_i^2$, we get
\begin{align} \displaystyle
  \left(\tradeVal(u)\right)^2
    &= \notional^2\left(-\indicator{u > T_0} + \sum_{k=\beta(u)}^m \overline{w}_k  \expPower{-y_{\shortRate}(t,u)B_{\shortRate}(u,T_k)}\right)^2 \nonumber \\
    &\leq \notional^2\left(\indicator{u > T_0} -2\cdot\indicator{u > T_0}\sum_{k=\beta(u)}^m \overline{w}_k  \expPower{-y_{\shortRate}(t,u)B_{\shortRate}(u,T_k)} + \left(m - \beta(u)\right)\sum_{k=\beta(u)}^m \overline{w}_k^2  \expPower{-2y_{\shortRate}(t,u)B_{\shortRate}(u,T_k)}\right)\nonumber \\
    &=\notional^2 \sum_i \gamma_i \expPower{-\alpha_i \cdot y_{\shortRate}(t,u)}. \label{eq:swapSquared1}
\end{align}
where in the last step we have written the expression in a generic form for ease of the analysis that follows.
Here, $\gamma_i$ and $\alpha_i$ are deterministic, and the sum over $i$ is finite.

Furthermore, we know that $y_{\shortRate}(t,u)\sim \N\left(0, \condVarSmall{y_{\shortRate}(t,u)}{t}\right)$.
Recall that $\E[\expPower{X}] = \expPower{\mu + \half \sigma^2}$ for $X\sim\N(\mu,\sigma^2)$.
This, combined with Equation~\eqref{eq:swapSquared1}, yields:
\begin{align} \displaystyle
  \condExpSmall{\left(\maxOperator{\tradeVal(u)}\right)^2}{t}
    &\leq \notional^2 \sum_i \gamma_i \condExpSmall{\expPower{-\alpha_i \cdot y_{\shortRate}(t,u)}}{t}
    = \notional^2 \sum_i \gamma_i \expPower{\half \alpha_i^2 \condVarSmall{y_{\shortRate}(t,u)}{t}}
    \rdef C_{\tradeVal}, \label{eq:swapSquared2App}
\end{align}
where constant $C_{\tradeVal}$ can be computed analytically.

\section{Moments of (truncated) normal random variables} \label{sec:moments}

Here, some results on the moments of (truncated) normal random variables are given, which are used in Section~\ref{sec:approxWWRIRSwap} for the IR swap WWR exposure approximation.
In particular, consider $\condExpSmall{y_{\shortRate}^l(t,u)}{t}$ and $\condExpSmall{ y_{\shortRate}^{l}(t,u)\indicator{y_{\shortRate}(t,u) \leq y_{\shortRate}^{*}(t,u)} }{t}$, where $y_{\shortRate}(t,u)\sim \N\left(0,\condVarSmall{y_{\shortRate}(t,u)}{t}\right)$ and $y_{\shortRate}^{*}(t,u)$ is computed numerically as explained in~\ref{app:derivationsMaxProb}.
The moments of a normal and generic truncated normal random variable are given in Result~\ref{res:normMoment} and~\ref{res:truncNormMoment}, respectively.

\begin{result} \label{res:normMoment}
Recall the result about the central moments of a normally distributed random variable $y_{\shortRate}(t,u)\sim \N\left(0,\condVarSmall{y_{\shortRate}(t,u)}{t}\right)$:
\begin{align} \displaystyle
  \moment_l
    &\ldef \condExpSmall{y_{\shortRate}^l(t,u)}{t}
    =  \left\{
  \begin{array}{ll}
    0, & l \text{ is odd,} \\
    (l-1)!! \left(\sqrt{\condVarSmall{y_{\shortRate}(t,u)}{t}}\right)^l,  & l \text{ is even,}
  \end{array}
  \right. \label{eq:expNormPow}
\end{align}
where $(\cdot)!!$ is the double factorial function.
\end{result}

\begin{result} \label{res:truncNormMoment}
Let $X\sim\N\left(\mu,\sigma^2\right)$ and $a,b\in[-\infty,\infty]$ where $a<b$.
Define $\momentTruncNorm_l \ldef \E \left[\left.X^l \right| a \leq X \leq b \right]$.
Then, there is a recursive formulation for $\momentTruncNorm_l$, with $\momentTruncNorm_{-1} = 0$ and $\momentTruncNorm_0 = 1$:
\begin{align}
  \momentTruncNorm_l = (l-1) \sigma^2 \momentTruncNorm_{l-2} + \mu \momentTruncNorm_{l-1} - \sigma \frac{b^{l-1} f_X(b) - a^{l-1} f_X(a) }{F_X(b) - F_X(a)}, \ \ l=1,2,\ldots, \label{eq:truncMom1}
\end{align}
where $f_X(x) = \normPDF\left(\frac{x-\mu}{\sigma}\right)$ and $F_X(x) = \normCDF\left(\frac{x-\mu}{\sigma}\right)$ are the normal PDF and CDF, respectively.

Furthermore, moments $\momentTruncNorm_l$ can be written as
\begin{align}
  \momentTruncNorm_l
    &= \E \left[\left.X^l \right| a \leq X \leq b \right]
    = \frac{\E\left[X^l \indicator{a \leq X \leq b} \right]}{\P\left(a \leq X \leq b\right)}
    = \frac{\E\left[X^l \indicator{a \leq X \leq b} \right]}{F_X(b) - F_X(a)}, \nonumber \\
  \Rightarrow \E\left[X^l \indicator{a \leq X \leq b} \right]
    &= \momentTruncNorm_l \left(F_X(b) - F_X(a)\right). \label{eq:mom1}
\end{align}

To conclude, apply this to $\condExpSmall{ y_{\shortRate}^{l}(t,u)\indicator{y_{\shortRate}(t,u) \leq y_{\shortRate}^{*}(t,u)} }{t}$, where $y_{\shortRate}(t,u)\sim \N\left(0,\condVarSmall{y_{\shortRate}(t,u)}{t}\right)$.
In terms of the above, $\mu = 0$ and $a=-\infty$, such that $F_{y_{\shortRate}}(a) = \normCDF\left(\frac{a}{\sigma}\right) = 0$ and $f_{y_{\shortRate}}(a) = \normPDF\left(\frac{a}{\sigma}\right) = 0$.
Now use Equations~\eqref{eq:truncMom1} and~\eqref{eq:mom1} to obtain the final result:
\begin{align}
  &\condExpSmall{ y_{\shortRate}^{l}(t,u)\indicator{y_{\shortRate}(t,u) \leq y_{\shortRate}^{*}(t,u)} }{t}
    = \momentTruncNorm_{l}   F_{y_{\shortRate}}\left(y_{\shortRate}^{*}(t,u)\right), \label{eq:truncNormMoment2a} \\
  &\momentTruncNorm_{l}
    = (l-1) \cdot \condVarSmall{y_r(t,u)}{t} \momentTruncNorm_{l-2} - \sqrt{\condVarSmall{y_r(t,u)}{t}} \frac{\left(y_{\shortRate}^{*}(t,u)\right)^{l-1} f_{y_\shortRate}\left(y_{\shortRate}^{*}(t,u)\right) }{F_{y_{\shortRate}}\left(y_{\shortRate}^{*}(t,u)\right)}. \label{eq:truncNormMoment2b}
\end{align}
\end{result}

\section{Extension for linear FX derivatives}\label{app:fxExtension}
\subsection{FX forward contract}
Consider a linear FX payoff, e.g., an FX forward contract with payoff at time $T$:
\begin{align} \displaystyle
 \tradeVal(T)
   &= \notional \left( \FX_f^d(T) - \strike \right). \nonumber
\end{align}
Value $\tradeVal$ is in the domestic currency $d$.
Define the FX forward $\FXForward_f^d(t, T)$, which is a martingale under the $T_d$ forward measure:
\begin{align} \displaystyle
 \FXForward_f^d(t, T)
   &\rdef \E_t^{T_d}\left[ \FX_f^d(T)\right]
   = \FX_f^d(t) \frac{\zcb_f(t,T)}{\zcb_d(t,T)}. \nonumber
\end{align}
Using standard risk-neutral pricing, and assuming unit notional, i.e., $\notional = 1$, write:
\begin{align} \displaystyle
 \tradeVal(t)
   &= \zcb_d(t,T) \left(\FXForward_f^d(t, T) - \strike\right)
   = \zcb_f(t,T) \FX_f^d(t) - \zcb_d(t,T)\strike. \nonumber
\end{align}

\subsection{$\left(\tradeVal(u)\right)^+$ in terms of stochastic processes}
Recall that in the current setup we can write:
\begin{align} \displaystyle
 \FX_f^d(u)
   &= \expPower{\mu_{\FX}(t,u) - \overline{y}_{\FX}(t,u)}, \nonumber \\
 \zcb_{d}(u,T)
   &= \expPower{A_{\shortRate_d}(u,T) - x_{\shortRate_d}(u)B_{\shortRate_d}(u,T) - \int_{u}^{T} b_{\shortRate_d}(v) \dv} \nonumber \\
   &= \expPower{\overline{A}_{\shortRate_d}(u,T) - \mu_{\shortRate_d}(t,u)B_{\shortRate_d}(u,T) - y_{\shortRate_d}(t,u)B_{\shortRate_d}(u,T)}, \nonumber \\
 \zcb_{f}(u,T)
   &= \expPower{\overline{A}_{\shortRate_f}(u,T) - \mu_{\shortRate_f}(t,u)B_{\shortRate_f}(u,T) - y_{\shortRate_f}(t,u)B_{\shortRate_f}(u,T)}. \nonumber
\end{align}

Using the above, the following expression for the FX forward is obtained:
\begin{align} \displaystyle
 \FXForward_f^d(u, T)
   &= \FX_f^d(u) \frac{\zcb_f(u,T)}{\zcb_d(u,T)} \nonumber \\
   &= \expPower{\mu_{\FX}(t,u) - \overline{y}_{\FX}(t,u)} \frac{\expPower{\overline{A}_{\shortRate_f}(u,T) - \mu_{\shortRate_f}(t,u)B_{\shortRate_f}(u,T) - y_{\shortRate_f}(t,u)B_{\shortRate_f}(u,T)}}{\expPower{\overline{A}_{\shortRate_d}(u,T) - \mu_{\shortRate_d}(t,u)B_{\shortRate_d}(u,T) - y_{\shortRate_d}(t,u)B_{\shortRate_d}(u,T)}} \nonumber \\
   &= \expPower{\mu_{\FX}(t,u) + \overline{A}_{\shortRate_f}(u,T)   - \mu_{\shortRate_f}(t,u)B_{\shortRate_f}(u,T) - \overline{A}_{\shortRate_d}(u,T) + \mu_{\shortRate_d}(t,u)B_{\shortRate_d}(u,T)}
   \expPower{- \overline{y}_{\FX}(t,u) - y_{\shortRate_f}(t,u)B_{\shortRate_f}(u,T) + y_{\shortRate_d}(t,u)B_{\shortRate_d}(u,T)} \nonumber \\
   &\rdef \Delta \cdot \expPower{- \overline{y}_{\FX}(t,u) - y_{\shortRate_f}(t,u)B_{\shortRate_f}(u,T) + y_{\shortRate_d}(t,u)B_{\shortRate_d}(u,T)}, \nonumber
\end{align}
where $\Delta > 0$ is defined such that equality holds, and contains all deterministic terms.

The next step is to write $\left(\tradeVal(u)\right)^+ = \tradeVal(u) \indicator{\tradeVal(u) \geq 0}$ in terms of stochastic processes $\overline{y}_{\FX}(t,u)$, $y_{\shortRate_d}(t,u)$ and $y_{\shortRate_f}(t,u)$.
Hence, first look at the case where $\tradeVal(u) \geq 0$:
\begin{align} \displaystyle
 &\tradeVal(u) \geq 0, \nonumber \\
 \Leftrightarrow &\FXForward_f^d(u, T) \geq \strike, \nonumber \\
 \Leftrightarrow & \expPower{- \overline{y}_{\FX}(t,u) - y_{\shortRate_f}(t,u)B_{\shortRate_f}(u,T) + y_{\shortRate_d}(t,u)B_{\shortRate_d}(u,T)} \geq \frac{\strike}{\Delta}, \nonumber \\
 \Leftrightarrow & - \overline{y}_{\FX}(t,u) - y_{\shortRate_f}(t,u)B_{\shortRate_f}(u,T) + y_{\shortRate_d}(t,u)B_{\shortRate_d}(u,T) \geq \ln \frac{\strike}{\Delta}, \nonumber \\
 \Leftrightarrow & y_{\FX}(t,u) - Y_{\shortRate_f}(t,u) - y_{\shortRate_f}(t,u)B_{\shortRate_f}(u,T) + Y_{\shortRate_d}(t,u) + y_{\shortRate_d}(t,u)B_{\shortRate_d}(u,T) \geq \ln \frac{\strike}{\Delta}. \nonumber
\end{align}

Using approximation $y_z(t,u)\approx  \corr_{\shortRate_d,z}\Sigma(y_z(t,u)) y_{\shortRate_d}(t,u)$, this can be rewritten as
\begin{align} \displaystyle
 &\tradeVal(u) \geq 0, \nonumber \\
 \Leftrightarrow & y_{\shortRate_d}(t,u)\underbrace{\left[\corr_{\shortRate_d,\FX}\Sigma(y_{\FX}(t,u))  - \corr_{\shortRate_d,\shortRate_f}\left[\Sigma(Y_{\shortRate_f}(t,u))  + B_{\shortRate_f}(u,T)\Sigma(y_{\shortRate_f}(t,u))\right]  + \Sigma(Y_{\shortRate_d}(t,u)) + B_{\shortRate_d}(u,T)\right]}_{\rdef \eta(t,u)} \geq \ln \frac{\strike}{\Delta}, \nonumber \\
 \Leftrightarrow & \left\{ \begin{matrix}
                              y_{\shortRate_d}(t,u) \geq \frac{1}{\eta(t,u)} \ln \frac{\strike}{\Delta}, & \eta(t,u) > 0, \\
                              y_{\shortRate_d}(t,u) \leq \frac{1}{\eta(t,u)} \ln \frac{\strike}{\Delta}, & \eta(t,u) < 0.
                            \end{matrix} \right.\nonumber
\end{align}
Here, $\eta(t,u)$ can only be zero if $\ln \frac{\strike}{\Delta} \leq 0 \ \Leftrightarrow \ \strike\leq \Delta$.
For simplicity, from now on assume that $\eta(t,u) \neq 0$.
Denoting $y_{\shortRate_d}^*(t,u) = \frac{1}{\eta(t,u)} \ln \frac{\strike}{\Delta}$ and combining with the above yields
\begin{align} \displaystyle
  \indicator{\tradeVal(u) \geq 0}
    &= \indicator{\eta(t,u) > 0} \indicator{y_{\shortRate_d}(t,u) \geq y_{\shortRate_d}^*(t,u)} + \indicator{\eta(t,u) < 0} \indicator{y_{\shortRate_d}(t,u) \leq y_{\shortRate_d}^*(t,u)} \nonumber \\
    &= \indicator{\eta(t,u) > 0} \indicator{y_{\shortRate_d}(t,u) \geq y_{\shortRate_d}^*(t,u)} + \indicator{\eta(t,u) < 0} \left( 1 - \indicator{y_{\shortRate_d}(t,u) \geq y_{\shortRate_d}^*(t,u)}\right) \nonumber \\
    &= \left(\indicator{\eta(t,u) > 0} - \indicator{\eta(t,u) < 0}\right) \indicator{y_{\shortRate_d}(t,u) \geq y_{\shortRate_d}^*(t,u)} + \indicator{\eta(t,u) < 0} \nonumber \\
    &= \sign{\eta(t,u)} \indicator{y_{\shortRate_d}(t,u) \geq y_{\shortRate_d}^*(t,u)} + \indicator{\eta(t,u) < 0}, \nonumber \\
  \sign{x}
    &= \left\{ \begin{array}{ll}
                 -1, & x < 0, \\
                 0, & x = 0, \\
                 1, & x > 0.
               \end{array} \right. \nonumber
\end{align}
This is a convenient formulation to work with, as $\eta(t,u)$ is deterministic and given in terms of the model parameters.

As a result, $\tradeVal(u)$ can be written as
\begin{align} \displaystyle
 \tradeVal(u)
   &= \zcb_f(u,T) \FX_f^d(u) - \zcb_d(u,T)\strike \nonumber \\
   &= \underbrace{\expPower{\mu_{\FX}(t,u) + \overline{A}_{\shortRate_f}(u,T) - \mu_{\shortRate_f}(t,u)B_{\shortRate_f}(u,T) }}_{\rdef w_1}
   \expPower{- \overline{y}_{\FX}(t,u) - y_{\shortRate_f}(t,u)B_{\shortRate_f}(u,T)} -
   \underbrace{\expPower{\overline{A}_{\shortRate_d}(u,T) - \mu_{\shortRate_d}(t,u)B_{\shortRate_d}(u,T)}\strike}_{\rdef w_2}
    \expPower{-  y_{\shortRate_d}(t,u)B_{\shortRate_d}(u,T)} \nonumber \\
   &= w_1 \expPower{Y_{\shortRate_d}(t,u) - Y_{\shortRate_f}(t,u) + y_{\FX}(t,u) - y_{\shortRate_f}(t,u)B_{\shortRate_f}(u,T)} - w_2 \expPower{-  y_{\shortRate_d}(t,u)B_{\shortRate_d}(u,T)} \nonumber \\
   &\approx  w_1 \cdot \expPower{-y_{\shortRate_d}(t,u)\left(-\eta(t,u) + B_{\shortRate_d}(u,T)\right)} - w_2 \expPower{-  y_{\shortRate_d}(t,u)B_{\shortRate_d}(u,T)} \nonumber \\
   &= w_1 \taylor\left(y_{\shortRate_d}(t,u)\left(-\eta(t,u) + B_{\shortRate_d}(u,T)\right)\right) - w_2 \cdot \taylor\left(  y_{\shortRate_d}(t,u)B_{\shortRate_d}(u,T)\right).  \nonumber
\end{align}
Thus, $\tradeVal(u)$ is written in terms of $y_{\shortRate_d}(t,u)$, where we used the proposed stochastic process approximation technique.
For non-pure linear IR derivatives, this approximation step will be necessary.

Finally, the previous derivations are used to obtain an expression for $\left(\tradeVal(u)\right)^+$ in terms of $y_{\shortRate_d}(t,u)$:
\begin{align} \displaystyle
 \left(\tradeVal(u)\right)^+
   &= \tradeVal(u) \indicator{\tradeVal(u) \geq 0} \nonumber \\
   &\approx \Big( w_1 \taylor\left(y_{\shortRate_d}(t,u)\left(-\eta(t,u) + B_{\shortRate_d}(u,T)\right)\right) - w_2 \cdot \taylor\left(  y_{\shortRate_d}(t,u)B_{\shortRate_d}(u,T)\right)\Big) \nonumber \\
   &\quad \cdot \left(\sign{\eta(t,u)} \indicator{y_{\shortRate_d}(t,u) \geq y_{\shortRate_d}^*(t,u)} + \indicator{\eta(t,u) < 0}\right). \nonumber
\end{align}

From this, the expressions $\condExpSmall{ y_{\shortRate}^{l}(t,u) \maxOperator{\tradeVal(u)} }{t}$ are again decomposed in terms of moments of a normal and truncated normal distribution.
The current example is for an FX forward contract, but this approach can easily be extended for all linear FX derivatives.
Recall that the current example has been illustrated for HW1F for foreign and domestic IR processes, and GBM for the FX process.

\section{Specific truncation error example} \label{app:approxErrorTruncSpecificForm}

Here, we apply the error bound given in Equation~\eqref{eq:genericErrorBoundC3a} to $\error{1}(x)$ from Equation~\eqref{eq:error1}.
This error term is chosen as this is the main driver of $\errorWWRPartOne$.
We split $\error{1}(x)$ as follows
\begin{align} \displaystyle
  \error{1}(x)
    &= H_{\shortRate,I,C}(t,u)\condExpSmall{\expPower{-Y_{\shortRate}(t,u)}\maxOperator{\tradeVal(u)} \taylorTrunc{2}{\infty}(Y_I(t,u) + Y_C(t,u)) \cdot x}{t} \nonumber \\
    &\quad -  H_{I,C}(t,u) \condExpSmall{\expPower{-\int_{t}^{u} \shortRate(v)\dv}\maxOperator{\tradeVal(u)}}{t} \condExpSmall{\taylorTrunc{2}{\infty}(Y_I(t,u) + Y_C(t,u)) \cdot x}{t}, \label{eq:error1Bound1}
\end{align}
where $x \in \{ 1, y_I\}$.

\paragraph{First term}
For the first expectation in Equation~\eqref{eq:error1Bound1}, we apply the bound from Equation~\eqref{eq:genericErrorBoundC3a} with $\overline{x} = \expPower{-Y_{\shortRate}(t,u)} \cdot x$, $n=n_{\intensity}=1$ and $Y_z(t,u) = Y_I(t,u) + Y_C(t,u)$.
Thus, we first need to look at the bound for expectation $\condExpSmall{\left(Y_z(t,u)\right)^{2(n+1)} \cdot \overline{x}^2}{t}$ from Equation~\eqref{eq:genericErrorBoundC1}.

For $Y_z(t,u) = Y_I(t,u) + Y_C(t,u)$, using the binomial formula $\displaystyle (a + b)^m = \sum_{k=0}^{m} \binom{m}{k} a^{m-k} b^k$, and the independence of $I$ and $C$, we write for $x \in \{ 1, y_I\}$:
\begin{align}\displaystyle
  \condExpSmall{\left(Y_z(t,u)\right)^{m} \cdot x}{t}
    &=  \sum_{k=0}^{m} \binom{m}{k} \condExpSmall{\left(Y_I(t,u)\right)^{m-k} \cdot x}{t} \condExpSmall{\left(Y_C(t,u)\right)^k}{t}
    \rdef C_2(m, x), \label{eq:error1Bound3c}
\end{align}
where for now we only need the case where $x=1$.

To compute $\condExpSmall{\overline{x}^m}{t} = \condExpSmall{\expPower{-mY_{\shortRate}(t,u)} x^m}{t}$, where $x \in \{ 1, y_I\}$, we apply the bound~\eqref{eq:corrDefBound}:
\begin{align}\displaystyle
  \condExpSmall{\overline{x}^m}{t}
    &\leq \sqrt{\condExpSmall{\expPower{-2mY_{\shortRate}(t,u)}}{t} - \left(\condExpSmall{\expPower{-mY_{\shortRate}(t,u)}}{t}\right)^2} \sqrt{\condExpSmall{x^{2m}}{t} - \left(\condExpSmall{x^m}{t}\right)^2} + \condExpSmall{\expPower{-mY_{\shortRate}(t,u)}}{t}\condExpSmall{x^m}{t} \nonumber \\
    &= \sqrt{\expPower{2m^2 \cdot \condVarSmall{Y_{\shortRate}(t,u)}{t}}- \expPower{m^2 \cdot \condVarSmall{Y_{\shortRate}(t,u)}{t}}} \sqrt{\condExpSmall{x^{2m}}{t} - \left(\condExpSmall{x^m}{t}\right)^2} + \expPower{\half m^2 \cdot \condVarSmall{Y_{\shortRate}(t,u)}{t}}\condExpSmall{x^m}{t} \nonumber \\
    &\rdef C_1(m,x) \geq 0. \label{eq:error1Bound3a}
\end{align}
For $x = 1$, the above holds with equality, so going forward we will use this inequality.
From Equation~\eqref{eq:error1Bound3a} we have $0 \leq \condExpSmall{\overline{x}^m}{t} \leq C_1(m,x)$ as all even moments are non-negative.
Using $\sqrt{a+b} \leq \sqrt{2} \sqrt{|a| + |b|}$, we can write
\begin{align}\displaystyle
  \sqrt{\condExpSmall{\overline{x}^4}{t} - \left(\condExpSmall{\overline{x}^2}{t}\right)^2}
    &\leq \sqrt{2} \sqrt{\condExpSmall{\overline{x}^4}{t} + \left(\condExpSmall{\overline{x}^2}{t}\right)^2}
    \leq \sqrt{2\left(C_1(4,x) + C_1^2(2,x)\right)}. \label{eq:error1Bound3b}
\end{align}

Using Equations~\eqref{eq:error1Bound3c} and~\eqref{eq:error1Bound3b}, the bound in Equation~\eqref{eq:genericErrorBoundC1} reads:
\begin{align}\displaystyle
  \condExpSmall{\left(Y_z(t,u)\right)^{2(n+1)} \cdot \overline{x}^2}{t}
    &\leq \sqrt{C_2(4(n+1),1) - C_2^2(2(n+1),1)} \sqrt{2\left(C_1(4,x) + C_1^2(2,x)\right)}\nonumber \\
    &\quad +  C_2(2(n+1),1) \cdot C_1(2,x).  \label{eq:error1Bound3e}
\end{align}

Now we go back to the first expectation in Equation~\eqref{eq:error1Bound1}, and use the bound from Equation~\eqref{eq:genericErrorBoundC3a}, together with the result from Equation~\eqref{eq:error1Bound3e} for $n=n_{\intensity}=1$:
\begin{align}\displaystyle
  &\left| \condExpSmall{\expPower{-Y_{\shortRate}(t,u)}\maxOperator{\tradeVal(u)} \taylorTrunc{2}{\infty}(Y_I(t,u) + Y_C(t,u)) \cdot x}{t}\right| \nonumber \\
    &\qquad \leq  \frac{\sqrt{C_{\tradeVal}} \cdot C_{T,2}}{2} \sqrt{\condExpSmall{\left(Y_I(t,u) + Y_C(t,u)\right)^{4} \cdot \left(\expPower{-Y_{\shortRate}(t,u)} x\right)^2}{t}} \nonumber \\
    &\qquad \leq \frac{\sqrt{C_{\tradeVal}} \cdot C_{T,2}}{2} \sqrt{\sqrt{C_2(8,1) - C_2^2(4,1)} \sqrt{2\left(C_1(4,x) + C_1^2(2,x)\right)}  + C_2(4,1) \cdot C_1(2,x)} \nonumber \\
    &\qquad \rdef \frac{\sqrt{C_{\tradeVal}} \cdot C_{T,2} \cdot C_3(x) }{2}, \label{eq:error1Bound4}
\end{align}
where $C_3(x)$ is defined such that equality holds.

\paragraph{Second term}
For the second expectation in Equation~\eqref{eq:error1Bound1}, i.e., $\condExpSmall{\taylorTrunc{2}{\infty}(Y_z(t,u)) \cdot x}{t}$, $Y_z(t,u) = Y_I(t,u) + Y_C(t,u)$, we can write the following using the definition of $\taylorTrunc{2}{\infty}(\cdot)$, the independence of $Y_I(t,u)$ and $Y_C(t,u)$ and the result from Equation~\eqref{eq:error1Bound3c} for $x \in \{ 1, y_I\}$:
\begin{align}\displaystyle
  \condExpSmall{\taylorTrunc{2}{\infty}(Y_z(t,u)) \cdot x}{t}
    &= \sum_{i=2}^{\infty}\frac{(-1)^i}{i!} \condExpSmall{ (Y_z(t,u))^i\cdot x}{t}
    =  \sum_{i=2}^{\infty}\frac{(-1)^i}{i!}  C_2(i, x)
    \rdef C_4(x). \label{eq:error1Bound1c}
\end{align}

\paragraph{Overall error bound}
Finally, we combine the results from Equations~(\ref{eq:error1Bound4}--\ref{eq:error1Bound1c}) to write the following expression for $\error{1}(x)$ from Equation~\eqref{eq:error1Bound1}, where $x \in \{ 1, y_I\}$:
\begin{align} \displaystyle
  \error{1}(x)
    &\leq  H_{\shortRate,I,C}(t,u)\frac{\sqrt{C_{\tradeVal}} \cdot C_{T,2} \cdot C_3(x)}{2}  -  H_{I,C}(t,u)  \condExpSmall{\expPower{-\int_{t}^{u} \shortRate(v)\dv}\maxOperator{\tradeVal(u)}}{t} \cdot C_4(x) , \label{eq:error1Bound5}
\end{align}
Constants $C_3(x)$ and $C_4(x)$, which are in turn functions of $C_1(m,x)$ and $C_2(m,x)$, are finite if variance $\condVarSmall{Y_{\shortRate}(t,u)}{t}$ and higher order (cross)moments $\condExpSmall{\left(y_I(t,u)\right)^k}{t}$, $\condExpSmall{\left(Y_I(t,u)\right)^k}{t}$, $\condExpSmall{\left(Y_C(t,u)\right)^k}{t}$ and $\condExpSmall{  (Y_I(t,u))^{k} \cdot y_I(t,u)}{t}$ are finite.

\end{document}